\documentclass[twocolumn,english,superscriptaddress,amsmath,amssymb,floats,prb]{revtex4-2}
\usepackage[latin9]{inputenc}
\usepackage{color}
\usepackage{mathtools}
\usepackage{amsmath}
\usepackage{graphicx}
\usepackage{multirow}
\usepackage{array}
\usepackage{xcolor,colortbl}
\usepackage[normalem]{ulem}

\makeatletter

\usepackage{comment}

\@ifundefined{textcolor}{}{%
 \definecolor{BLACK}{gray}{0}
 \definecolor{WHITE}{gray}{1}
 \definecolor{RED}{rgb}{1,0,0}
 \definecolor{GREEN}{rgb}{0,1,0}
 \definecolor{BLUE}{rgb}{0,0,1}
 \definecolor{CYAN}{cmyk}{1,0,0,0}
 \definecolor{MAGENTA}{cmyk}{0,1,0,0}
 \definecolor{YELLOW}{cmyk}{0,0,1,0}
}

\definecolor{mcolor}{rgb}{0.89,0.17,0.18}
\definecolor{lcolor}{rgb}{0.34, 0.74, 0.34}
\definecolor{mmmpcolor}{rgb}{0.87, 0.73, 0.87}
\definecolor{mmmmcolor}{rgb}{0.545, 0, 0.545}
\definecolor{lllcolor}{rgb}{0.23, 0.48, 0.47}
\definecolor{mmmlllposcolor}{rgb}{0.2, 0.28, 0.89}
\definecolor{mmmlllnegcolor}{rgb}{0.5, 0.58, 0.89}
\definecolor{mllposcolor}{rgb}{0.94, 0.63, 0.14}
\definecolor{mllnegcolor}{rgb}{0.627, 0.32, 0.176}
\definecolor{m1m23l2color}{rgb}{0.2, 0.32, 0.16}
\definecolor{m1m23l1l23color}{rgb}{0.60,0.47,0.26}
\definecolor{m1l1l2color}{rgb}{0.52,0.55,0.29}

\usepackage{epstopdf}
\usepackage{array}
\usepackage{mathrsfs}
\usepackage{dcolumn}
\usepackage{bm}

\usepackage{xcolor}
\newcommand\crule[3][black]{\textcolor{#1}{\rule{#2}{#3}}}
\newcolumntype{M}[1]{>{\centering\arraybackslash}m{#1}}
\newcolumntype{P}[1]{>{\centering\arraybackslash}p{#1}}

\newcolumntype{L}[1]{>{\raggedright\let\newline\\\arraybackslash\hspace{0pt}}m{#1}}
\newcolumntype{C}[1]{>{\centering\let\newline\\\arraybackslash\hspace{0pt}}m{#1}}
\newcolumntype{R}[1]{>{\raggedleft\let\newline\\\arraybackslash\hspace{0pt}}m{#1}}

\graphicspath{{../Inkscape/},{../Mathematica/},{./Figures/}}

\newcommand{\mbf}[1]{\mathbf{#1}}

\usepackage{tikz}
\usetikzlibrary{calc,intersections}
\usetikzlibrary{shadings}
\usepgflibrary{shadings}

\newcommand{\timessmall}{{\mkern-2mu\times\mkern-2mu}}

\usepackage{babel}

\usepackage{babel}

\makeatother

\usepackage{babel}
\begin{document}

\title{Theory of the charge-density wave in $A$V$_3$Sb$_5$ kagome metals}

\author{Morten H. Christensen}
\email{mchriste@nbi.ku.dk}
\affiliation{Niels Bohr Institute, University of Copenhagen, 2100 Copenhagen, Denmark}

\author{Turan Birol}
\affiliation{Department of Chemical Engineering and Materials Science, University of Minnesota, MN 55455, USA}

\author{Brian M. Andersen}
\affiliation{Niels Bohr Institute, University of Copenhagen, 2100 Copenhagen, Denmark}

\author{Rafael M. Fernandes}
\affiliation{School of Physics and Astronomy, University of Minnesota, Minneapolis,
MN 55455, USA}

\date{\today}
\begin{abstract}

The family of metallic kagome compounds $A$V$_3$Sb$_5$ ($A$=K, Rb, Cs) was recently discovered to exhibit both superconductivity and charge order. The nature of the charge-density wave (CDW) phase is presently unsettled, which complicates the interpretation of the superconducting ground state. In this paper, we use group-theory and density-functional theory (DFT) to derive and solve a phenomenological Landau model for this CDW state. The DFT results reveal three unstable phonon modes with the same in-plane momentum but different out-of-plane momenta, whose frequencies depend strongly on the electronic temperature. This is indicative of an electronically-driven CDW, stabilized by features of the in-plane electronic dispersion. Motivated by the DFT analysis, we construct a Landau free-energy expansion for coupled CDW order parameters with wave-vectors at the $M$ and $L$ points of the hexagonal Brillouin zone. We find an unusual trilinear term coupling these different order parameters, which can promote the simultaneous condensation of both CDWs even if the two modes are not nearly-degenerate. We classify the different types of coupled multi-$\bf{Q}$ CDW orders, focusing on those that break the sixfold rotational symmetry and lead to a unit-cell doubling along all three crystallographic directions, as suggested by experiments. We determine a region in parameter space, characterized by large nonlinear Landau coefficients, where these phases -- dubbed staggered tri-hexagonal and staggered Star-of-David -- are the leading instabilities of the system. Finally, we discuss the implications of our results for the kagome metals.
\end{abstract}
\maketitle

\section{Introduction}

The kagome lattice offers a powerful framework to investigate several intriguing physical phenomena, such as frustrated magnetism and spin liquids~\cite{Balents2010}, flat bands~\cite{Kang2019,Kang2020,Meier2020}, and Dirac points~\cite{Guo2009,Mazin2014}. An exotic chiral $d$-wave superconducting phase promoted by van Hove singularities has also been predicted for certain carrier concentrations ~\cite{Nandkishore2012,Kiesel2012,Kiesel2013,Nandkishore2014}, providing a possible route to realize an intrinsic topological superconductor. The synthesis and subsequent discovery of superconductivity in a family of metallic kagome compounds, $A$V$_3$Sb$_5$ ($A$=K, Rb, Cs), provides the opportunity of potentially realizing this prediction in a real material~\cite{Ortiz2019,Ortiz2020,Ortiz2021,Yin2021}. The superconducting critical temperature varies between $T_c \sim 1-3$~K depending on the alkali metal and can be increased upon either hole doping~\cite{Song2021} or application of pressure~\cite{Du2021,Zhao2021a,Zhang2021_pressure,Chen_pressure_2021,Zhu_pressure_2021}. In addition, near $T_{\rm CDW} \sim 80-100$~K, these materials exhibit a kink in the specific heat~\cite{Ortiz2019} that has been widely interpreted as a signature of a charge-density wave (CDW) ~\cite{Ortiz2019,Ortiz2020,Jiang2020,Yin2021,Uykur_2021,Zhou_HHWen_2021}. This ordered state is also observed in scanning tunneling microscopy (STM) measurements at low temperatures, which report static charge modulations consistent with a doubling of the unit cell along both the $a$- and $b$-directions~\cite{Jiang2020,Zhao2021b,Liang2021,Chen2021}. 
It is also supported by x-ray diffraction which, in addition to the doubling along the $a$- and $b$-directions, find a unit cell increase along the $c$-direction~\cite{Jiang2020,Li2021_observation,Ortiz2021b}. While some experiments report a simple doubling along the $c$-direction~\cite{Liang2021}, corresponding to $2 \times 2 \times 2$ order, a recent work reported a structure consistent with a $2 \times 2 \times 4$ increase of the unit cell~\cite{Ortiz2021b}. Several spectroscopic probes have also reported evidence for the three-dimensionality of the CDW order~\cite{Liang2021,Luo2021}. Finally, recent transport measurements under uniaxial strain indicates that the CDW state competes strongly with superconductivity~\cite{Ni2021}.

The nature of the CDW state is currently under intense scrutiny. Since superconductivity sets in at much lower temperatures than charge order, elucidating the symmetries and properties of the CDW phase is essential to understand e.g. the symmetry of the superconducting order parameter. In this regard, while there is spectroscopic evidence for both Star-of-David and tri-hexagonal (also known as inverse Star-of-David) configurations in the plane~\cite{Jiang2020,Zhao2021b,Li2021_rotation} (see Fig.~\ref{fig:distortionpatterns}), the character of the inter-layer modulation responsible for the unit-cell doubling along the $c$-direction remains under debate. The observation of threefold rotational symmetry-breaking by STM experiments~\cite{Jiang2020,Zhao2021b,Chen2021,Li2021_rotation} offers important clues about the three-dimensional character of the CDW pattern. In particular, recent density functional theory (DFT) calculations~\cite{Tan2021} and coherent phonon spectroscopy~\cite{Ratcliff2021} suggest that the system has unstable phonon modes at both the $M$ and $L$ points of the hexagonal Brillouin zone [BZ, illustrated in Fig.~\ref{fig:distortionpatterns}(a)]. Under these conditions, the only way to obtain a charge-ordered state that doubles the unit cell in every direction \emph{and} lowers the sixfold rotational symmetry to twofold is by a combination of wave-vectors from both the $M$ and $L$ points. Indeed, DFT analyses find that the energy of the system is minimized by a configuration that intertwines wave-vectors from both the $M$ and $L$ points~\cite{Tan2021,Ratcliff2021}. We note that STM data has also been interpreted in terms of a chiral charge order~\cite{Jiang2020,Shumiya2021,Setty2021} whereas $\mu$SR experiments have been interpreted in favor of time-reversal symmetry-breaking associated with orbital currents in the CDW state~\cite{Mielke2021}.

The condensation of CDW order parameters associated with two different BZ momenta, $M$ and $L$, which is necessary to explain the $2 \times 2 \times 2$ unit-cell expansion and the breaking of sixfold rotational symmetry, raises several important questions. Why would two different types of CDW order condense? Do the two order parameters onset simultaneously or at two different temperatures? What is the mechanism responsible for the intertwining of these CDW orders? In this paper, we employ a phenomenological approach, combined with DFT calculations, to address these issues. Our DFT analysis reveals a strong dependence of the unstable $M$ and $L$ phonon frequencies on Fermi surface smearing -- a proxy of the electronic temperature -- which is indicative of an electronic rather than a structural mechanism for the formation of the charge order. We also find another unstable mode along the $U$-line connecting the $M$ and $L$ momenta of the BZ, see Fig.~\ref{fig:distortionpatterns}(a). The fact that three charge-order configurations with different $c$-axis periodicity but identical in-plane periodicity are viable instabilities suggests that the electronic mechanism is dominated by in-plane processes. Such a mechanism could be connected, for example, to the van Hove singularities of the in-plane electronic dispersion of the kagome lattice, as found by a recent renormalization group calculation~\cite{Park2021}. 

For the phenomenological analysis, we write down and minimize the most general Landau free-energy expansion for the CDW order parameters associated with the $M$ and $L$ wave-vectors. Generally, we find that a direct transition to a coupled $M$-$L$ state requires a trilinear coupling that compensates the energy difference between the two ``pure" states. This opens the possibility of a coupled state appearing even if the ``pure" instabilities are not nearly-degenerate. Over a wide range of parameters, the leading coupled-state instability is the superimposed tri-hexagonal Star-of-David charge order, which corresponds to a triple-$\mathbf{Q}_M$/triple-$\mathbf{Q}_L$ order that does not break sixfold rotational symmetry. The latter is broken by the single-$\mathbf{Q}_M$/double-$\mathbf{Q}_L$ orders dubbed staggered tri-hexagonal and staggered Star-of-David orders. These orders are illustrated in Fig.~\ref{fig:mixed_order_examples}. For most of the parameter space analyzed here, these two states are not the leading instabilities of the system, but onset at temperatures below the superimposed tri-hexagonal Star-of-David order or below the pure triple-$\mathbf{Q}_M$ planar Star-of-David (or planar tri-hexagonal, see Fig.~\ref{fig:order_examples_uncoupled}) order. Only when the nonlinear quartic terms are large enough we find a direct, single phase transition to the single-$\mathbf{Q}_M$/double-$\mathbf{Q}_L$ orders that break sixfold rotational symmetry and results in a $2\times 2 \times 2$ increase of the unit cell. Our results point to the key role played by the nonlinear couplings between the $M$ and $L$ CDWs in the $A$V$_3$Sb$_5$ kagome metals, and call for further experimental studies to determine whether a single or multiple charge-order transitions are realized in these compounds.

The paper is organized as follows: We present our DFT analysis in Sec.~\ref{sec:DFT}, and in Sec.~\ref{sec:free_energy} we introduce and motivate the Landau free energy describing a system of coupled CDW orders. The uncoupled free energies are minimized in Secs.~\ref{sec:only_M} and \ref{sec:only_L}. In Sec.~\ref{sec:phase_diagram} we minimize the fully coupled free energy and elucidate the various phases that emerge. Sec.~\ref{sec:conclusions} is devoted to the conclusions.

\section{Phonon instabilities in $\mathbf{\emph{A}V_3Sb_5}$: DFT and symmetry analysis}
\label{sec:DFT}

We start by identifying the unstable lattice modes associated with the CDW transition. We performed lattice response calculations in CsV$_3$Sb$_5$ using DFT as implemented in the Vienna Ab Initio Simulation package~\cite{Kresse1993,Kresse1996CMS,Kresse1996PRB}. We employed the Perdew-Burke-Ernzerhof exchange correlation functional~\cite{Perdew2008}, and used unshifted k-grids with a density of a point per $\sim0.012\times 2\pi$\AA$^{-1}$. The plane wave energy cutoff was set to 350~eV and a Gaussian smearing scheme was used for the electronic occupations.

\begin{figure}
\includegraphics[width=0.75\columnwidth]{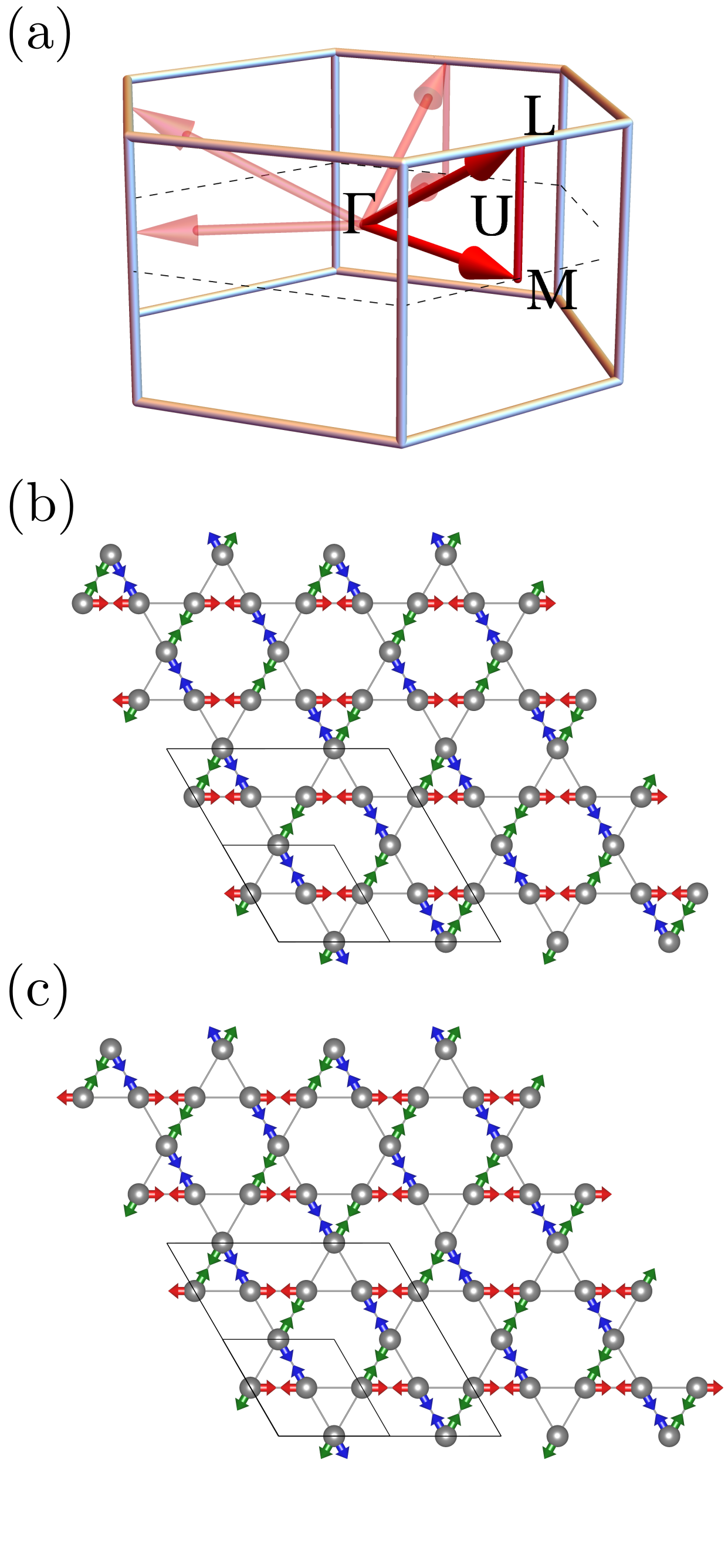}
\caption{\label{fig:distortionpatterns} (a) Illustration of the three-dimensional hexagonal Brillouin zone (BZ) corresponding to the space group $P6/mmm$ with the $M$ and $L$ points highlighted, as well as the $U$ line connecting them. The other two wave-vectors in the stars of $M$ and $L$ are obtained by three-fold rotations. (b)-(c) The displacement pattern of the V ions on a kagome layer according to the unstable $M_1^+$ and $L_2^-$ modes. Both of these modes lead to similar distortions on a single kagome layer. Different components ($M_1$, $M_2$, and $M_3$) are shown with arrows of different colors. Here (b) corresponds to the bond displacements of the tri-hexagonal charge order, described by  $|M_1| = |M_2| = |M_3|$ and $M_1 M_2 M_3 > 0$. In (c) the bond displacements that give rise to the Star-of-David order are shown. In this case, $|M_1| = |M_2| = |M_3|$ but $M_1 M_2 M_3 < 0$.}
\end{figure}

In agreement with earlier results in the literature~\cite{Ratcliff2021}, our DFT calculations find two unstable phonon modes that transform as the $M_1^+$ and $L_2^-$ irreducible representations (irreps) of the space group $P6/mmm$ (\#191). Formally, $M_1^{+}$ and $L_2^{-}$ are one-dimensional irreps of the little group of the wave-vectors $M$ and $L$. However, there are three vectors in the stars of both the $M$ and $L$ points, as shown in Fig.~\ref{fig:distortionpatterns}(a). The vectors in the star of $M$ are
\begin{align}
    \mbf{Q}_M^{(1)}&=(\tfrac{1}{2} \ 0 \ 0)\,, \
    \mbf{Q}_M^{(2)}=(0 \ \tfrac{1}{2} \ 0)\,, \
    \mbf{Q}_M^{(3)}=(-\tfrac{1}{2} \ -\tfrac{1}{2} \ 0)\,, \label{eq:M_Q_vectors}
\end{align}
and those in the star of $L$ are
\begin{align}
    \mbf{Q}_L^{(1)}&=(\tfrac{1}{2} \ 0 \ \tfrac{1}{2})\,, \
    \mbf{Q}_L^{(2)}=(0 \ \tfrac{1}{2} \ \tfrac{1}{2})\,, \
    \mbf{Q}_L^{(3)}=(-\tfrac{1}{2} \ -\tfrac{1}{2} \ \tfrac{1}{2})\,, \label{eq:L_Q_vectors}
\end{align}
in the basis ($\mbf{G}_1$, $\mbf{G}_2$, $\mbf{G}_3$) with
\begin{equation}
    \mbf{G}_1 = \frac{2\pi}{a}\begin{pmatrix}
        1 \\
        -\frac{1}{\sqrt{3}} \\
        0
    \end{pmatrix},\,
    \mbf{G}_2 = \frac{2\pi}{a}\begin{pmatrix}
        0 \\
        \frac{2}{\sqrt{3}} \\
        0
    \end{pmatrix},\,
    \mbf{G}_3 = \frac{2\pi}{c}\begin{pmatrix}
        0 \\
        0 \\
        1
    \end{pmatrix}\,.
\end{equation}
Hence, for the $M_1^{+}$ irrep, we denote the components of the CDW order parameter with different wave-vectors in the star by $M_i$, which corresponds to a displacement with wave-vector $\mbf{Q}_M^{(i)}$ ($i=1,2,3$). A similar notation is defined for $L_i$ and $\mbf{Q}_L^{(i)}$.

In real space, the in-plane displacements of the V atoms, see Figs.~\ref{fig:distortionpatterns}(b) and \ref{fig:distortionpatterns}(c), account for more than 90\% of the total displacements associated with the $M$ and $L$ modes. The only significant difference between them is in the relative phase of the displacements in neighboring V layers along the $c$ direction. For the $M$ mode, the V atoms are displaced in-phase between the layers, while for the $L$ mode, the displacement is out-of-phase between the layers. 

Figs.~\ref{fig:distortionpatterns}(b) and \ref{fig:distortionpatterns}(c)
show the in-plane displacement patterns of the V atoms corresponding to the three wave-vectors of the star of either the $M$ or the $L$ point. The red, green, and blue displacement patterns denote different periodicities corresponding to each of the three distinct order parameters $M_i$ (or $L_i$). Note that the predominant displacement is a shortening of certain nearest neighbor V--V bonds. As a result, we associate the CDW ordered states to a pattern of shorter V--V bonds, i.e. a bond-order pattern. 

For an isolated layer, the equal-weight superposition of the three types of in-plane bond-order can give rise to two distinct sixfold-symmetric patterns. If all V--V bond displacements have the same phase, or more generally for any ground state with $|M_1| = |M_2| = |M_3|$ and $M_1 M_2 M_3 > 0$, the resulting pattern is the tri-hexagonal (or inverse Star-of-David) bond-order configuration shown in Fig. \ref{fig:distortionpatterns}(b). In this state, there are short and long V--V bond-loops forming triangles and hexagons. On the other hand, if we shift the phase of one of the three bond displacements by $\pi$, or more generally for any ground state with $|M_1| = |M_2| = |M_3|$ and $M_1 M_2 M_3 < 0$, the resulting configuration is the so-called Star-of-David bond-order pattern of Fig. \ref{fig:distortionpatterns}(c). In this state, twelve V atoms form a bond-loop in the shape of a six-pointed star. We note that even though the Star-of-David and the inverse Star-of-David phases have different bond-order patterns, these phases break the same symmetries of the high temperature structure. In other words, their space groups and unit cells are identical.

Going beyond an isolated layer, the bond pattern in consecutive layers depends on whether the wave-vector is in the star of $M$ or $L$. In the former case, the triple-$\mathbf{Q}_M$ bond patterns of Figs. \ref{fig:distortionpatterns}(b)-(c) are the same for all layers. We refer to these two different states as planar tri-hexagonal and planar Star-of-David, respectively. On the other hand, in the case of a triple-$\mathbf{Q}_L$ bond-order, consecutive layers will alternate between the tri-hexagonal and Star-of-David patterns, regardless of the relative phases between the $L_i$ order parameters. Therefore, there is only one triple-$\mathbf{Q}_L$ state that we dub alternating tri-hexagonal Star-of-David state (see Fig.~\ref{fig:order_examples_uncoupled}). We will further discuss this type of bond-order in Sec.~\ref{sec:only_L}. Note that any combination of the order parameters $M_i$ can be represented in terms of a $2\times2\times 1$ supercell, which is shown in the lower left corner of Fig.~\ref{fig:distortionpatterns}(b), whereas combinations of $L_i$ require a $2\times2 \times 2$ supercell.

\begin{figure}
\includegraphics[width=0.9\columnwidth]{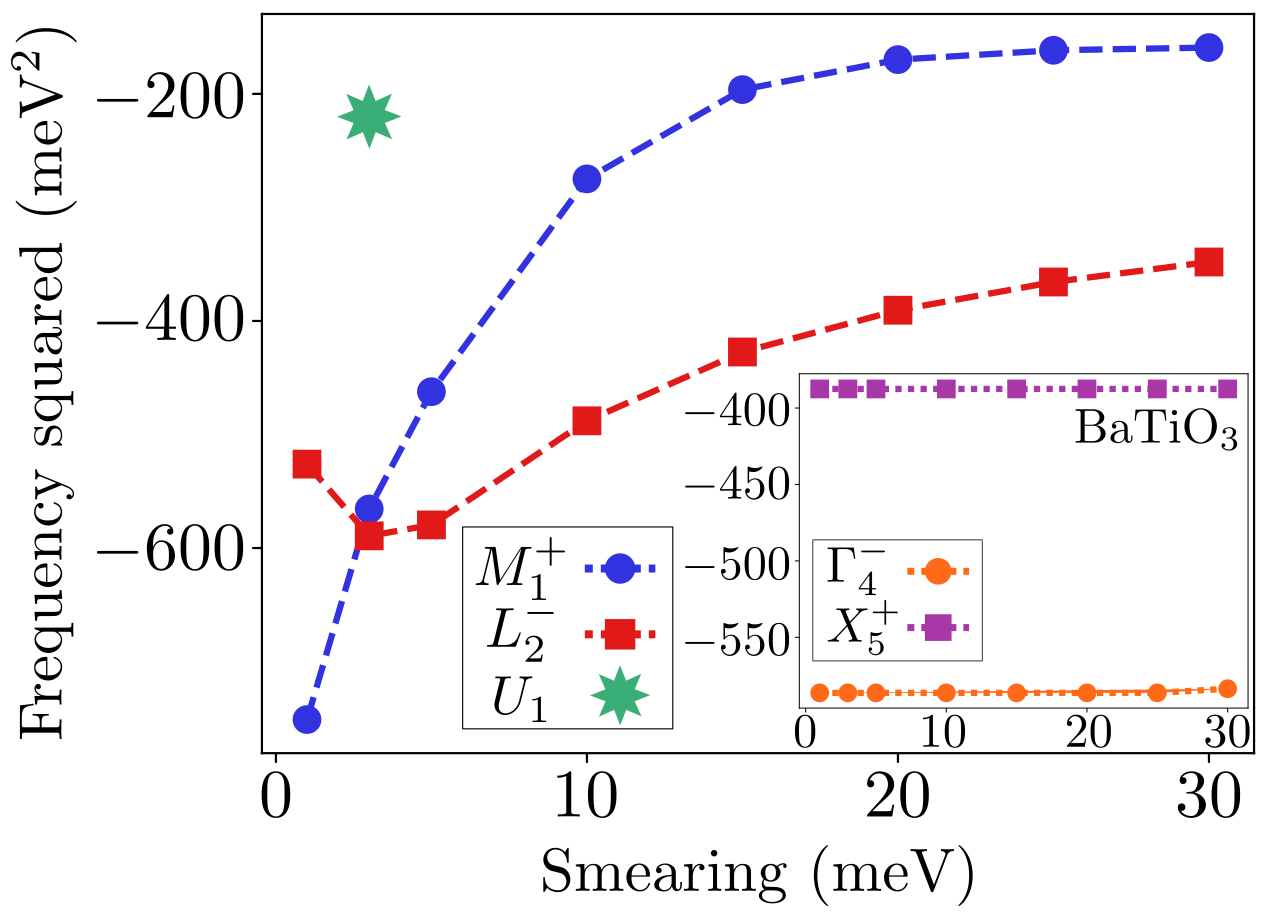}
\caption{\label{fig:dft_phonons} Squared frequency of the unstable phonon modes of CsV$_3$Sb$_5$ as a function of Fermi surface smearing (in meV), which is a proxy for the electronic temperature. The inset shows the unstable $\Gamma$ and $X$ point phonons in cubic insulating BaTiO$_3$ as a function of electronic smearing. In such insulators with large band gaps, thermal smearing has no effect on the phonon frequencies. In contrast, for the kagome material, there is a significant change of the frequency as the electronic temperature increases.}
\end{figure}

While no phonon instabilities have been reported at other high symmetry points of the BZ, our DFT calculations also found an instability on the $U$ line that connects the $M$ and $L$ points [see Fig.~\ref{fig:distortionpatterns}(a)]. A point on this line is parametrized by $\mbf{Q}_U^{(3)}=(-\tfrac{1}{2} \ -\tfrac{1}{2} \ q_z)$ (and its three-fold symmetric partners), with $q_z=0$ corresponding to the $M$ point and $q_z=\tfrac{1}{2}$, to the $L$ point. The corresponding bond-order patterns are generally incommensurate along the $c$-axis. For our DFT calculations, we considered the commensurate case $q_z=\tfrac{1}{4}$, resulting in a CDW state with a $2\times2 \times 4$ unit cell. It is interesting to note that such a periodicity has been proposed in recent experiments~\cite{Ortiz2021b}. The fact that the three instabilities uncovered here have the same in-plane wave-vector suggests that the driving force behind them are not the phonons themselves, but their interaction with the electronic degrees of freedom -- and, in particular, the in-plane electronic dispersion, which may display van Hove singularities~\cite{Ortiz2021b,Cho2021,Kang2021,Hu2021}. This is supported by the fact that the displacements associated with the $M$ and $L$ phonon modes occur primarily on the V atoms, which are the dominant contributors to the density of states at the Fermi level.

To further examine the effects of the electronic degrees of freedom on the lattice instabilities, we repeat our DFT calculations as a function of Fermi surface smearing. Fermi-surface induced lattice instabilities depend sensitively on the electronic temperature of the system. While the DFT calculations are in principle performed at zero temperature, it is possible to consider Fermi surface smearing when calculating the occupations of the electronic energy levels. We employ a Gaussian smearing scheme, where the occupation probability of a state depends on a Gaussian function of its energy with respect to the Fermi level. The width of this Gaussian, $\sigma$, is a proxy for the electronic temperature. Consequently, electronically-driven instabilities are expected to weaken or disappear as $\sigma$ is increased. 

To assess the evolution of the strength of phonon instabilities as the electronic temperature is changed, in Fig.~\ref{fig:dft_phonons} we present the square of the $M_1^{+}$ and $L_2^{-}$ phonon frequencies as function of the smearing width $\sigma$. Because unstable phonon modes have imaginary-valued frequencies, the smaller the absolute value of the squared frequency is, the weaker the corresponding instability is expected to be. Consistent with an instability for which the electrons play a substantial role, the absolute value of the squared frequencies of the unstable phonon modes decrease significantly when the electronic temperature is increased, particularly for the $M_1^{+}$ mode. In the same plot, the unstable $U_1$ phonon mode, which was computed for a single smearing value, is shown by the green symbol. As a comparison, in the inset of Fig.~\ref{fig:dft_phonons} we plot the squared frequency of the unstable phonon modes (labeled $\Gamma_4^{-}$ and $X_5^{+}$) of the insulator BaTiO$_3$ as a function of Fermi-surface smearing. In this case, where the phonon instability is not related to a Fermi-surface effect, the frequencies barely change as the electronic temperature increases.

\section{Landau free-energy expansion for CDW order}\label{sec:free_energy}

\begin{figure}
\includegraphics[width=0.48\columnwidth]{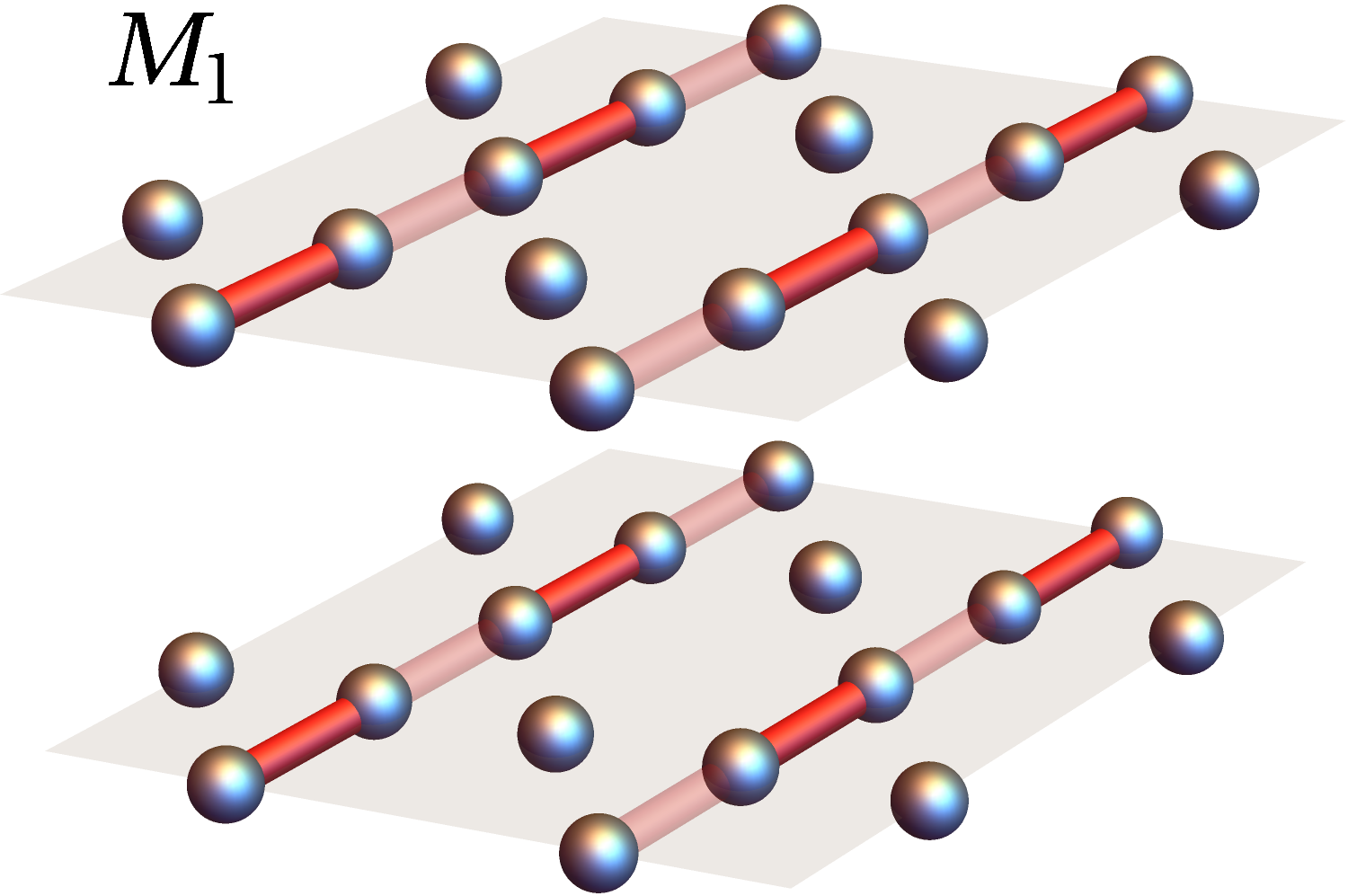}\includegraphics[width=0.48\columnwidth]{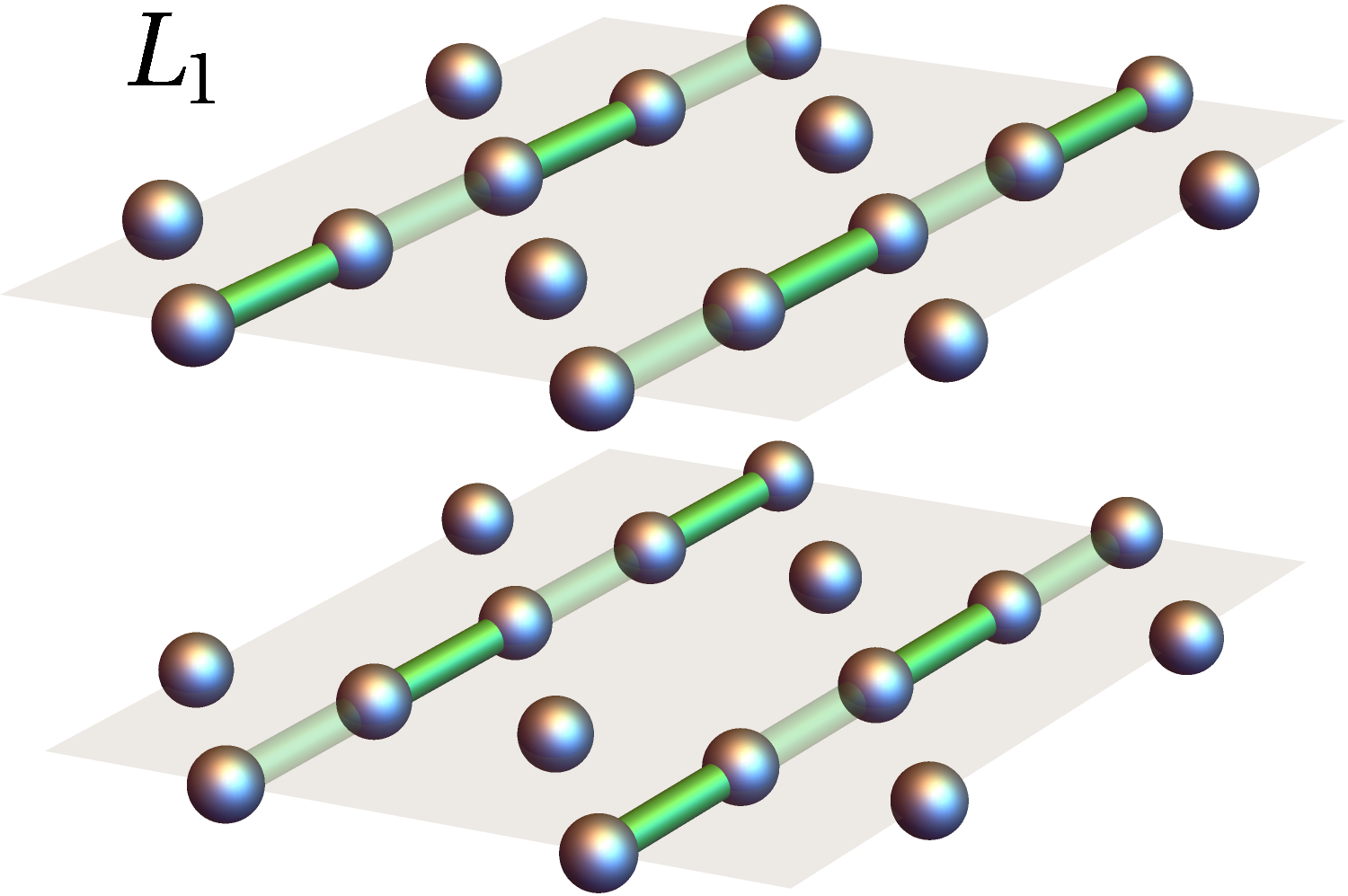}\\
\vspace{6mm}
\includegraphics[width=0.48\columnwidth]{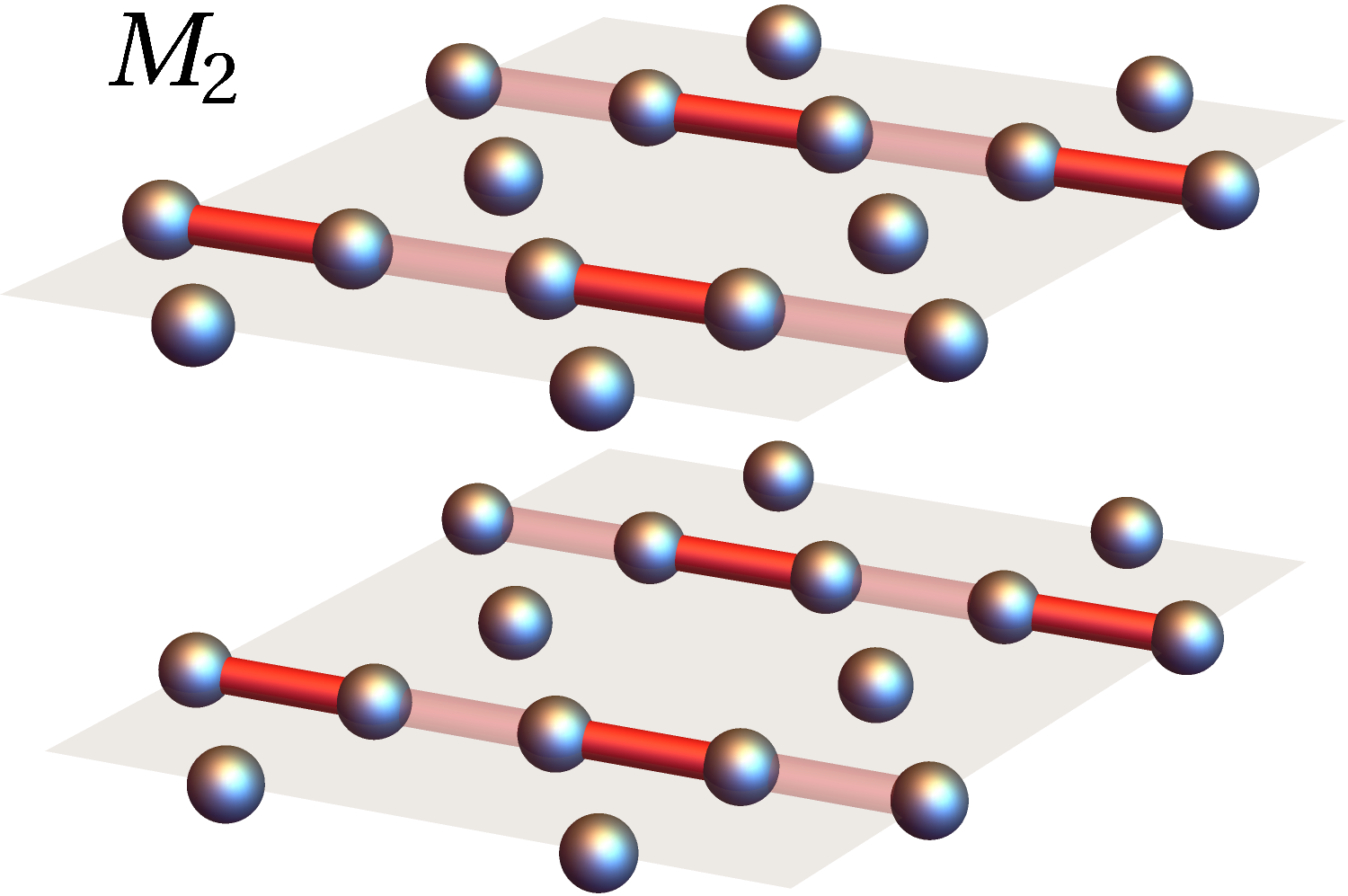}\includegraphics[width=0.48\columnwidth]{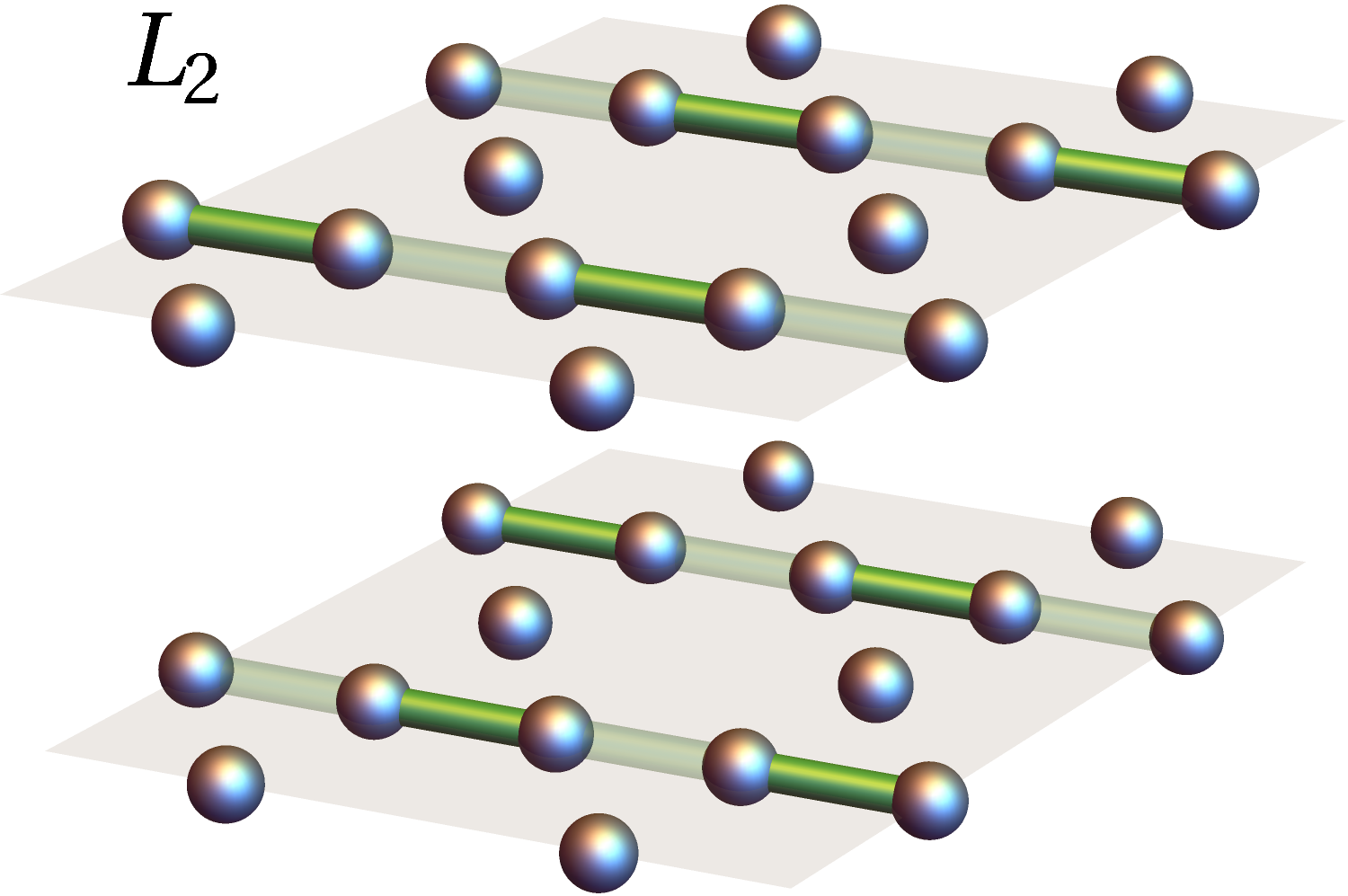}\\
\vspace{6mm}
\includegraphics[width=0.48\columnwidth]{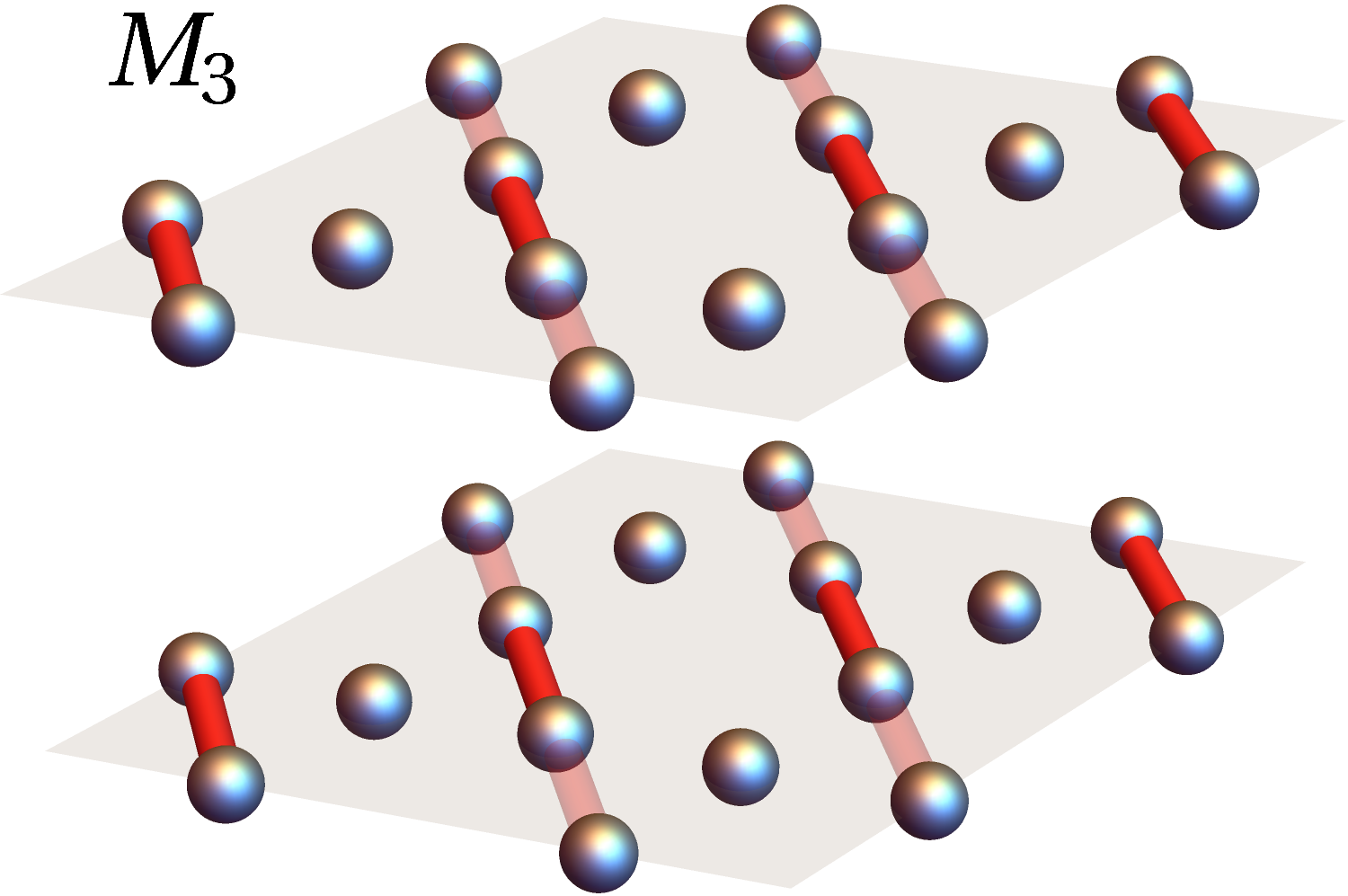}\includegraphics[width=0.48\columnwidth]{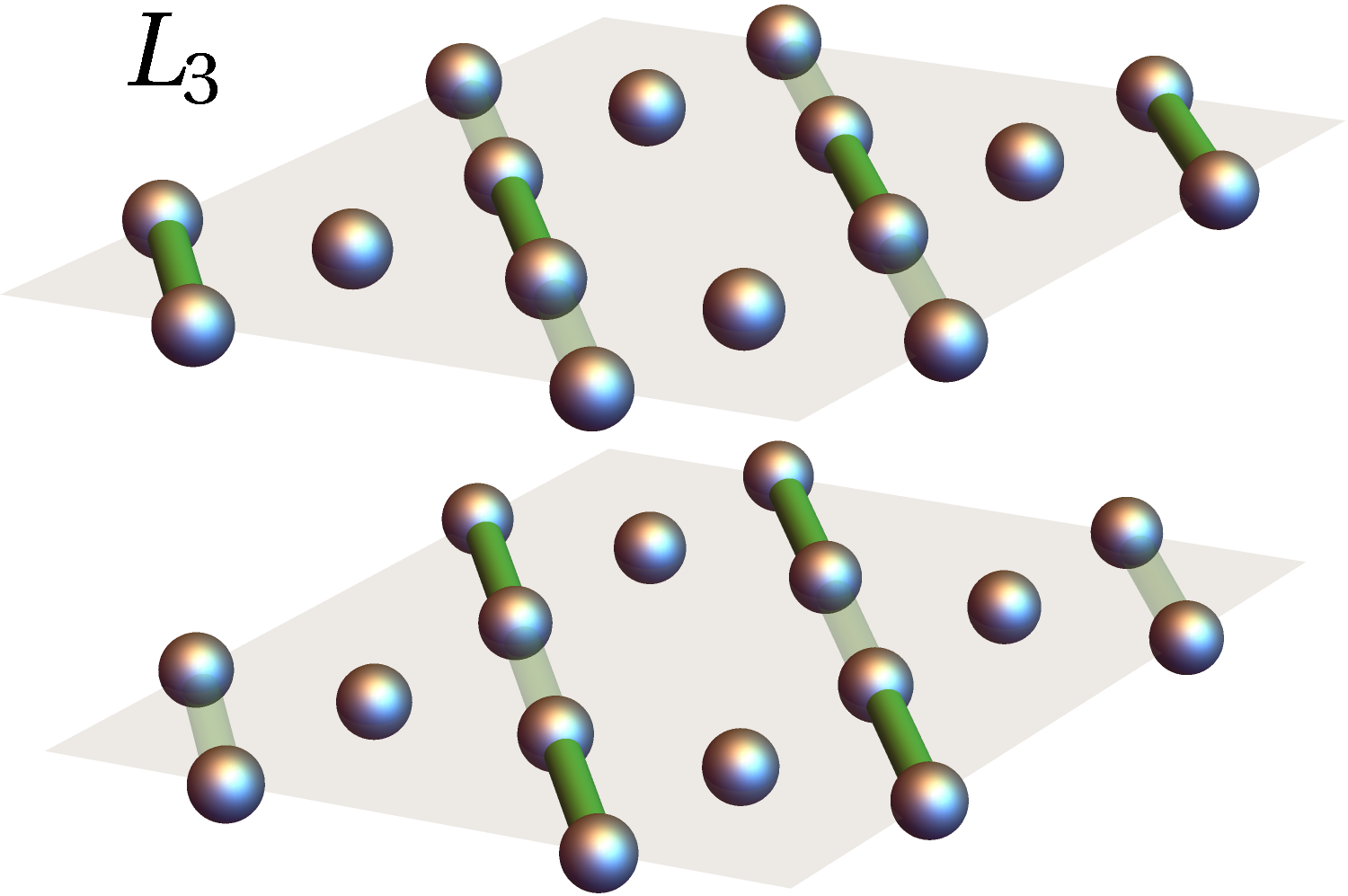}
\caption{\label{fig:stripes} Bond-order configurations corresponding to the order parameters $M_i$ and $L_i$. We use the notation $\overline{M}_i \equiv -M_i$ and $\overline{L}_i \equiv -L_i$ to refer to the configurations in which the strong and weak bonds are interchanged.}
\end{figure}

To elucidate the possible CDW instabilities of the system with unstable $M_1^+$ and $L_2^-$ phonon modes, we derive the Landau free-energy expansion for the coupled $M_i$ and $L_i$ order parameters, with $i=1, \, 2, \, 3$. For our phenomenological treatment, it is unimportant whether $M_i$ and $L_i$ are lattice distortions or electronic CDWs, since all of these orders transform as the same irreducible representations of the space group. The important point is that each component of these order parameters has a different wave-vector, since they come from different symmetry-equivalent points in the star of the $M$ and $L$ points of the BZ. As stated above, the wave-vector of $M_i$ is $\mbf{Q}_M^{(i)}$ whereas $L_i$ has wave-vector $\mbf{Q}_L^{(i)}$ [see Eqs.~\eqref{eq:M_Q_vectors} and \eqref{eq:L_Q_vectors}]. In Fig. \ref{fig:stripes}, we represent each of the three components of these order parameters separately as bond-order patterns in two consecutive layers. Clearly, each  $M_i$ or $L_i$ component corresponds to a stripe pattern of bond order that doubles the unit cell along the stripe direction. In the case of the $L_i$ order parameter, the unit cell is also doubled along the $c$-axis.

We use the INVARIANTS tool~\cite{Hatch2003} to scan all combinations of $M_i$ and $L_i$ that transform trivially under the space group, up to quartic order in the order parameters. We obtain the Landau free-energy:
\begin{align}
	\mathcal{F}_{\rm tot} &= \mathcal{F}_M + \mathcal{F}_L + \mathcal{F}_{ML}, \label{eq:free_energy_tot} \\
	\mathcal{F}_M &= \frac{\alpha_M}{2}M^2 + \frac{\gamma_M}{3}M_1 M_2 M_3 \label{eq:FM} \\ &+ \frac{u_M}{4}M^4 + \frac{\lambda_M}{4}\left( M_1^2 M_2^2 + M_1^2 M_3^2 + M_2^2 M_3^2 \right),  \nonumber \\
	\mathcal{F}_L &= \frac{\alpha_L}{2}L^2  + \frac{u_L}{4}L^4  \label{eq:FL} \\ &+ \frac{\lambda_L}{4}\left( L_1^2 L_2^2 + L_1^2 L_3^2 + L_2^2 L_3^2 \right), \nonumber \\
	\mathcal{F}_{ML} &= \frac{\gamma_{ML}}{3} \left( M_1 L_2 L_3 + L_1 M_2 L_3 + L_1 L_2 M_3 \right)  \label{eq:FML} \\
	&+ \frac{\lambda^{(1)}_{ML}}{4}\left( M_1 M_2 L_1 L_2 + M_1 M_3 L_1 L_3 + M_2 M_3 L_2 L_3 \right) \nonumber \\
	& + \frac{\lambda^{(2)}_{ML}}{4}\left( M_1^2 L_1^2 + M_2^2 L_2^2 + M_3^2 L_3^2 \right)  + \frac{\lambda^{(3)}_{ML}}{4}M^2 L^2\,, \nonumber
\end{align}
where, e.g., $M^2=M_1^2 + M_2^2 + M_3^2$ and $M^4 = (M^2)^2$. As $M_i$ and $L_i$ belong to different irreducible representations, $\alpha_M$ and $\alpha_L$ will in general be different, which in turn is manifested in different transition temperatures, at least in the absence of a coupling between the two. Thus, we will assume $\alpha_M=\alpha (T - T_M)$ and $\alpha_L = \alpha(T - T_L)$ and, in the numerical calculations presented in Sec.~\ref{sec:phase_diagram}, we will set $\alpha=1$.

Note the asymmetry between $\mathcal{F}_M$ and $\mathcal{F}_L$: while $\mathcal{F}_M$ has a trilinear term, $\mathcal{F}_L$ does not. This is a direct consequence of the different wave-vectors of the order parameters $M_i$ and $L_i$ described above, since $\sum_{i} \mbf{Q}_M^{(i)} = 0$ but $\sum_{i} \mbf{Q}_L^{(i)} \neq 0$. 
\begin{table*}
\begin{tabular}{M{0.35\textwidth}M{0.08\textwidth}M{0.1\textwidth}M{0.1\textwidth}M{0.1\textwidth}M{0.05\textwidth}M{0.15\textwidth}}
\hline\hline
         Phase & Color & $\mbf{Q}$-vector & OP & Unit cell & $C_6$ & Space group \\
         \hline
         Planar stripe & \crule[mcolor]{0.09\columnwidth}{0.3cm} & $\mbf{Q}_M$ & $(M00)$ & $2 \timessmall 1 \timessmall 1$ & No & $Pmmm$ (\#47) \\
         Planar tri-hexagonal & \crule[mmmpcolor]{0.09\columnwidth}{0.3cm} & $3\mbf{Q}_M$ & $(MMM)$ & $2 \timessmall 2 \timessmall 1$ & Yes & $P6/mmm$ (\#191) \\
         Planar Star-of-David & \crule[mmmmcolor]{0.09\columnwidth}{0.3cm} & $3\mbf{Q}_M$ & $(\overline{M}MM)$ & $2 \timessmall 2 \timessmall 1$ & Yes & $P6/mmm$ (\#191) \\
         Alternating stripe & \crule[lcolor]{0.09\columnwidth}{0.3cm} & $\mbf{Q}_L$ & $(L00)$ & $2 \timessmall 1 \timessmall 2$ & No & $Immm$ (\#71) \\
         Alternating tri-hexagonal Star-of-David & \crule[lllcolor]{0.09\columnwidth}{0.3cm} & $3\mbf{Q}_L$ & $(LLL)$ & $2 \timessmall 2 \timessmall 2$ & Yes & $P6/mmm$ (\#191) \\
         \hline\hline
\end{tabular}
\caption{\label{tab:phase_summary_uncoupled} Summary of the phases that minimize the uncoupled free energies $\mathcal{F}_M$ or $\mathcal{F}_L$. The single-$\mathbf{Q}$ stripe phases are illustrated in Fig. \ref{fig:stripes} whereas the triple-$\mathbf{Q}$ phases are shown in Fig.~\ref{fig:order_examples_uncoupled}. The order parameter (OP) column refers to the type of order parameters $(M_1 M_2 M_3)$ and $(L_1 L_2 L_3)$ that are condensed in the corresponding phase. For instance, in the planar Star-of-David phase, all three $M_{i}$ are finite and identical in magnitude, but one has the opposite sign of the other two. In the fifth column we report the increase in the unit cell brought on by the condensation of the respective phase. This is reported relative to the disordered phase. The sixth column denotes whether the associated phase respects six-fold rotational symmetry ($C_6$), and the seventh column displays the space group of the ordered CDW phase. Even though there are multiple phases which have the same space group as the parent structure ($P6/mmm$), these phases have larger unit cells and hence lower translational symmetry.}
\end{table*}
\begin{figure}
\includegraphics[width=0.95\columnwidth]{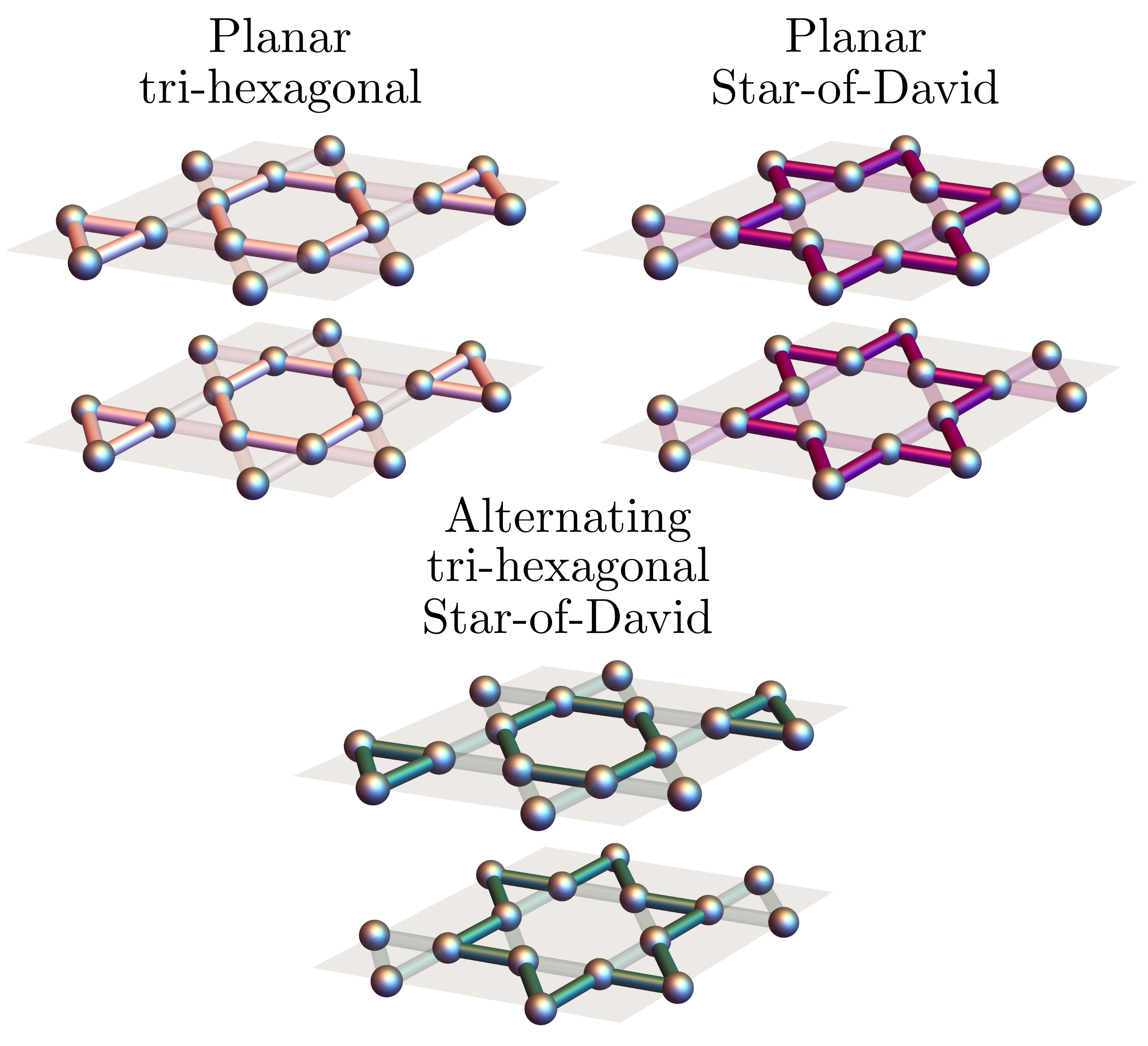}
\caption{\label{fig:order_examples_uncoupled} Triple-$\mathbf{Q}$ bond-order configurations obtained from minimizing the uncoupled free energies $\mathcal{F}_M$ or $\mathcal{F}_L$. The upper panels correspond to the two types of triple-$\mathbf{Q}_M$ order, whereas the lower panel illustrated the triple-$\mathbf{Q}_L$ order. They correspond to linear combinations of the three $M_i$ or three $L_i$ shown in Fig.~\ref{fig:stripes}, as explained in Table \ref{tab:phase_summary_uncoupled}.}
\end{figure}
An uncommon feature of the free energy expansion in the equations above is the presence of a trilinear coupling $\gamma_{ML}$ between components of two order parameters that transform as different irreps. While cubic terms involving the components of a single order parameter are rather common~\cite{Toledano1987}, for example in isostructural transitions or in a transition described by the four-state clock model, trilinear terms involving different order parameters are rare. The closest example to the free energy expansion discussed in this study is possibly observed in the layered perovskite ferroelectrics where there are trilinear couplings between two octahedral rotation modes and the electrostatic polarization~\cite{Bousquet2008, Benedek2011}. In that case, this couplings was shown to give rise to different types of phase transitions; for instance, a large enough trilinear coupling by itself could induce a first-order transition where multiple order parameters condense simultaneously~\cite{Etxebarria2010}. We will argue in Sec.~\ref{sec:phase_diagram} that a similar phenomenon may occur in the CDW kagome metals. In particular, the trilinear term with coefficient $\gamma_{ML}$ promotes phases in which both $M_i$ and $L_i$ are finite, playing a leading role in shaping the phase diagram.

In the remainder of this section, we restrict our analysis to the two terms $\mathcal{F}_M$ and $\mathcal{F}_L$, which are considered separately. Later on, in Sec.~\ref{sec:phase_diagram}, we tackle the full free-energy for the coupled CDW order parameters. The possible minima of $\mathcal{F}_M$ and $\mathcal{F}_L$ that will be derived below are summarized in Table~\ref{tab:phase_summary_uncoupled}, and schematically illustrated in Figs.~\ref{fig:stripes} and \ref{fig:order_examples_uncoupled}.

\subsection{CDW order at the $M$ point}\label{sec:only_M}

We start by analyzing the case where only the $M_i$ CDW order parameters are allowed to condense. The trilinear term in $\mathcal{F}_M$ in Eq. (\ref{eq:FM}) plays a primary role in selecting the leading instability. Since the individual $M_i$ can always be chosen in a configuration for which the contribution from the trilinear term to the free energy is negative, regardless of the sign of $\gamma_M$, this term will always lower the energy of the triple-$\mbf{Q}_M$ configuration, in which all $|M_i|=M$. For $\gamma_M < 0$, the bond-order configuration corresponds to the planar tri-hexagonal of Fig.~\ref{fig:order_examples_uncoupled}, with $\mathrm{sign}(M_1 M_2 M_3) > 0$, whereas for $\gamma_M > 0$, it corresponds to the planar Star-of-David of Fig.~\ref{fig:order_examples_uncoupled}, with $\mathrm{sign}(M_1 M_2 M_3) < 0$. Importantly, the trilinear term contribution vanishes for the single-$\mbf{Q}_M$ stripe configurations of Fig. \ref{fig:stripes}, where only one $M_i$ is non-zero in the CDW phase. Hence, a triple-$\mbf{Q}_M$ phase will always be favored, no matter the sign of $\lambda_M$.

To derive these results, we start by calculating the value of the free energy in the triple-$\mbf{Q}_M$ phase:
\begin{equation}
	\mathcal{F}_M^{3Q} = \frac{\mathcal{K}^2}{2592 u_{M \ast}^3}\left[  54 \alpha (T- T_M) u_{M \ast} - |\gamma_M| \mathcal{K}  \right]\,,
\end{equation}
where
\begin{align}
	u_{M \ast} &= 3u_{M} + \lambda_M\,, \\
	\mathcal{K} &= |\gamma_M| + \sqrt{\gamma_M^2 - 36 \alpha (T-T_M) u_{M\ast}}\,,
\end{align}
Note that, for the free energy to remain bounded, we must require $u_{M\ast}>0$. For the single-$\mbf{Q}_M$ phase, the free energy is given by:
\begin{equation}
	\mathcal{F}^{1Q}_M = -\frac{\alpha^2 (T-T_M)^2 }{4u_M}\,,
\end{equation}
Therefore, in addition to $u_{M\ast}>0$, we must also require that $u_M >0$ for the free energy to remain bounded. While $\mathcal{F}_M^{1Q}$ crosses zero at $T=T_M$, $\mathcal{F}_M^{3Q}$ does so at
\begin{equation}
	T_M^{3Q} = \frac{2\gamma_M^2}{81 \alpha u_{M\ast}} + T_M\,, \label{eq:first_order_M}
\end{equation}
which is always larger than $T_M$ if the free energy is bounded and the trilinear coefficient $\gamma_M$ is not zero. Hence, the trilinear term induces a transition to a triple-$\mbf{Q}_M$ phase. The order parameters are given by:
\begin{equation}
	M_i = \pm\frac{\mathcal{K}}{6u_{M\ast}}\,,
\end{equation}
with the relative signs chosen such that $\text{sgn}(\gamma_M M_1 M_2 M_3 )<0$. Thus, at the transition to the triple-$\mbf{Q}_M$ phase, the order parameter exhibits a jump proportional to $\gamma_M$:
\begin{equation}
	|\Delta M_i| = \frac{2|\gamma_M|}{9u_{M\ast}}\,.
\end{equation}
The triple-$\mbf{Q}_M$ phase in which $\text{sgn}(M_1 M_2 M_3 )>0$ takes place for $\gamma_M < 0$ and corresponds to the planar tri-hexagonal phase (see Table~\ref{tab:phase_summary_uncoupled} and Fig.~\ref{fig:order_examples_uncoupled}). There are four equivalent ground states corresponding to four different ways of arranging the enhanced $2\times2\times1$ unit cell, given by the combinations of order parameters $(M_1 M_2 M_3 ) = (M M M)$, $(M \overline{M} \overline{M})$, $(\overline{M} M \overline{M})$, and $(\overline{M} \overline{M} M)$. Here, we denote $\overline{M}_i \equiv - M_i$. Analogously, for $\gamma_M > 0$, the resulting configuration is the planar Star-of-David phase shown in Fig.~\ref{fig:order_examples_uncoupled}. It is characterized by $\text{sgn}(M_1 M_2 M_3 )<0$ and consists of four equivalent states $(M_1 M_2 M_3 ) = (M M \overline{M})$, $(M \overline{M} M)$, $(\overline{M} M M)$, and $(\overline{M} \overline{M} \overline{M})$. The space group of both planar tri-hexagonal and planar Star-of-David phases is $P6/mmm$, which is the same as the space group of the compound without charge order. 

\begin{figure}[t]
\includegraphics[width=0.49\columnwidth]{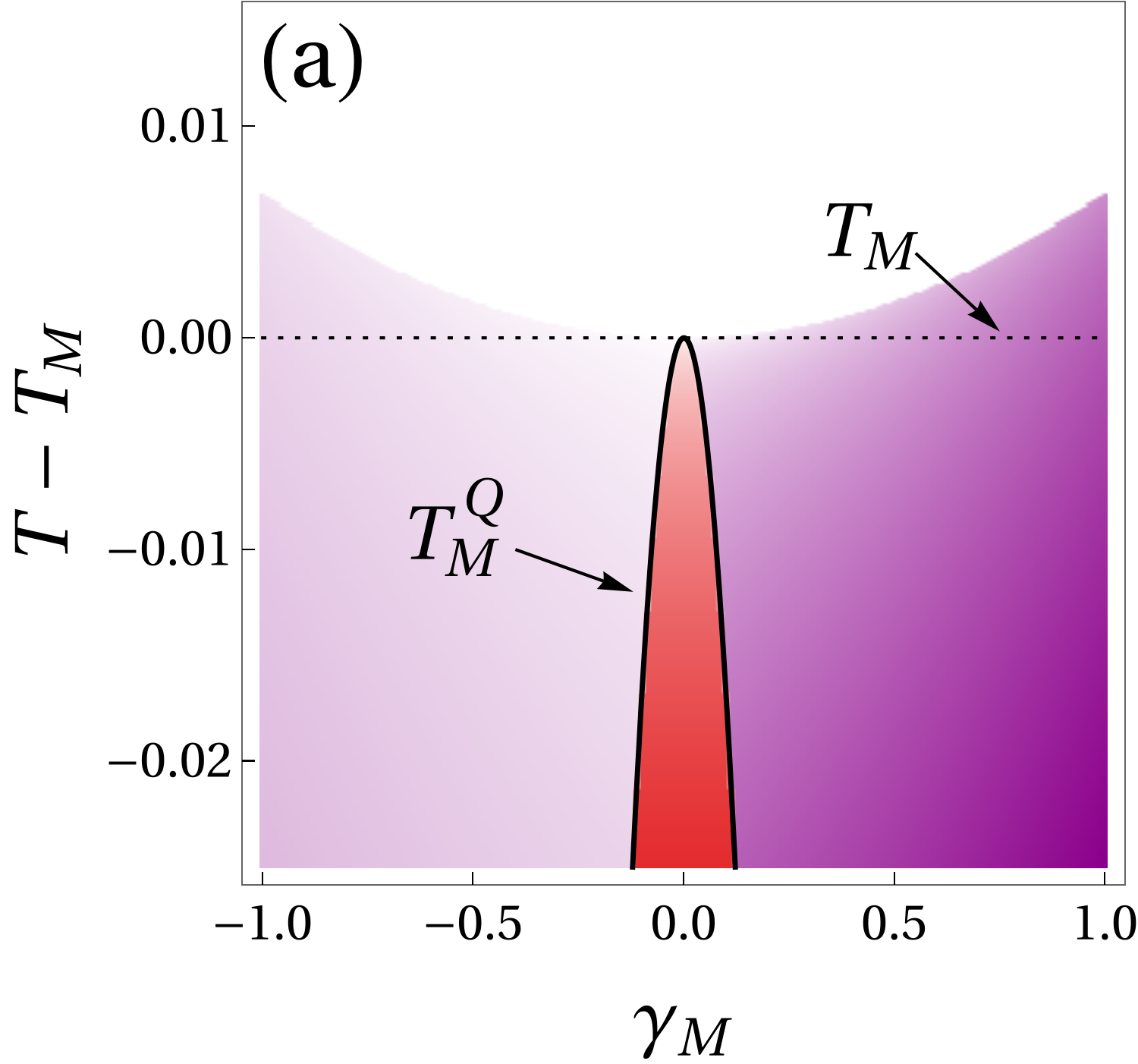}\includegraphics[width=0.49\columnwidth]{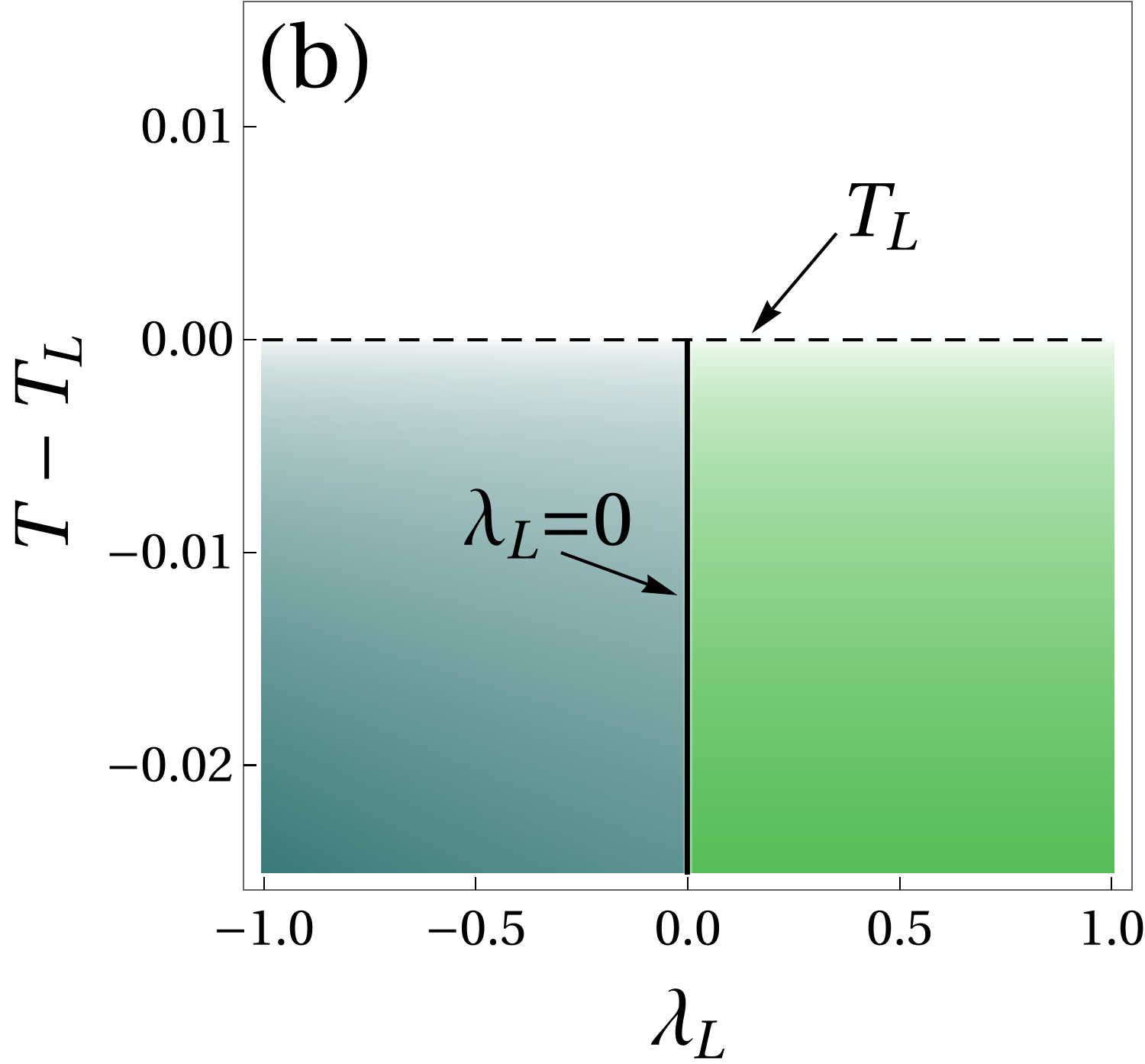}
\caption{\label{fig:phase_diagram_M} (a) Phase diagram for CDW order at the $M$ point only, obtained from minimizing the free energy $\mathcal{F}_M$. Here, $\lambda_M=0.6$ and $u_M=1.0$. These values only affect the precise shape of the single-$\mbf{Q}_M$ region [see Eq.~\eqref{eq:triple_to_single_trans}]. Colors correspond to the phases listed in Table~\ref{tab:phase_summary_uncoupled} and illustrated in Figs.~\ref{fig:stripes} and \ref{fig:order_examples_uncoupled}. As explained in the text, the leading instability is always to one of the two triple-$\mbf{Q}_M$ phases. The sign of $\gamma_M$ decides between the planar tri-hexagonal ($\gamma_{M} <0$, light purple) and the planar Star-of-David ($\gamma_M > 0$, dark purple) phases. For positive $\lambda_M$, a secondary transition to a single-$\mbf{Q}_M$ phase occurs at lower temperatures (red), although for $\gamma_M=0$ this can occur at $T=T_M$. For negative $\lambda_M$, the triple-$\mbf{Q}_M$ phase is always favored. The upwards curvature of the transition temperature is a consequence of the cubic term as discussed in the text [see Eq.~\eqref{eq:first_order_M}]. (b) Phase diagram for CDW order at the $L$ point only, obtained from minimizing $\mathcal{F}_L$. Here, for concreteness, we chose $u_L=1.0$, although it has no impact on the shape of the phase diagram. The sign of $\lambda_L$ selects between the single-$\mbf{Q}_L$ ($\lambda_L >0$, green) and triple-$\mbf{Q}_L$ ($\lambda_L < 0$, petroleum) phases. No secondary transitions occur at lower temperatures.}
\end{figure}

These considerations are valid for the leading instability of $\mathcal{F}_M$ in Eq. (\ref{eq:FM}). When the quartic coefficient $\lambda_M > 0$, however, the corresponding quartic term penalizes the triple-$\mbf{Q}_M$ phase and vanishes for the single-$\mbf{Q}_M$ phase. As a result, because the quartic term is sub-leading with respect to the cubic term, the single-$\mbf{Q}_M$ phase corresponding to the in-plane stripes shown in Fig. \ref{fig:stripes} can occur for $\lambda_M > 0$ as a sub-leading instability that onsets at a temperature $T_M^{Q} < T_M^{3Q}$ given by:
\begin{align}
	T_M^{Q} = -\frac{\gamma_M^2}{9\alpha \lambda_M^2 u_{M\ast}}&\big[\sqrt{u_M(4u_M+\lambda_M)^3} \nonumber \\  & \quad + u_M(8u_M + 3\lambda_M) \big] + T_M\,, \label{eq:triple_to_single_trans}
\end{align}
which, like the other quantities above, is independent of the sign of $\gamma_M$. This state has a three-fold degeneracy and is parametrized by $(M 0 0)$, $(0 M 0)$, and $(0 0 M)$. It lowers the sixfold rotational symmetry to twofold, and has space group $Pmmm$. Note that, if $\lambda_M <0$, the triple-$\mbf{Q}_M$ phase is always favored over the single-$\mbf{Q}_M$ stripe phase. The phase diagram for CDW order at the $M$ point is shown in Fig.~\ref{fig:phase_diagram_M}(a) (for $\lambda_M >0$). The colors correspond to those shown in Table \ref{tab:phase_summary_uncoupled}. As expected, a double-$\mbf{Q}_M$ phase is never favored.

\subsection{CDW order at the $L$ point}\label{sec:only_L}

We now proceed to analyze the free energy $\mathcal{F}_L$, which contains only the $L_i$ order parameters. The absence of a cubic (trilinear) term makes the analysis simpler than the case of $M_i$. Specifically, the sign of the quartic coefficient $\lambda_L$ completely determines whether the instability is towards the single-$\mbf{Q}_L$ or the triple-$\mbf{Q}_L$ phase. The former gives rise to the alternating pattern of stripes with opposite phases in consecutive layers shown in Fig.~\ref{fig:stripes}. As displayed in Table \ref{tab:phase_summary_uncoupled}, the space group of this alternating stripe CDW phase is $Immm$, which does not have $C_6$ rotational symmetry and is different from the space group of the single-$\mbf{Q}_M$ planar stripe-phase. 

As for the triple-$\mbf{Q}_L$ phase, it consists of a bond-order configuration of alternating tri-hexagonal and Star-of-David patterns in consecutive layers, as shown in Fig.~\ref{fig:order_examples_uncoupled}. In contrast to the case of CDW order at the $M$ point, there is only a single eight-fold degenerate triple-$\mbf{Q}_L$ phase, rather than two four-fold degenerate triple-$\mbf{Q}_M$ phases. This is a consequence of the fact that the unit cell in the triple-$\mbf{Q}_L$ phase is enhanced by a factor of eight. Consequently, the configurations $(L L L)$ and $(\overline{L} L L)$ are related by translational symmetry and correspond to different ``domains" of the same phase. Interestingly, as shown in Table \ref{tab:phase_summary_uncoupled}, the alternating tri-hexagonal Star-of-David phase shares the same space group as the planar tri-hexagonal and planar Star-of-David CDW states, which preserves sixfold rotational symmetry. 

Minimizing the free energy $\mathcal{F}_L$ in each of the two phases gives:
\begin{align}
	\mathcal{F}^{1Q}_L &= -\frac{\alpha^2 (T-T_L)^2}{4u_L}\,, \\
	\mathcal{F}^{3Q}_L &= -\frac{\alpha^2 (T-T_L)^2}{4u_L + \frac{4}{3}\lambda_L}\,.
\end{align}
Thus, for the free energy to remain bounded, both $u_L > 0$ and $3u_L + \lambda_L >0$. These are identical to the conditions for $u_M$ and $\lambda_M$ reported above. Comparison between $\mathcal{F}^{1Q}_L$ and $\mathcal{F}^{3Q}_L$ shows that a single-$\mbf{Q}_L$ phase is favored for $\lambda_L>0$ and a triple-$\mbf{Q}_L$ phase is favored for $\lambda_L < 0$, as anticipated. The resulting phase diagram is shown in Fig.~\ref{fig:phase_diagram_M}(b), with the colors corresponding to the CDW states outlined in Table \ref{tab:phase_summary_uncoupled}. As in the case of $\mathcal{F}_M$, a double-$\mbf{Q}_L$ phase is not favored in any region of the phase diagram. In contrast to that case, however, no sub-leading instabilities appear in the phase diagram.

\subsection{CDW order along the $U$ line}\label{sec:only_U}

While several experimental observations support a $2\times 2\times 2$ supercell for the CDW phase of CsV$_3$Sb$_5$, the possibility of a $2\times 2\times 4$ phase has been raised in Ref.~\onlinecite{Ortiz2021b}. Combinations of the $L_i$ and $M_i$ order parameters cannot lead to a supercell with a periodicity of 4 lattice constants along the $c$ direction. Instead, translational symmetry breaking that leads to a  $2\times 2\times 4$, or in general $2\times 2\times n$ (where $n>2$) unit cell, requires CDW order with wave-vectors on the $U$ line of the Brillouin zone, which connects the $M$ and $L$ points. A generic wave-vector along this line has six vectors in its star, described by the inversion-symmetry-related pairs $\pm \mbf{Q}^{(1)}_U= \pm (\tfrac{1}{2} \ 0 \ q_z)$, $\pm \mbf{Q}^{(2)}_U= \pm (0 \ \tfrac{1}{2} \ q_z)$, and $\pm \mbf{Q}^{(3)}_U= \pm (-\tfrac{1}{2} \ -\tfrac{1}{2} \ q_z)$, with $q_z \neq 0, \ \tfrac{1}{2}$. 

Our DFT calculations indeed find that, besides the $M_1^+$ and $L_2^-$ modes, a phonon mode that transforms as the $U_1$ irreducible representation and increases the unit cell by $2\times 2 \times 4$ is also unstable. As presented in Fig. \ref{fig:dft_phonons}, the absolute value of the squared frequency of this $U_1$ mode is not as large as those of the $M_1^+$ and $L_2^-$ modes. Whether this implies that the corresponding CDW is a subleading instability as compared to the other two requires further investigation.

While the analysis of the Landau free-energy expansion for the $U_i$ order parameters is beyond the scope of this work, we point out some general properties of these CDW order parameters. Since $\mbf{Q}^{(i)}_U \neq -\mbf{Q}^{(i)}_U$, there are 6 wavevectors in the star of $U$, and as a result, the order parameter $U_{\pm i}$ with $i=1,\,2,\,3$ is six-dimensional. Similar terms as those in $\mathcal{F}_L$, Eq.~\eqref{eq:FL}, and $\mathcal{F}_{ML}$, Eq.~\eqref{eq:FML}, will appear for $\mathcal{F}_U$ and $\mathcal{F}_{UM}$, but with $L_i L_j$ replaced by $U_i U_{-j}$. 
Similarly, for the specific case where $q_z = 1/4$, there will also be a trilinear coupling between $U_i$ and $L_i$ of the form $U_i U_j L_k$, with $i, \, j , \, k$ all different.
The full analysis of this type of $U$ CDW order is a topic for future studies. In the remainder of the paper, we will focus on the commensurate case $q_z = 0, \, 1/2$, for which the order parameters are real. Thus, in what follows, we only consider the coupled CDW orders with $M$ and $L$ wave-vectors.

\section{Phase diagrams for coupled CDW orders}\label{sec:phase_diagram}

Having determined the phase diagrams for the ``pure" CDW orders at the $M$ and $L$ points, we now investigate the phase diagram of the coupled case. The difference between $\mathcal{F}_M$ and $\mathcal{F}_L$ is evident from Secs.~\ref{sec:only_M} and \ref{sec:only_L}. While the trilinear term present in $\mathcal{F}_M$ ensures that a triple-$\mbf{Q}_M$ phase is always favored as the leading instability, its absence in $\mathcal{F}_L$ implies that a single-$\mbf{Q}_L$ phase is favored when $\lambda_L >0$. Hence, coupling the two terms can be expected to lead to phase diagrams exhibiting a multitude of additional phases. 

\begin{figure}
\includegraphics[width=0.95\columnwidth]{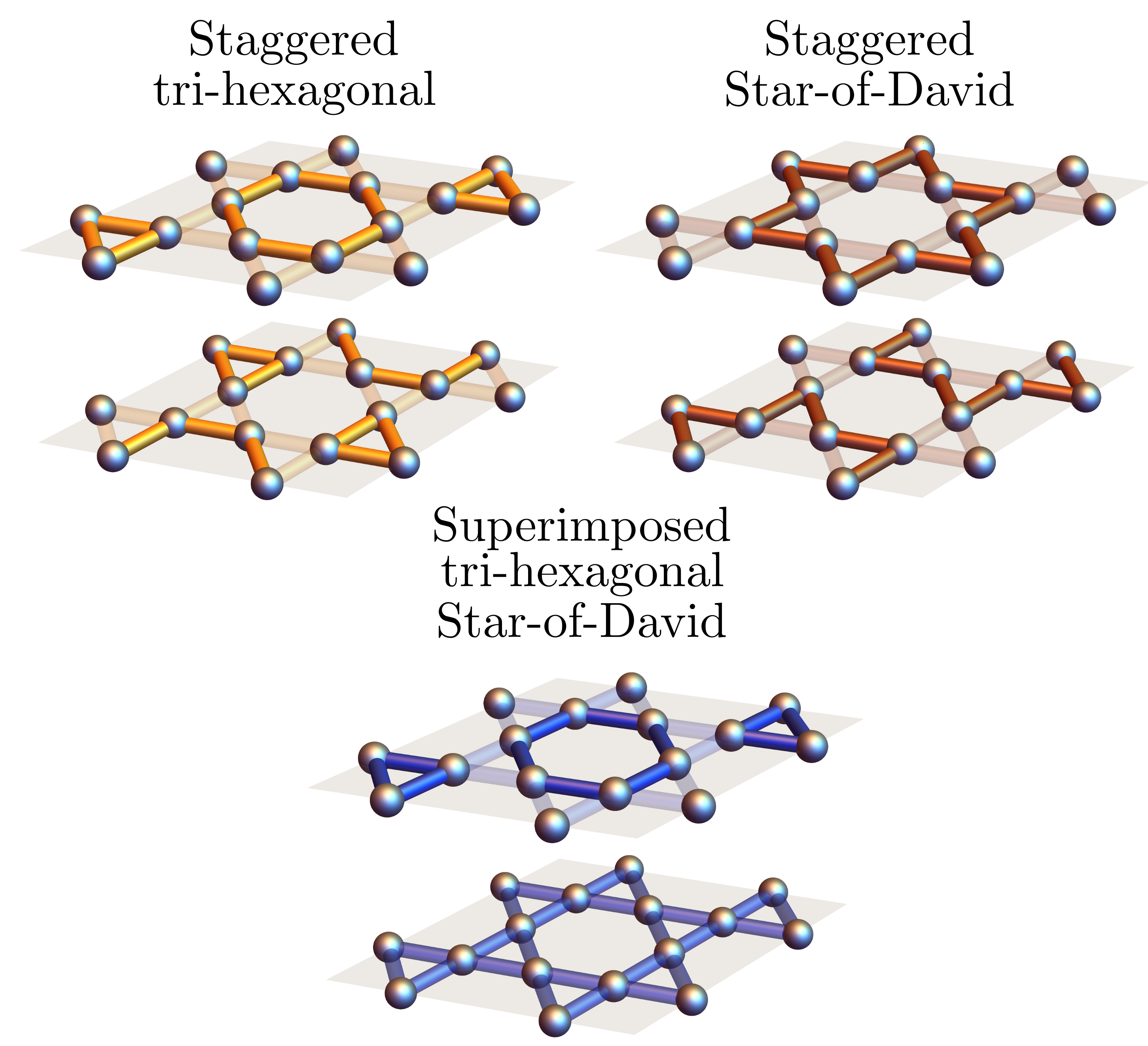}
\caption{\label{fig:mixed_order_examples} Three possible bond-order patterns arising from the simultaneous condensation of $M_i$ and $L_i$ order parameters, obtained from minimizing the coupled free energy, $\mathcal{F}_{\rm tot}$. The upper panels show different types of single-$\mbf{Q}_M$/double-$\mbf{Q}_L$ order, whereas the lower panel illustrates the triple-$\mbf{Q}_M$/triple-$\mbf{Q}_L$ order. These phases are described by linear combinations of the orders in Figs.~\ref{fig:stripes} and \ref{fig:order_examples_uncoupled}, as written in Table \ref{tab:phase_summary_coupled}.}
\end{figure}

While the full expression $\mathcal{F}_{\rm tot}$ is sufficiently complicated that the analytical solutions are no longer tractable, a few insights can be gained before we present the numerical results. Besides the ``pure" phases presented in the previous section, our numerical analysis reveals that three different coupled CDW phases, described in Table \ref{tab:phase_summary_coupled} and illustrated in Fig. \ref{fig:mixed_order_examples}, also appear over a wide range of parameters. The superimposed tri-hexagonal Star-of-David CDW state is a triple-$\mbf{Q}_M$/triple-$\mbf{Q}_L$ configuration of the form $(M M M)+(L L L)$; note that changing the sign of only one set of components, e.g. to $(\overline{M} \overline{M} \overline{M})+(L L L)$, does not change the symmetries of this phase. We emphasize that, in our notation, for a given $i$, $M_i$ and $L_i$ have the same sign, whereas $\overline{M}_i$ and $L_i$ have opposite signs. The resulting space group of the superimposed tri-hexagonal Star-of-David phase, $P6/mmm$, is the same as the disordered phase and all the ``pure" triple-$\mbf{Q}$ phases: the planar tri-hexagonal phase, the planar Star-of-David phase, and the alternating tri-hexagonal Star-of-David phase discussed in the previous section.

The staggered tri-hexagonal and staggered Star-of-David CDW states are single-$\mbf{Q}_M$/double-$\mbf{Q}_L$ phases described by $(M 0 0)+(0 L L)$ and $(\overline{M} 0 0)+(0 L L)$, respectively. This type of states has been studied previously using DFT in Ref.~\onlinecite{Ratcliff2021} and proposed to be realized in the kagome metal CsV$_3$Sb$_5$. As shown in Fig. \ref{fig:mixed_order_examples}, they break the sixfold rotational-symmetry of the system, and are described by the space group $Fmmm$. Like the superimposed tri-hexagonal Star-of-David phase, these states increase the unit cell by $2\times 2\times 2$. 
We found that these three coupled CDW configurations can become leading instabilities of the system, which compete with the ``pure" CDW orders at $M$ and $L$ over some range of parameters. In contrast, the mixed phases shown in Table \ref{tab:phase_summary_mixed}, which were also found in our numerical phase diagrams, only appeared as secondary instabilities inside other phases. They can be interpreted as the superposition of two or more of the phases presented in Tables~\ref{tab:phase_summary_uncoupled} and \ref{tab:phase_summary_coupled}.

Insight about the emergence of the three coupled CDW states of Table \ref{tab:phase_summary_coupled} can be obtained by a qualitative analysis of the different terms of the free-energy $\mathcal{F}_{ML}$ in Eq. (\ref{eq:FML}). The trilinear term, with coefficient $\gamma_{ML}$, lowers the total free energy for the triple-$\mbf{Q}_M$/triple-$\mbf{Q}_L$ state regardless of the sign of $\gamma_{ML}$. It also lowers the free energy of one of the single-$\mbf{Q}_M$/double-$\mbf{Q}_L$ phases, but by a smaller amount than for the triple-$\mbf{Q}_M$/triple-$\mbf{Q}_L$: either the staggered tri-hexagonal phase, if $\gamma_{ML} < 0$, or the staggered  Star-of-David phase, if $\gamma_{ML} > 0$. Notice also that any coupled phase with a single $\mbf{Q}_L$ component gains no energy from this cubic term -- or from any other term of $\mathcal{F}_{\rm tot}$, for that matter. 

\begin{table*}
\begin{tabular}{M{0.34\textwidth}M{0.07\textwidth}M{0.1\textwidth}M{0.15\textwidth}M{0.09\textwidth}M{0.04\textwidth}M{0.15\textwidth}}
\hline\hline
         Phase & Color & $\mbf{Q}$-vector & OP & Unit cell & $C_6$ & Space group \\
         \hline
         Superimposed tri-hexagonal Star-of-David & \crule[mmmlllposcolor]{0.09\columnwidth}{0.3cm} & $3\mbf{Q}_M+3\mbf{Q}_L$ & $(MMM)+(LLL)$ & $2 \timessmall 2 \timessmall 2$ & Yes & $P6/mmm$ (\#191) \\
         Staggered tri-hexagonal & \crule[mllposcolor]{0.09\columnwidth}{0.3cm} & $\mbf{Q}_M + 2\mbf{Q}_L$ & $(M00)+(0LL)$ & $2 \timessmall 2 \timessmall 2$ & No & $Fmmm$ (\#69) \\
         Staggered Star-of-David & \crule[mllnegcolor]{0.09\columnwidth}{0.3cm} & $\mbf{Q}_M+2\mbf{Q}_L$ & $(\overline{M}00)+(0LL)$ & $2 \timessmall 2 \timessmall 2$ & No & $Fmmm$ (\#69) \\
         \hline\hline
\end{tabular}
\caption{\label{tab:phase_summary_coupled} Summary of the phases that minimize the coupled free energy $\mathcal{F}_{\mathrm{tot}}$ in Eq.~\eqref{eq:free_energy_tot}. All of these phases can occur as leading instabilities of the system. They emerge as a consequence of the coupling terms in $\mathcal{F}_{ML}$ and intertwine the ordered phases with wave-vector $\mbf{Q}_M$ with those with wave-vector $\mbf{Q}_L$.}
\end{table*}

\begin{table*}
\begin{tabular}{M{0.25\textwidth}M{0.08\textwidth}M{0.13\textwidth}M{0.2\textwidth}M{0.09\textwidth}M{0.04\textwidth}M{0.15\textwidth}}
\hline\hline
         Phase & Color & $\mbf{Q}$-vector & OP & Unit cell & $C_6$ & Space group \\
         \hline
         $2M+M+2L$ & \crule[m1m23l2color]{0.09\columnwidth}{0.3cm} & $3\mbf{Q}_M+2\mbf{Q}_L$ & $(M_1\overline{M}_1M_2) +(L_1L_10)$ & $2 \timessmall 2 \timessmall 2$ & No & $Cccm$ (\#66) \\
         $2M+M+2L+L$ & \crule[m1m23l1l23color]{0.09\columnwidth}{0.3cm} & $3\mbf{Q}_M+3\mbf{Q}_L$ & $(M_1M_1\overline{M}_2) +(L_1L_1L_2)$ & $2 \timessmall 2 \timessmall 2$ & No & $Cmmm$ (\#65) \\
         $M+L+L$ & \crule[m1l1l2color]{0.09\columnwidth}{0.3cm} & $\mbf{Q}_M+2\mbf{Q}_L$ & $(M00)+(0 L_1L_2)$ & $2 \timessmall 2 \timessmall 2$ & No & $C2/m$ (\#12) \\
         \hline\hline
\end{tabular}
\caption{\label{tab:phase_summary_mixed} Additional phases that minimize the coupled free energy $\mathcal{F}_{\mathrm{tot}}$ in Eq.~\eqref{eq:free_energy_tot}. In contrast to the phases in Table \ref{tab:phase_summary_coupled}, these states do not occur as leading instabilities of the free energy. They can be described as mixing two or more of the phases in Tables~\ref{tab:phase_summary_uncoupled} and \ref{tab:phase_summary_coupled}.}
\end{table*}

The quartic terms of $\mathcal{F}_{ML}$ also impact the three coupled CDW states in distinct ways. For coefficients $\lambda_{ML}^{(1)}, \lambda_{ML}^{(2)} < 0$, these quartic terms further lower the energy of the triple-$\mbf{Q}_M$/triple-$\mbf{Q}_L$ phase, whereas for  $\lambda_{ML}^{(1)}, \lambda_{ML}^{(2)} > 0$ they increase its energy. In contrast, the energies of both single-$\mbf{Q}_M$/double-$\mbf{Q}_L$ phases are unaffected by these two quartic terms, since they vanish for the configurations $(M 0 0)+(0 L L)$ and $(\overline{M} 0 0)+(0 L L)$.

This qualitative analysis suggests that coupled CDW states may become energetically more favorable than the pure CDW states in the presence of a large (in magnitude) trilinear coefficient $\gamma_{ML}$. The triple-$\mbf{Q}_M$/triple-$\mbf{Q}_L$ phase seems generally favored over the single-$\mbf{Q}_M$/double-$\mbf{Q}_L$ phases, except when the  $\lambda_{ML}^{(1)}$, $\lambda_{ML}^{(2)}$ quartic coefficients are sizable and positive. While this general tendency is confirmed by our numerical calculations, we will see that other terms also play an important role, such as $\lambda_{ML}^{(3)}$ and the coefficients $\gamma_{M}$ and $\lambda_L$ of the ``pure" free energies. For instance, $\gamma_{ML}$ and $\gamma_M$ may be incompatible if they have opposite signs; for example, $\gamma_M < 0$ favors $(M M M)$ whereas $\gamma_{ML} > 0$ favors $(\overline{M} \overline{M} \overline{M})$. Similarly, $\lambda_L$ favors either a single-$\mbf{Q}_L$ or a triple-$\mbf{Q}_L$ phase, but not a double-$\mbf{Q}_L$ phase, which does not even appear in the phase diagram of the $L_i$ CDW states.

We finish by noting that, due to the sizable parameter space, we cannot rule out that other coupled CDW states may be stabilized outside the regimes we investigated. For instance, there are several possible mixed phases that can emerge inside the ordered states discussed above. A few of these mixed phases were seen in our numerical calculations (see Table \ref{tab:phase_summary_mixed}), all of which break sixfold rotational symmetry.

\subsection{Impact of the coupling between $M_i$ and $L_i$}

\begin{figure}
\includegraphics[width=0.49\columnwidth]{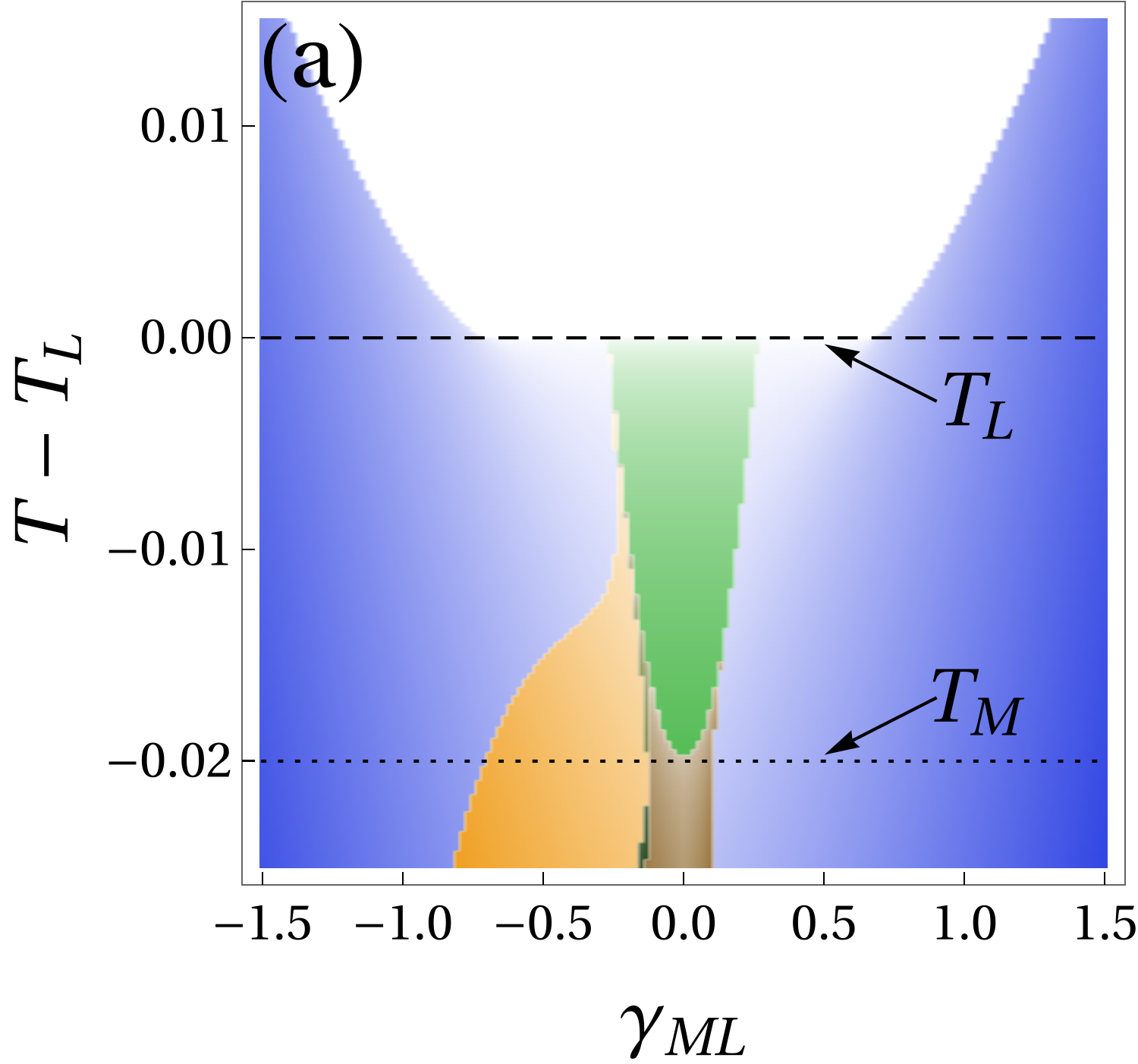}\includegraphics[width=0.49\columnwidth]{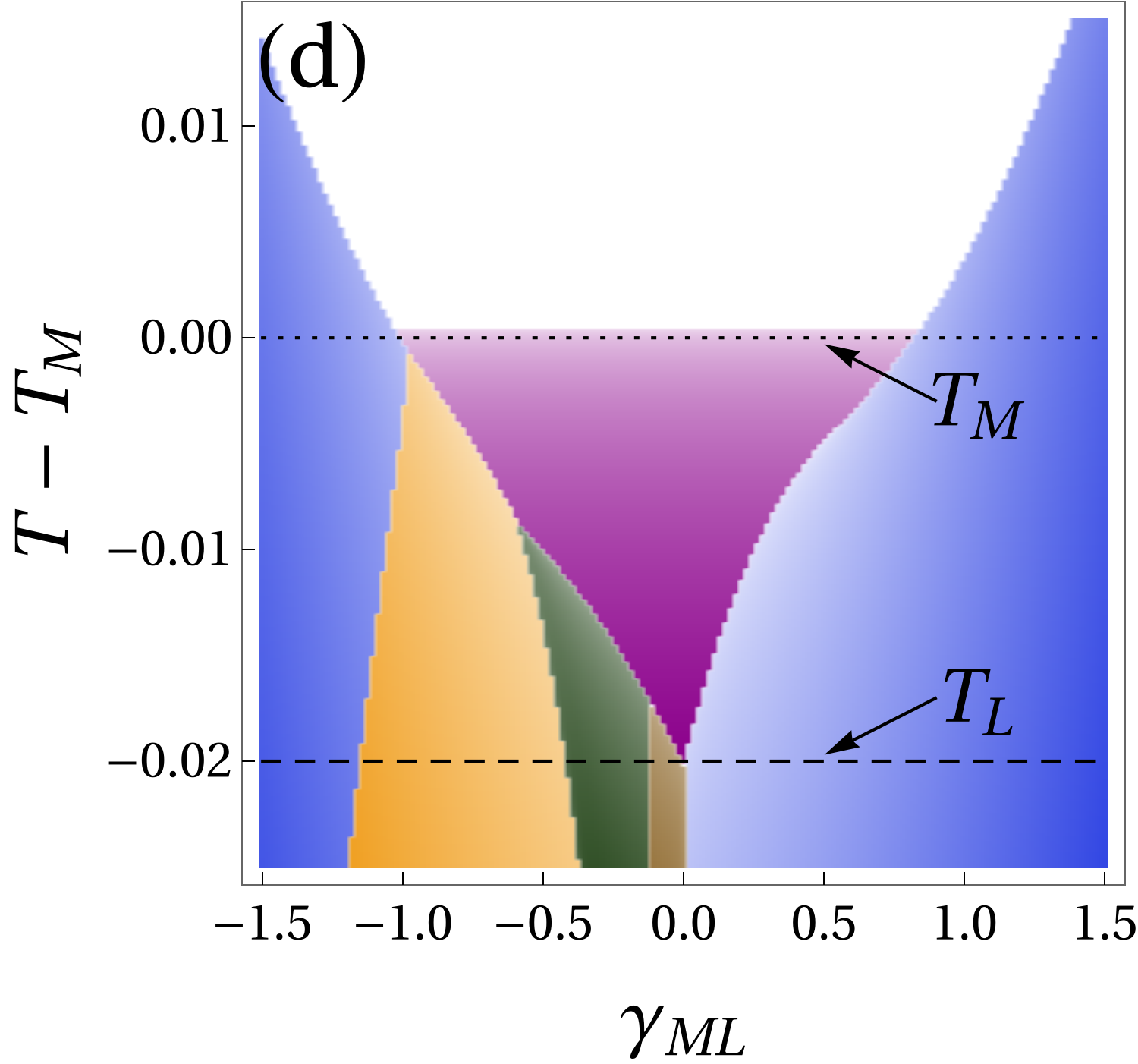}\\
\vspace{3mm}
\includegraphics[width=0.49\columnwidth]{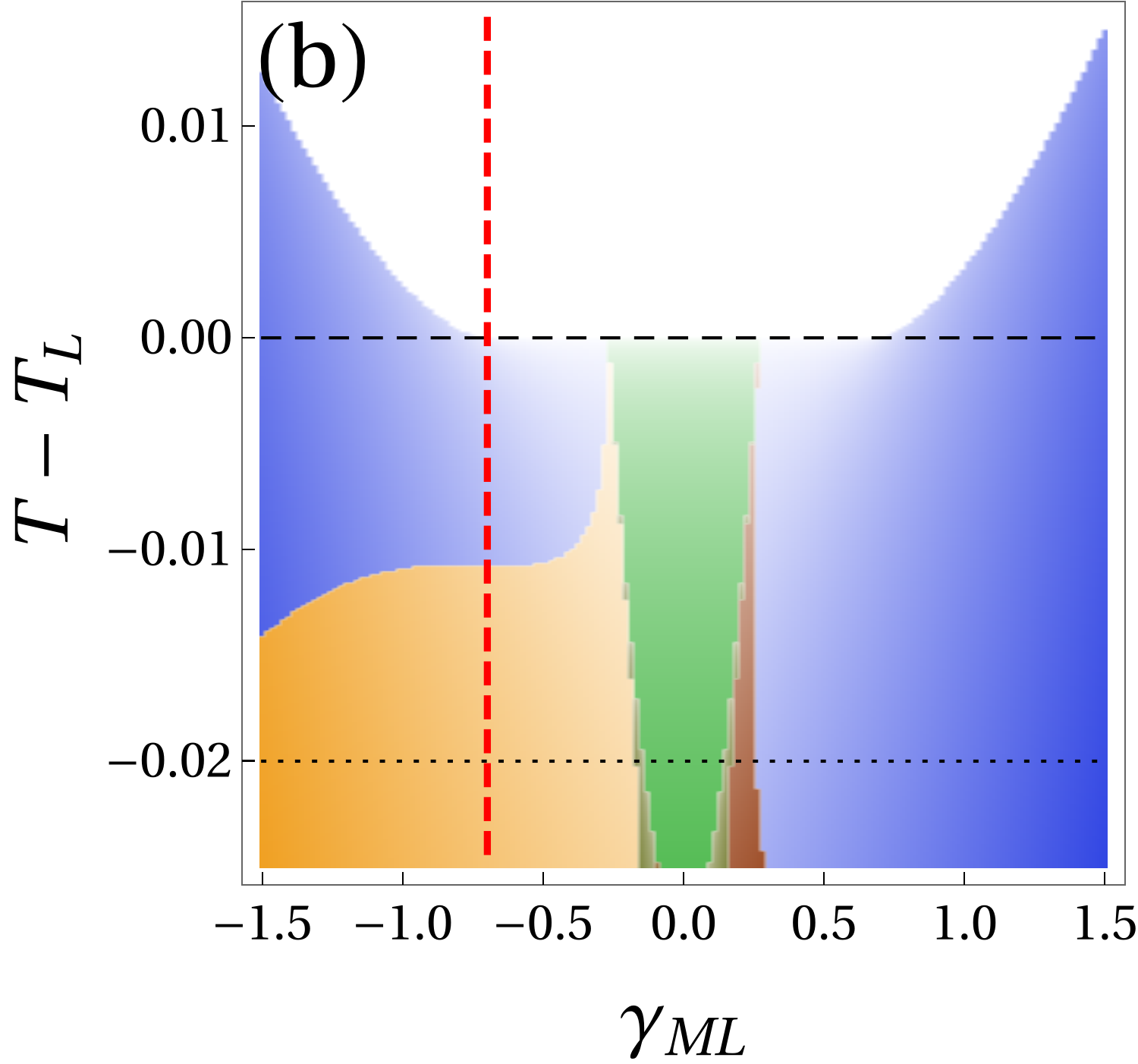}\includegraphics[width=0.49\columnwidth]{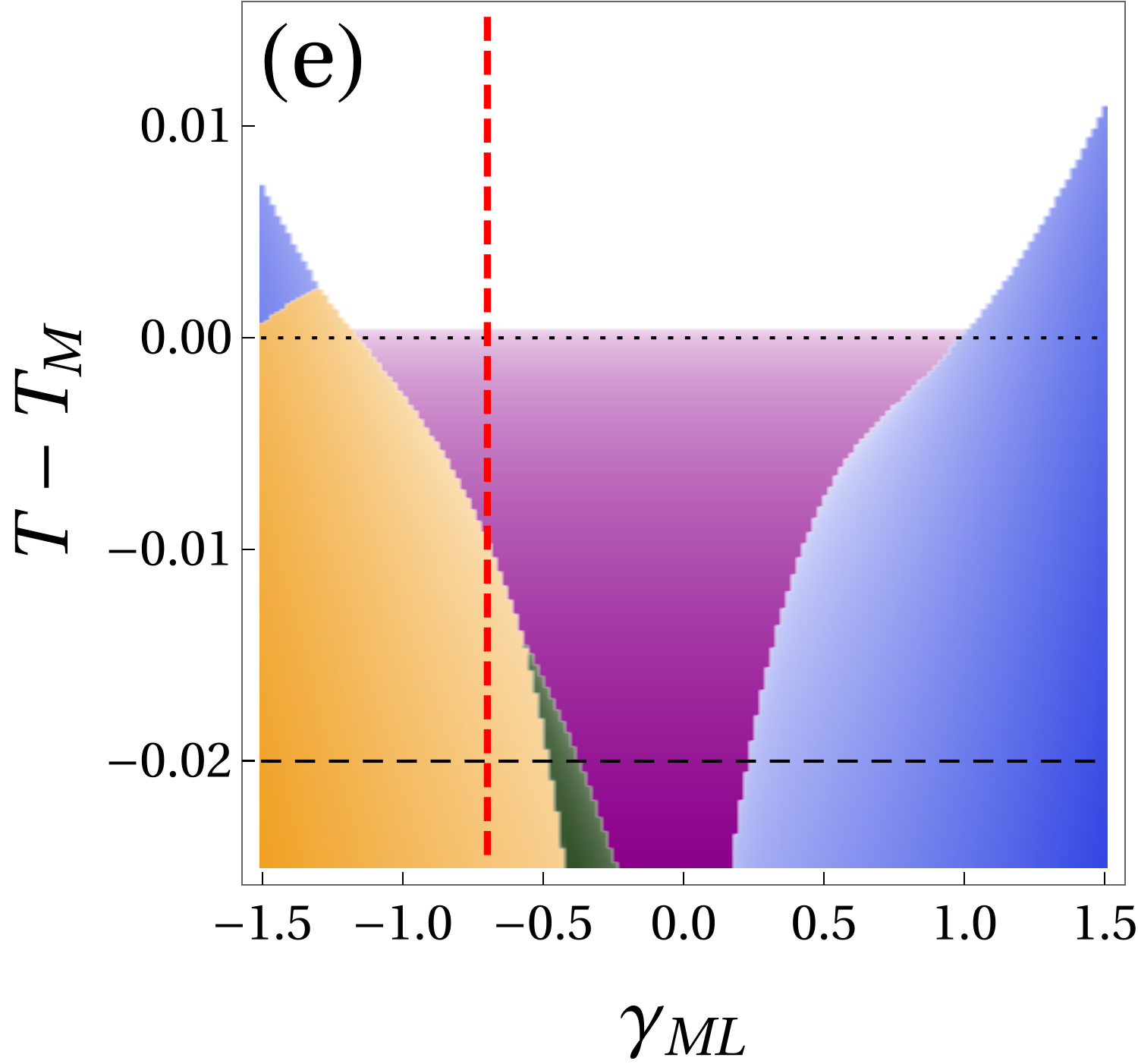}\\
\vspace{3mm}
\includegraphics[width=0.49\columnwidth]{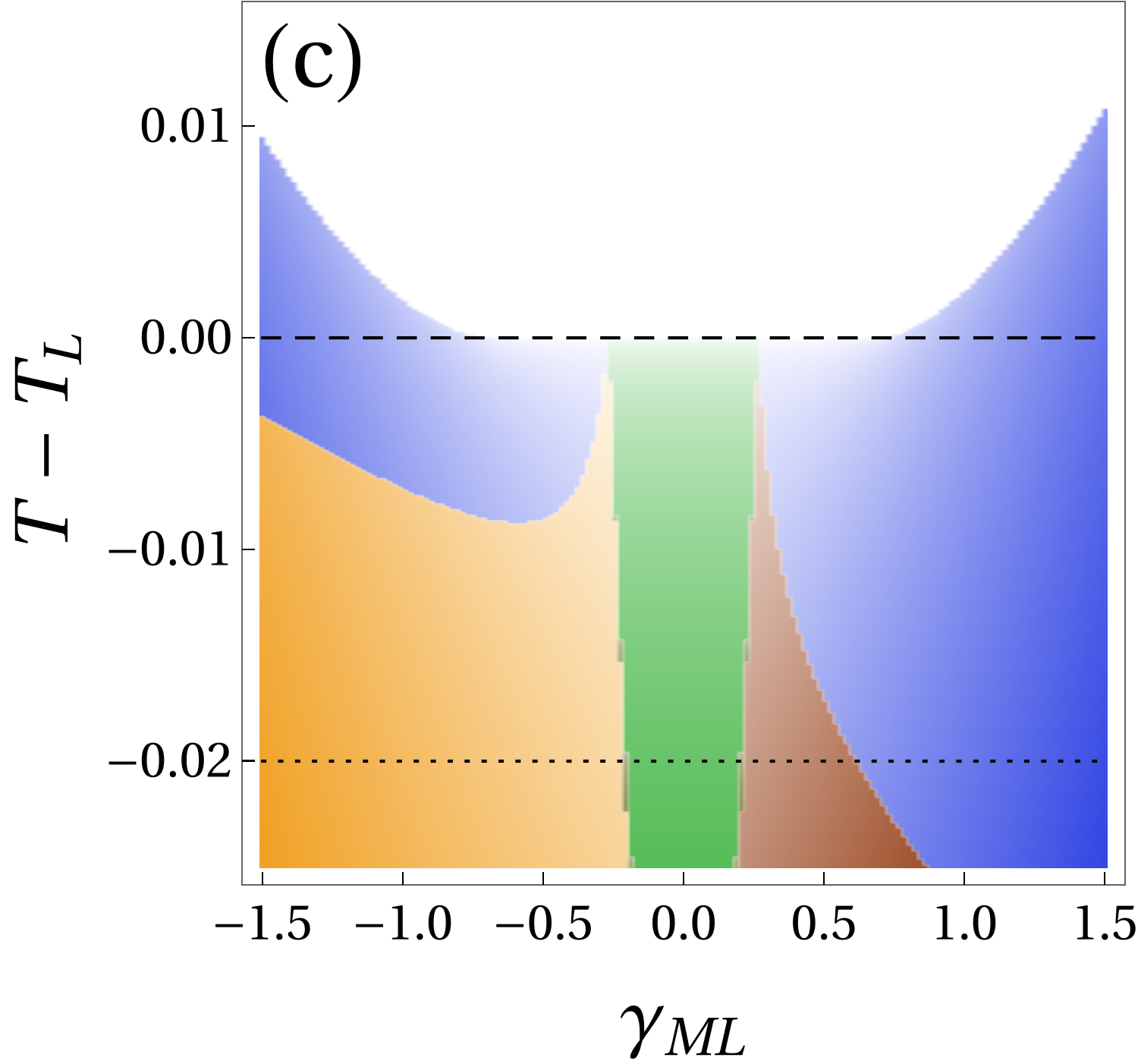}\includegraphics[width=0.49\columnwidth]{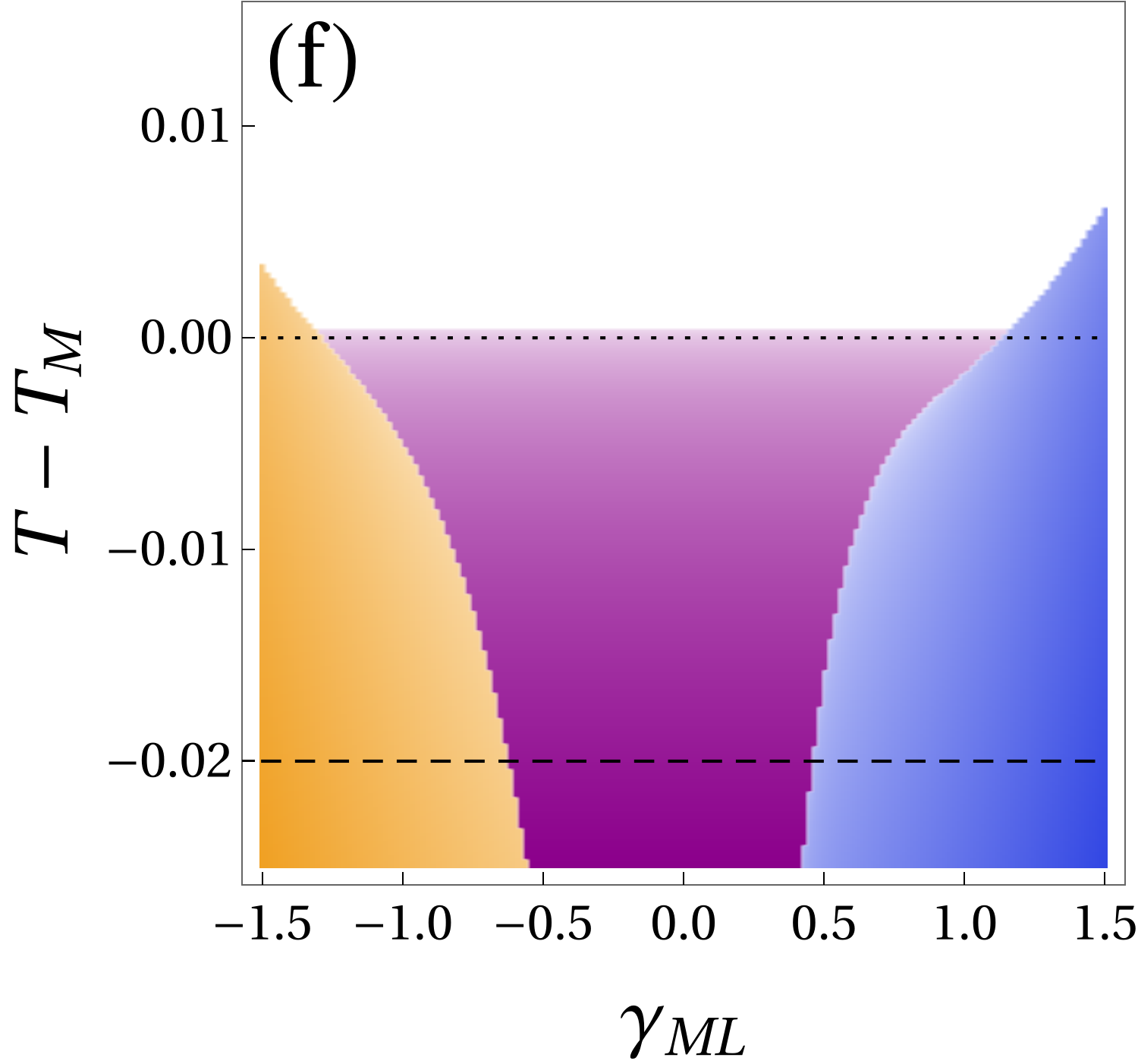}
\caption{\label{fig:increasing_coupling_phase_diagrams} Phase diagrams for coupled $M$ and $L$ CDW orders for progressively larger values of $\lambda_{ML}^{(i)}$ (top to bottom) as functions of $\gamma_{ML}$ and $T$. The parameters used are summarized in Table~\ref{tab:parameters}. The dashed black line denotes $T_L$ and the dotted black line denotes $T_M$. The color scheme is that summarized in Tables~\ref{tab:phase_summary_uncoupled}, \ref{tab:phase_summary_coupled}, and \ref{tab:phase_summary_mixed}. For panels (a)-(c), $T_L > T_M$ and in (d)-(f), $T_M > T_L$. The value of $\gamma_{ML}$ required to change the leading instability from a ``pure" CDW phase to a coupled CDW phase is much smaller for $T_L > T_M$ compared to $T_M > T_L$. While the alternating stripe phase (green) appearing for $T_L > T_M$ is rather unaffected by an increasing $\lambda_{ML}^{(i)}$, the planar Star-of-David phase (dark purple) appearing for $T_M > T_L$ becomes more prominent as $\lambda_{ML}^{(i)}$ increases. The red dashed line denotes the value of $\gamma_{ML}$ chosen in Figs.~\ref{fig:phase_diagrams_M_variation}--\ref{fig:phase_diagrams_ML_variation}.}
\end{figure}

Here, we illustrate how the coupling between $\mathcal{F}_M$ and $\mathcal{F}_L$ modifies the ``pure" phase diagrams in Fig. \ref{fig:phase_diagram_M}. In Fig.~\ref{fig:increasing_coupling_phase_diagrams} we show the numerically-calculated $\gamma_{ML}$-temperature phase diagrams for $T_L > T_M$ [panels (a)--(c)] and $T_M > T_L$ [panels (d)--(f)]. The phases are labeled according to the color scheme of Tables \ref{tab:phase_summary_uncoupled}, \ref{tab:phase_summary_coupled}, and \ref{tab:phase_summary_mixed}. The dashed (dotted) line denotes $T_L$ ($T_M$) in this and all subsequent figures. Upon moving from the top to the bottom panels, progressively larger values of the quartic couplings $\lambda_{ML}^{(i)}$ are considered, as shown in the last three columns of Table \ref{tab:parameters}. The coefficients of the uncoupled free energies are set by the values in the first five columns of the same table. Because we consider positive values for the ``pure" coefficients $\gamma_M$, $\lambda_M$, and $\lambda_L$, the leading instabilities in the uncoupled cases are an alternating stripe phase ($L_i$ case) and a planar Star-of-David phase ($M_i$ case). The addition of finite $\lambda_{ML}^{(i)}$ and $\gamma_{ML}$ changes this, and several additional phases appear. 

\begin{table}
\begin{tabular}{>{\centering}p{0.09\columnwidth}>{\centering}p{0.09\columnwidth}>{\centering}p{0.09\columnwidth}>{\centering}p{0.09\columnwidth}>{\centering}p{0.09\columnwidth}>{\centering}p{0.09\columnwidth}>{\centering}p{0.105\columnwidth}>{\centering}p{0.105\columnwidth}>{\centering\arraybackslash}p{0.105\columnwidth}}
\hline\hline
     $\gamma_M$ & $\lambda_M$ & $u_M$ & $\lambda_L$ & $u_L$ & $\gamma_{ML}$ & $\lambda_{ML}^{(1)}$ & $\lambda_{ML}^{(2)}$ & $\lambda_{ML}^{(3)}$ \\
     \hline
     \multirow{3}{*}{$0.25$} & \multirow{3}{*}{$0.6$} & \multirow{3}{*}{$1.2$} & \multirow{3}{*}{$0.8$} & \multirow{3}{*}{$1.5$} & \multirow{3}{*}{$-0.7$} & $0.01$ & $0.015$ & $0.0175$ \\
      &  &  &  & & & $0.5$ & $0.75$ & $0.875$ \\
      &  &  &  & & & $1.0$ & $1.5$ & $1.75$ \\
\hline\hline
\end{tabular}
\caption{\label{tab:parameters} Parameters used for the numerical minimization presented in Sec.~\ref{sec:phase_diagram} (unless the specific parameter is being varied as indicated in the text). The three rows of $\lambda_{ML}^{(i)}$ denote the three different sets used in Fig.~\ref{fig:increasing_coupling_phase_diagrams}. In Figs.~\ref{fig:phase_diagrams_M_variation}--\ref{fig:phase_diagrams_ML_variation} we use the middle row values for $\lambda_{ML}^{(i)}$.}
\end{table}

Starting with the case $T_L > T_M$ [panels (a)--(c)] in Fig.~\ref{fig:increasing_coupling_phase_diagrams}, for small values of $\gamma_{ML}$ the leading instability remains the alternating stripe phase (green). It gives way to the superimposed tri-hexagonal Star-of-David phase (blue) as the magnitude of $\gamma_{ML}$ increases. The staggered tri-hexagonal phase (orange) appears as a sub-leading instability in the $\gamma_{ML} < 0$ region, condensing at a lower temperature inside the superimposed tri-hexagonal Star-of-David phase. Other mixed phases presented in Table \ref{tab:phase_summary_mixed} appear for narrow ranges of $\gamma_{ML}$ at low temperatures. Upon increasing the positive-valued coefficients $\lambda_{ML}^{(i)}$ (i.e. moving from (a) to (c)), the staggered tri-hexagonal phase becomes more prominent. At the same time, the staggered Star-of-David phase (brown) emerges in the $\gamma_{ML} > 0$ region of the phase diagram, although it occupies a smaller area than its counterpart on the $\gamma_{ML} < 0$ region. The staggered tri-hexagonal and star-of-David phases do not become the leading instabilities for the parameters studied in panels (a)--(c).

Moving on to the case $T_M > T_L$ [panels (d)--(f)], the planar Star-of-David phase (dark purple) remains the leading instability for a large range of $\gamma_{ML}$ values -- larger than the range for which the alternating stripe phase appears in panels (a)--(c). This range of $\gamma_{ML}$ values is only weakly dependent on the magnitude of $\lambda_{ML}^{(i)}$, as one moves from panel (d) to panel (f). Larger absolute values of $\gamma_{ML}$ in panel (d) change the leading CDW instability to the superimposed tri-hexagonal Star-of-David phase (blue), although the staggered tri-hexagonal phase (orange) emerges at lower temperatures. Increasing $\lambda_{ML}^{(i)}$ as one moves from (d) to (f), the staggered tri-hexagonal phase expands and becomes the dominant and only leading instability for $\gamma_{ML}$ negative and large in magnitude. For large enough $\gamma_{ML}>0$, the superimposed tri-hexagonal Star-of-David phase remains the leading instability for all values of $\lambda_{ML}^{(i)}$ studied here. In contrast to the case $T_L > T_M$ [panels (a)--(c)], the staggered Star-of-David phase does not appear on the $\gamma_{ML}>0$ side of the phase diagrams (d)--(f). One possible explanation is that, because $\gamma_M > 0$ (see Table \ref{tab:parameters}), when $\gamma_{ML} > 0$ the superimposed tri-hexagonal Star-of-David phase (blue) becomes more robust as compared to the case when $\gamma_{ML} < 0$, since the two cubic terms favor the same $\mathrm{sign}(M_1 M_2 M_3)$ only when the two coefficients have the same sign.

\subsection{Robustness of the phase diagrams}

The behavior of the phase diagrams shown in Fig. \ref{fig:increasing_coupling_phase_diagrams} generally agrees with our qualitative analysis: a coupled CDW phase is stabilized and wins over the ``pure" CDW phase (green or purple) when the trilinear coefficient $\gamma_{ML}$ is relatively large (in magnitude). For most of the phase diagram, the dominant coupled CDW state is the superimposed tri-hexagonal Star-of-David phase (blue), although upon increasing the values of the positive quartic coefficients $\lambda_{ML}^{(i)}$, the staggered tri-hexagonal phase (orange) becomes more prominent in the $\gamma_{ML} < 0$ side of the phase diagram. 

We now consider the impact of further varying the various parameters of $\mathcal{F}_{\rm tot}$ to elucidate the robustness of the phase diagrams of Fig.~\ref{fig:increasing_coupling_phase_diagrams}. In particular, we are interested in establishing under which conditions the leading instability is towards one of the coupled CDW phases that break sixfold rotational symmetry -- i.e. the staggered tri-hexagonal (orange) and staggered Star-of-David (brown) phases. 

\begin{figure}
\includegraphics[width=0.48\columnwidth]{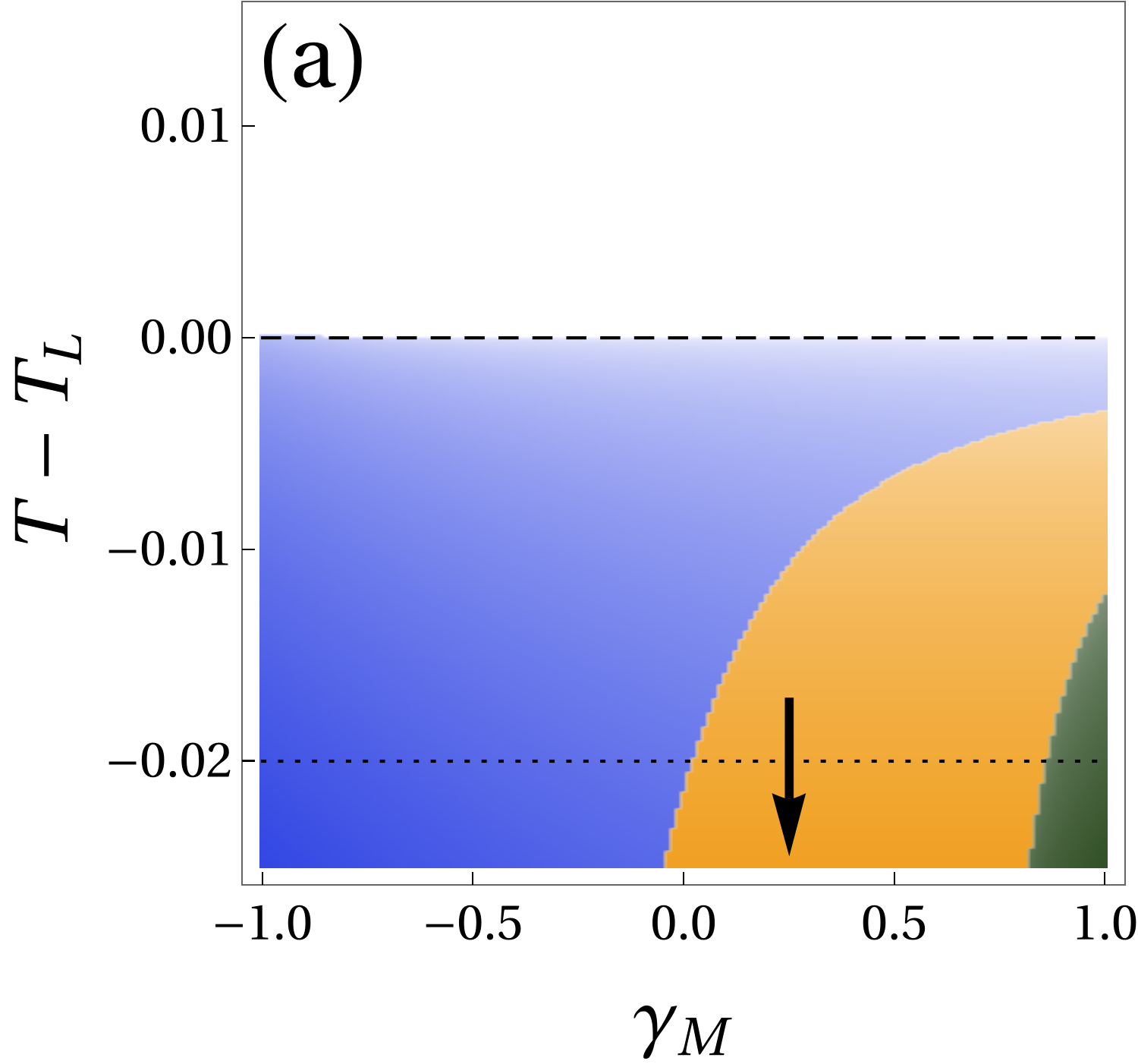}\includegraphics[width=0.48\columnwidth]{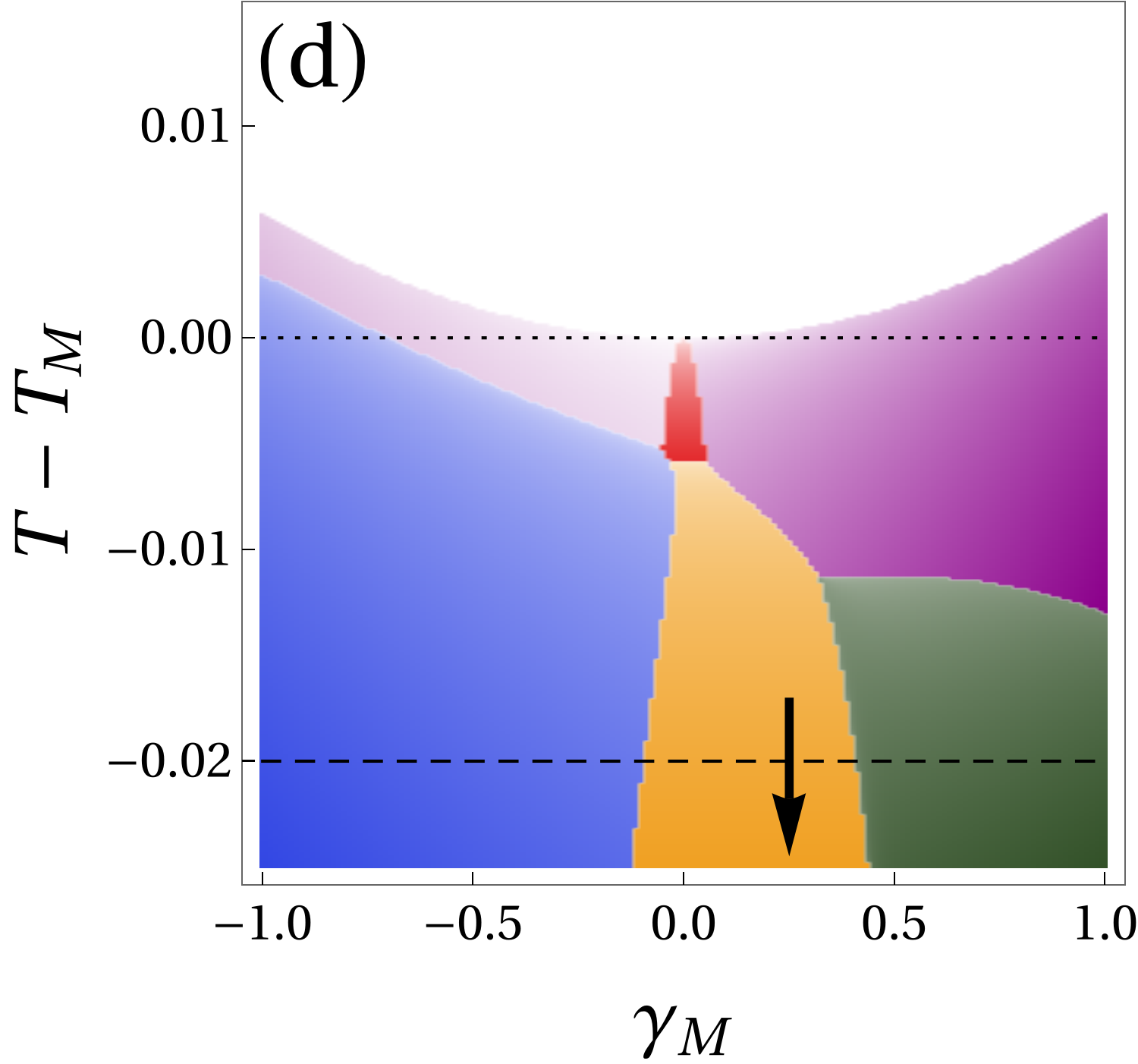}\\
\vspace{3mm}
\includegraphics[width=0.48\columnwidth]{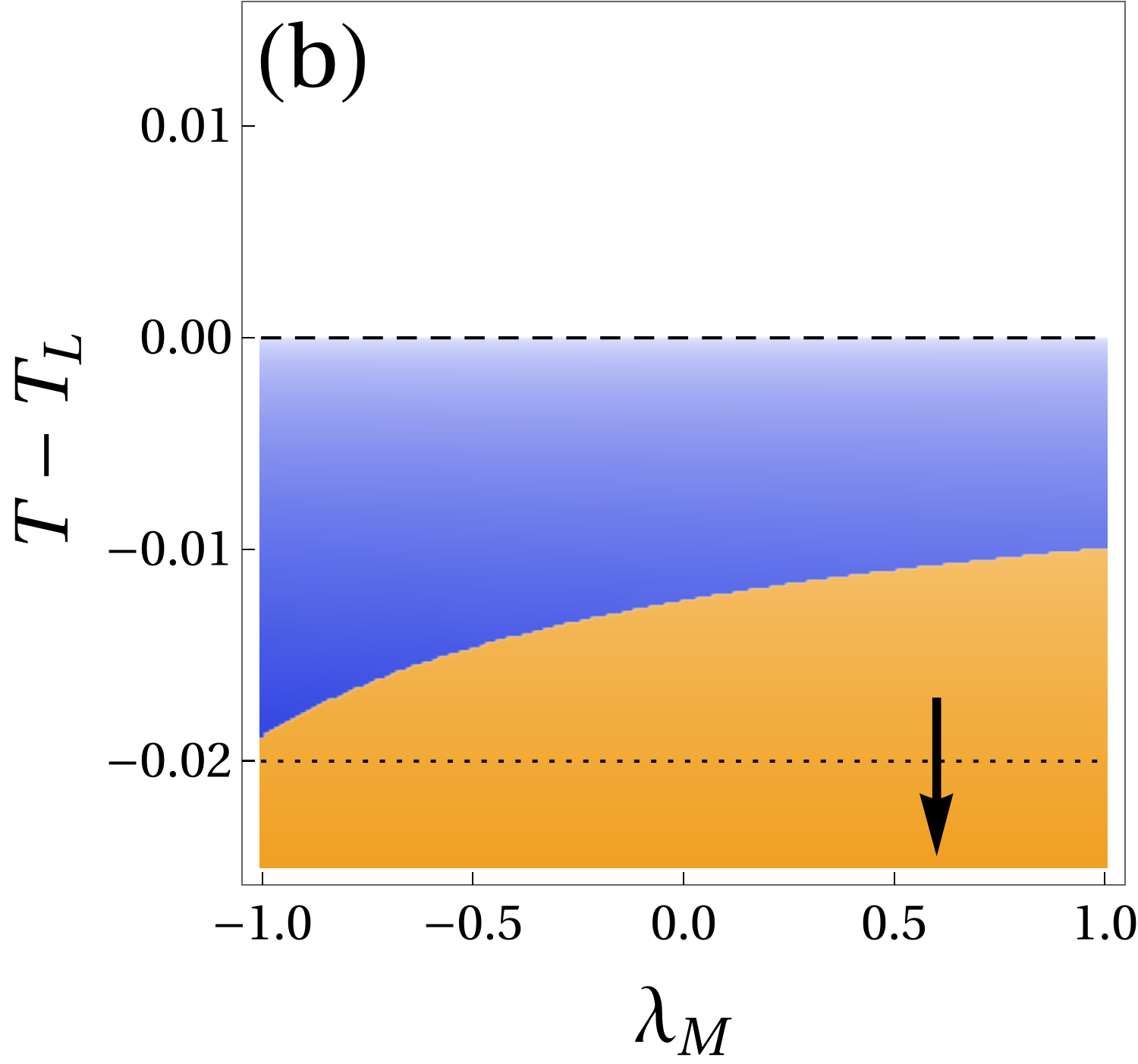}\includegraphics[width=0.48\columnwidth]{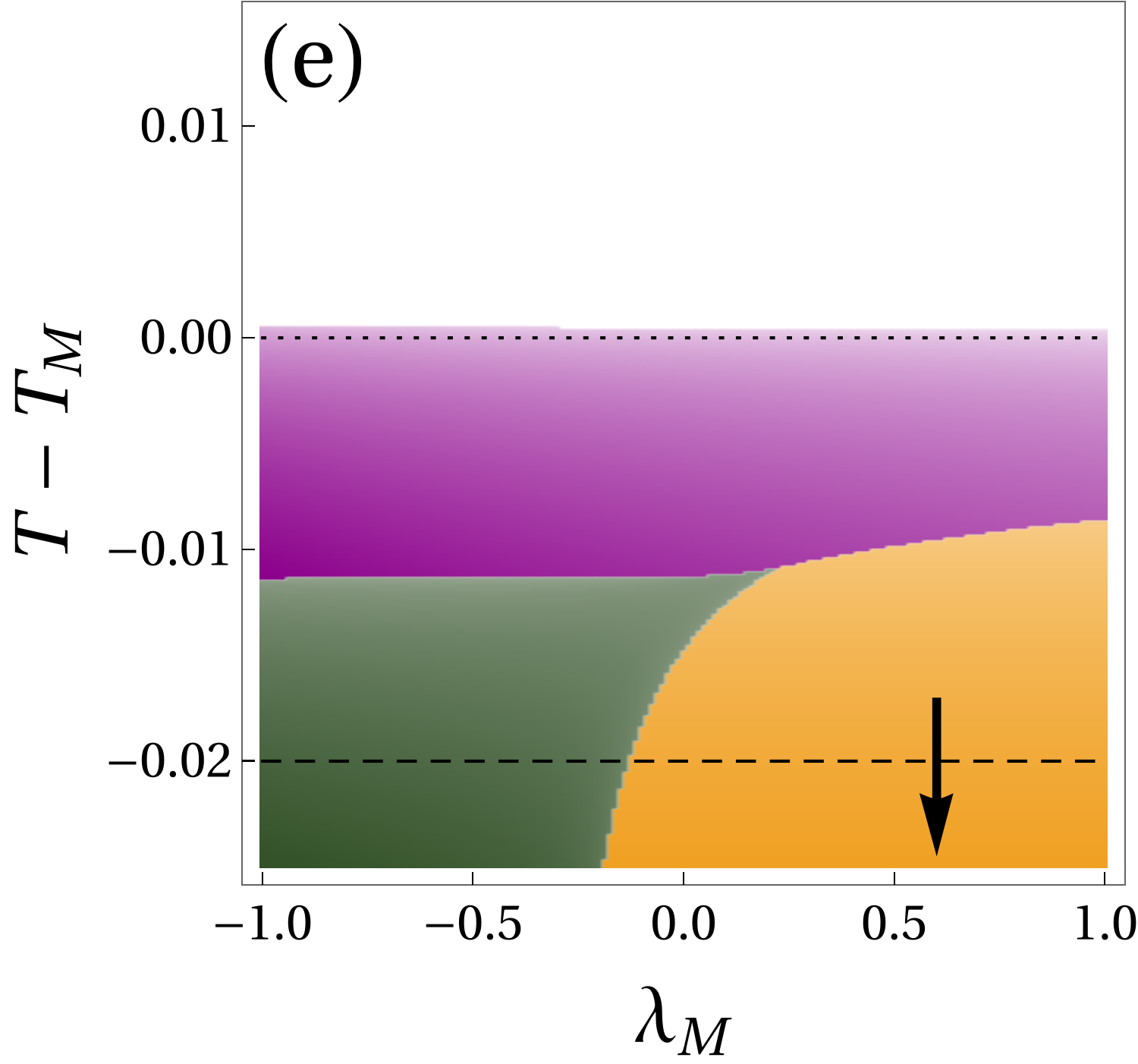}\\
\vspace{3mm}
\includegraphics[width=0.48\columnwidth]{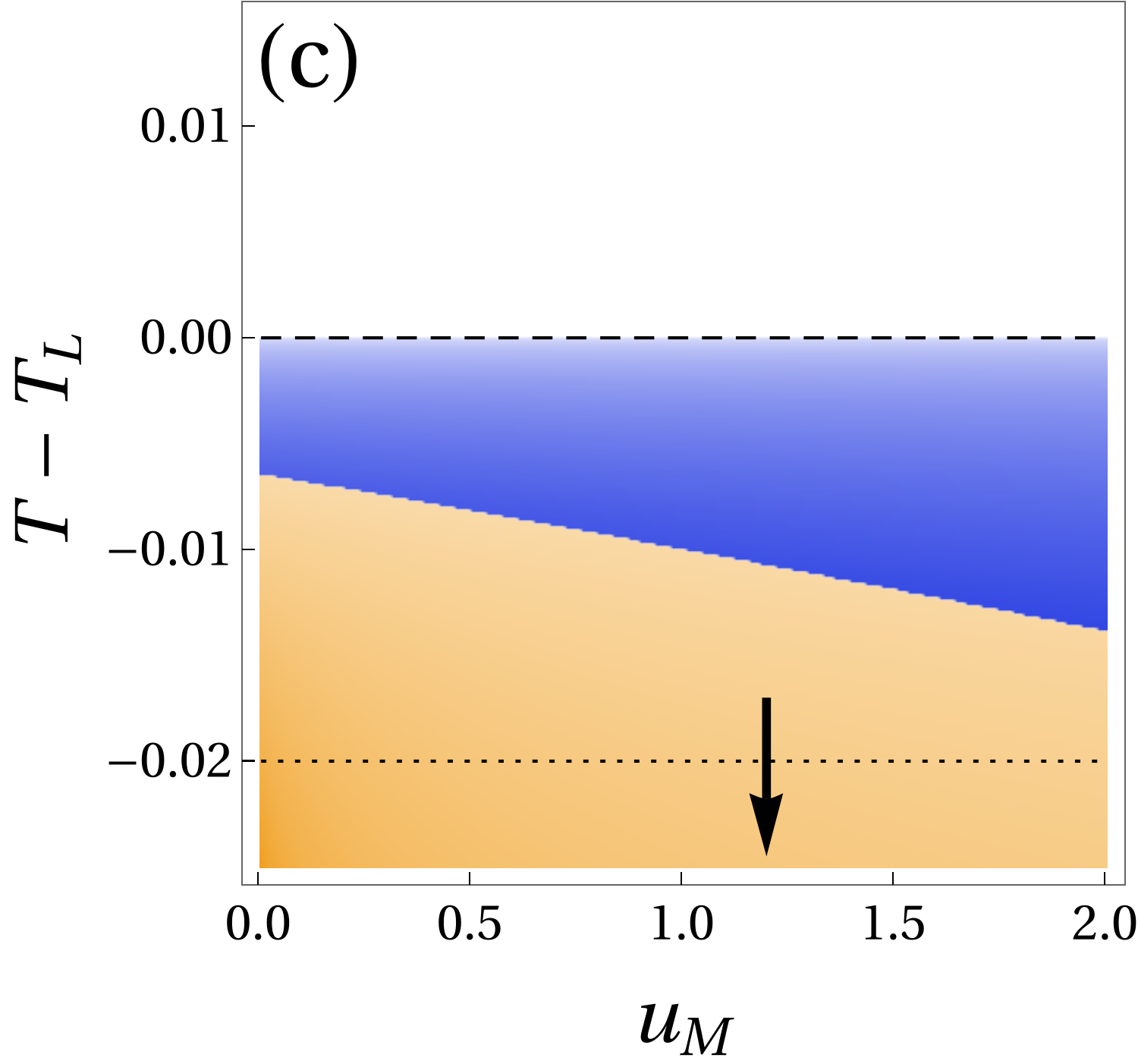}\includegraphics[width=0.48\columnwidth]{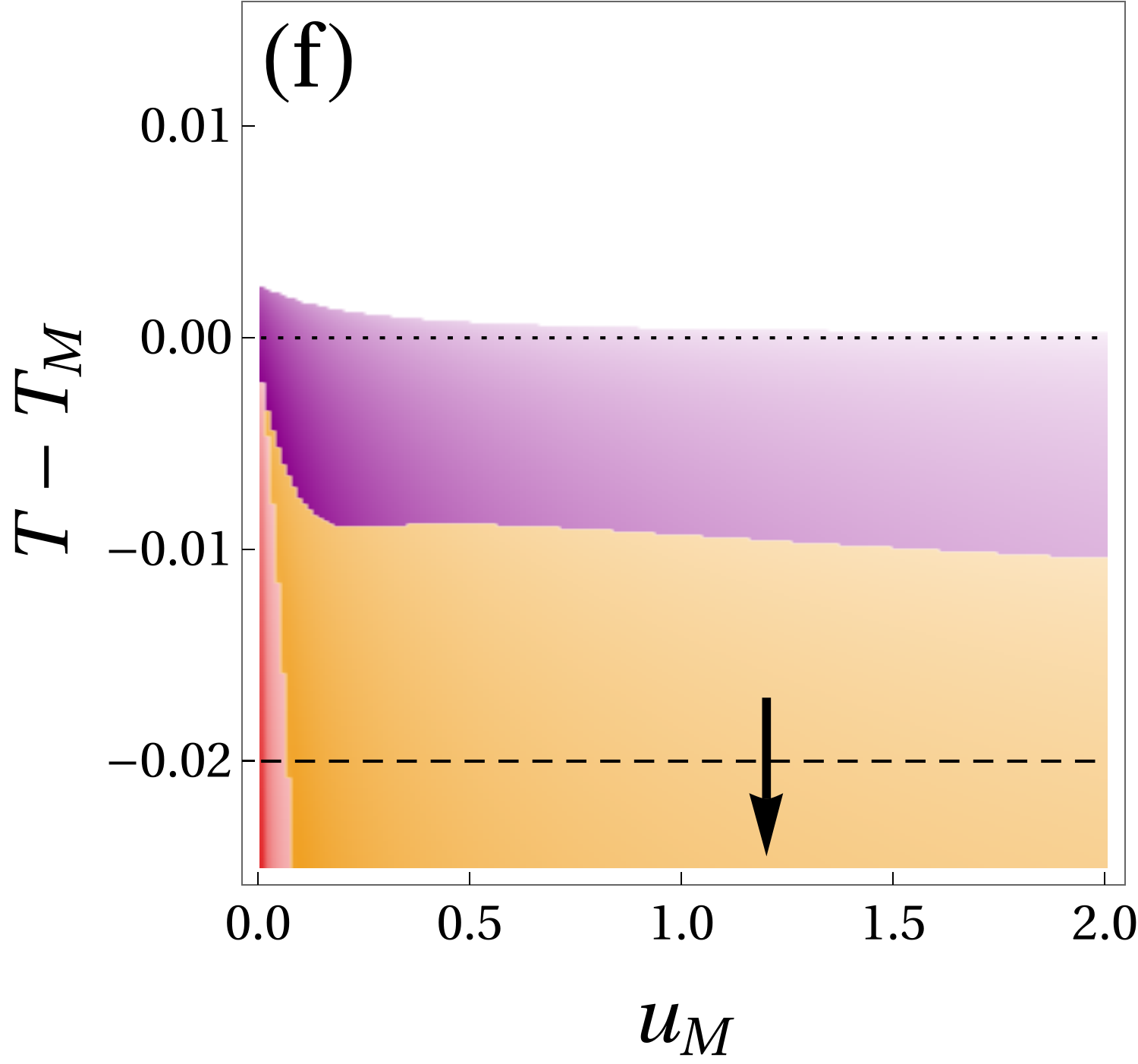}
\caption{\label{fig:phase_diagrams_M_variation} Impact of changing the coefficients $\gamma_M$, $\lambda_M$, and $u_M$ of $\mathcal{F}_M$, for fixed $\gamma_{ML}=-0.7$, on the phase diagrams of Figs.~\ref{fig:increasing_coupling_phase_diagrams}(b) and (e). Panels (a)--(c) correspond to $T_L > T_M$  and (d)--(f), to $T_M > T_L$. The arrows denote the parameters values corresponding to the dashed lines in Figs.~\ref{fig:increasing_coupling_phase_diagrams}(b) and (e). For the majority of cases, the leading instabilities are unaffected by changing these parameters. The only exception is the case of $\gamma_M$ which, for $T_M > T_L$ [panel (d)], causes a change from the planar tri-hexagonal to the planar Star-of-David phase. In this regime, the behavior is reminiscent of that observed in the ``pure" phase diagram Fig.~\ref{fig:phase_diagram_M}(a), indicating that $L_i$ plays nearly no role in this region of the phase diagram. In addition, changing $\gamma_M$ leads to a variety of additional phases at lower temperatures. In general, the impact of changing the parameters of $\mathcal{F}_M$ is the greatest when $T_M > T_L$, and the $M_i$ are the leading order parameters [panels (d)--(f)].}
\end{figure}

\begin{figure}
\includegraphics[width=0.48\columnwidth]{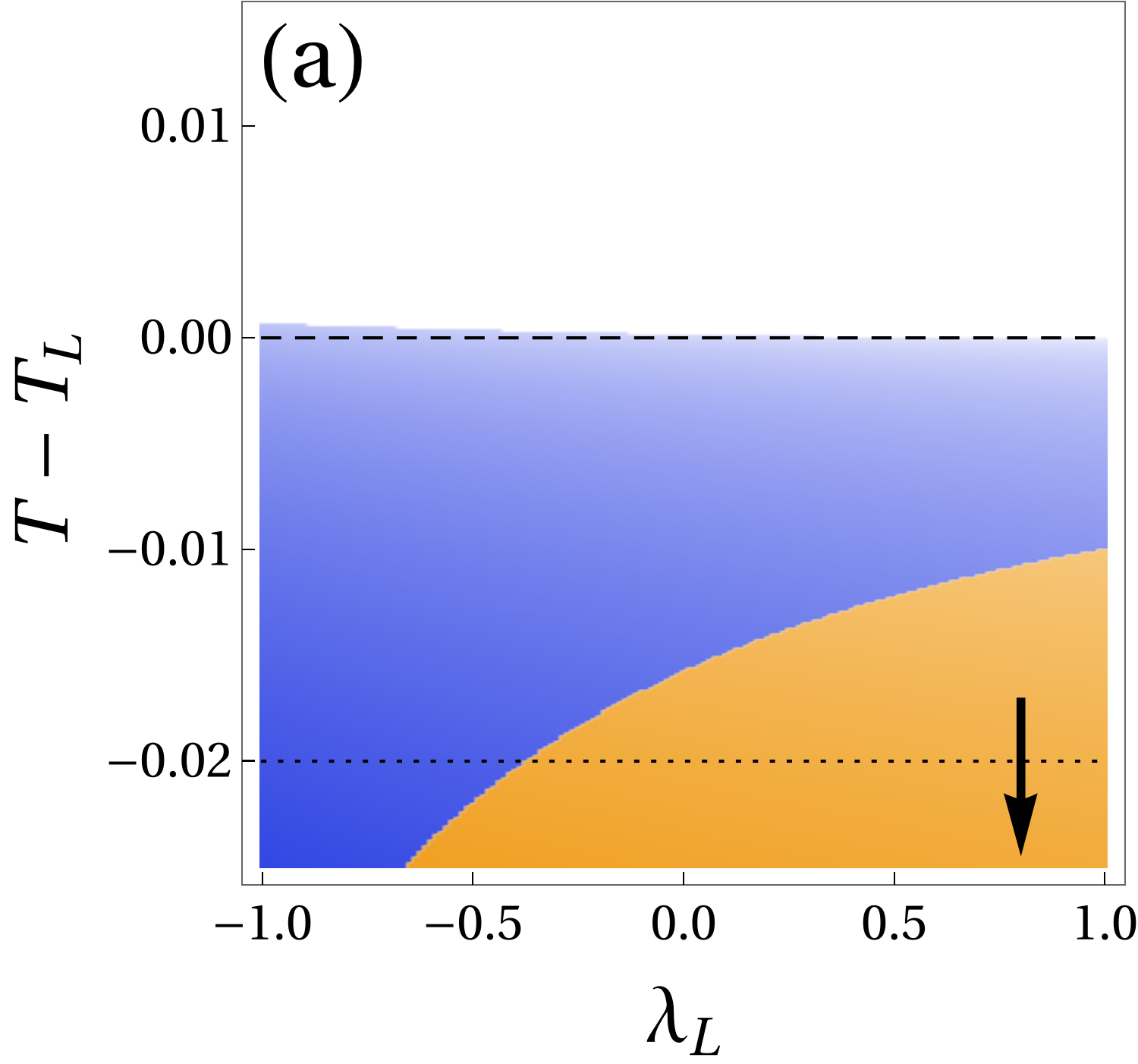}\includegraphics[width=0.48\columnwidth]{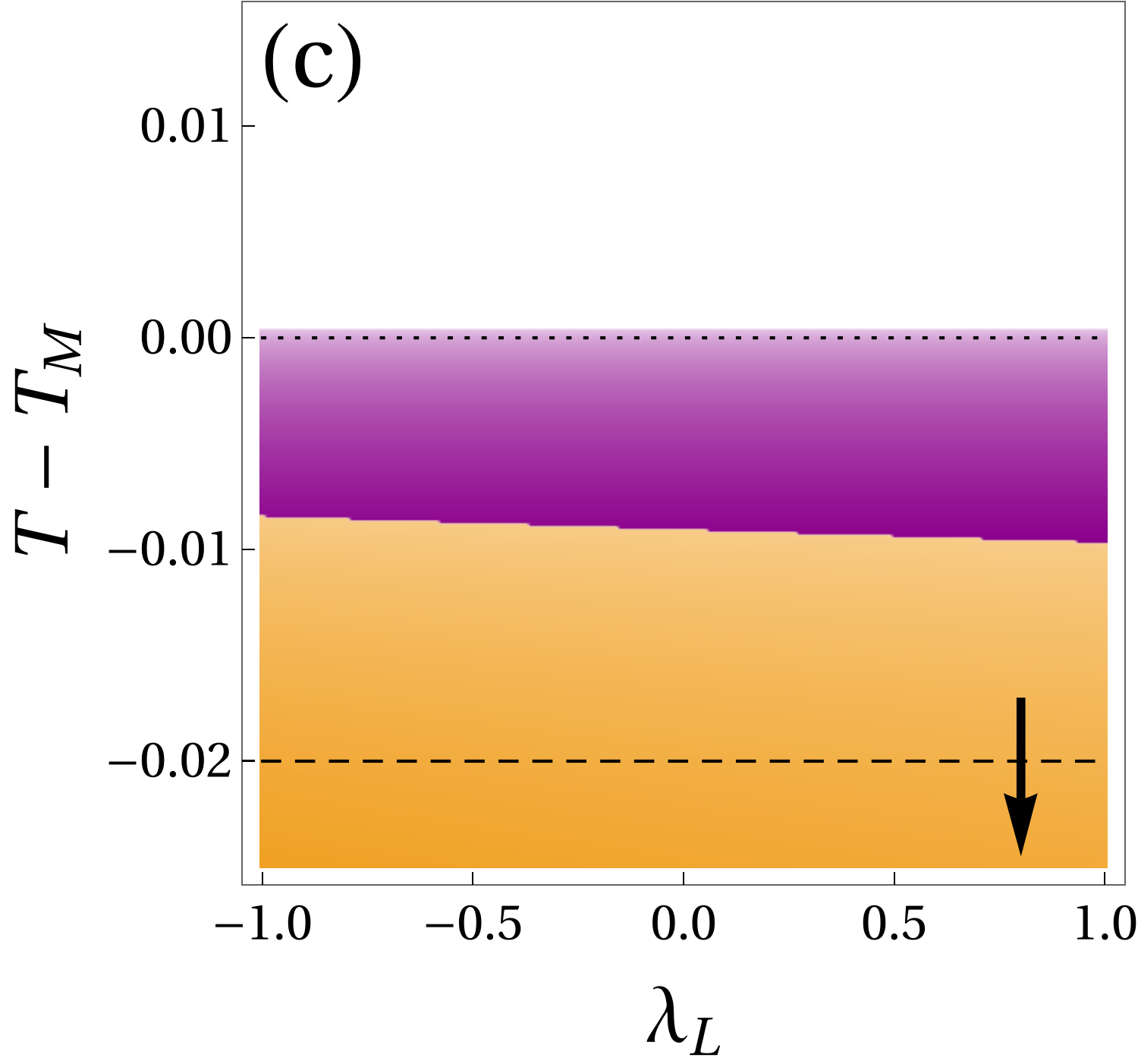}\\
\vspace{3mm}
\includegraphics[width=0.48\columnwidth]{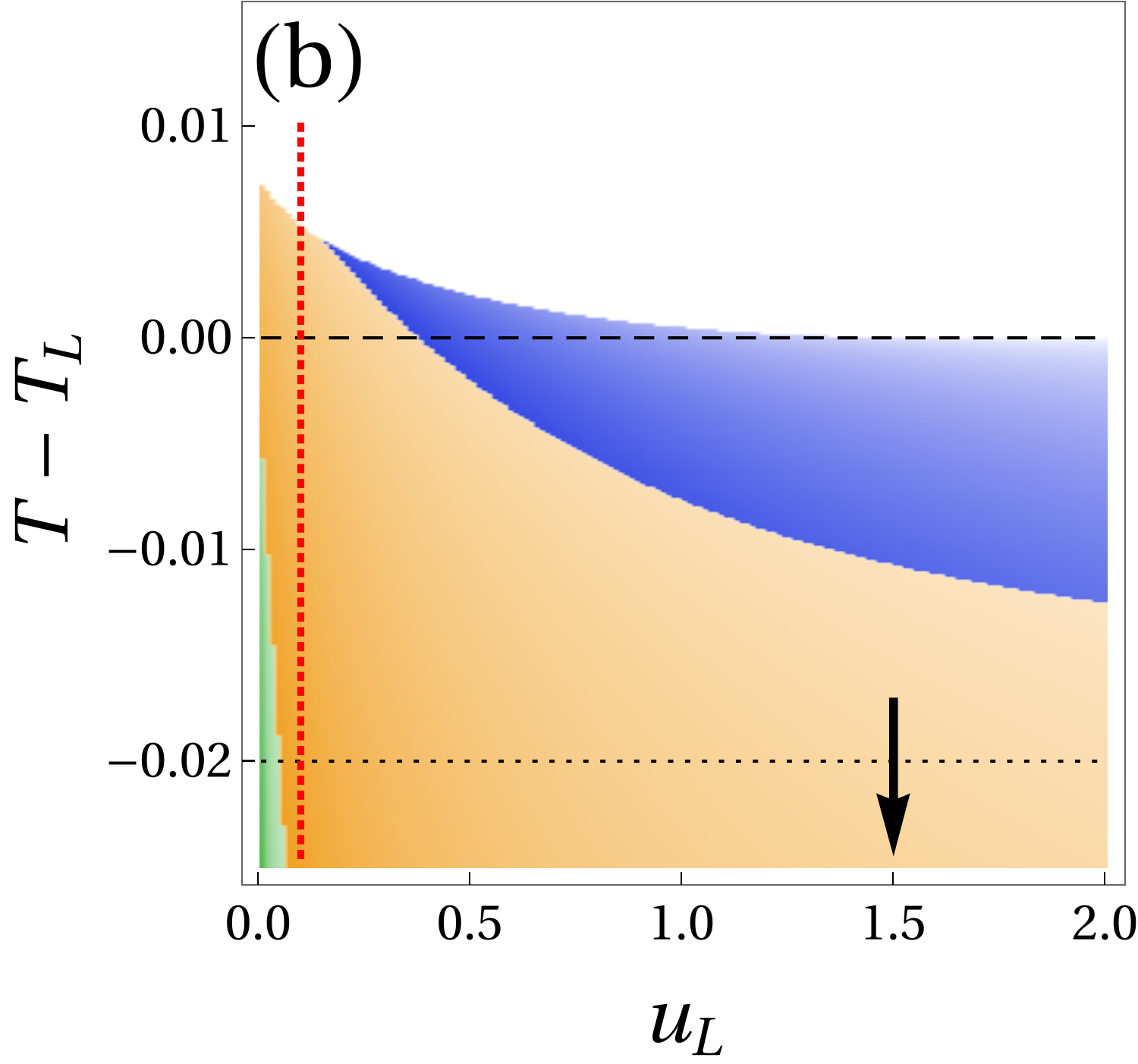}\includegraphics[width=0.48\columnwidth]{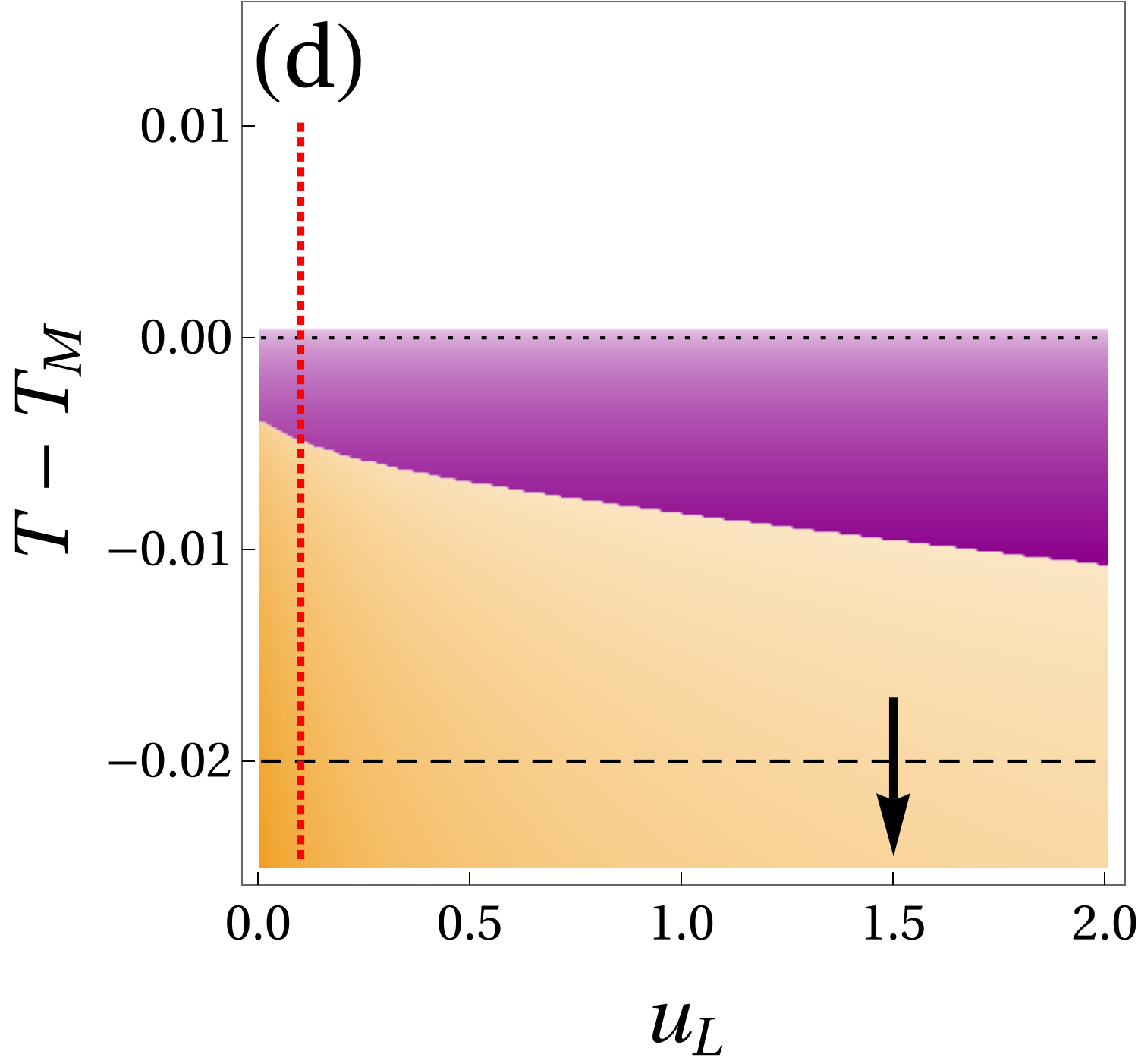}
\caption{\label{fig:phase_diagrams_L_variation} Impact of changing the coefficients $\lambda_L$ and $u_L$ of $\mathcal{F}_L$, for fixed $\gamma_{ML}=-0.7$, on the phase diagrams of Figs.~\ref{fig:increasing_coupling_phase_diagrams}(b) and (e). Panels (a)--(b) correspond to $T_L > T_M$  and (c)--(d), to $T_M > T_L$. The arrows denote the parameters values corresponding to the dashed lines in Figs.~\ref{fig:increasing_coupling_phase_diagrams}(b) and (e) while the red dotted line denotes the value of $u_L$ chosen in Fig.~\ref{fig:phase_diagrams_MLL_pref}(a)--(d). Overall, the impact of changing these parameters is minor, except for small values of $u_L$, where the leading instability becomes the staggered tri-hexagonal phase (orange) rather than the superimposed tri-hexagonal Star-of-David phase (blue).}
\end{figure}

\begin{figure}
\includegraphics[width=0.48\columnwidth]{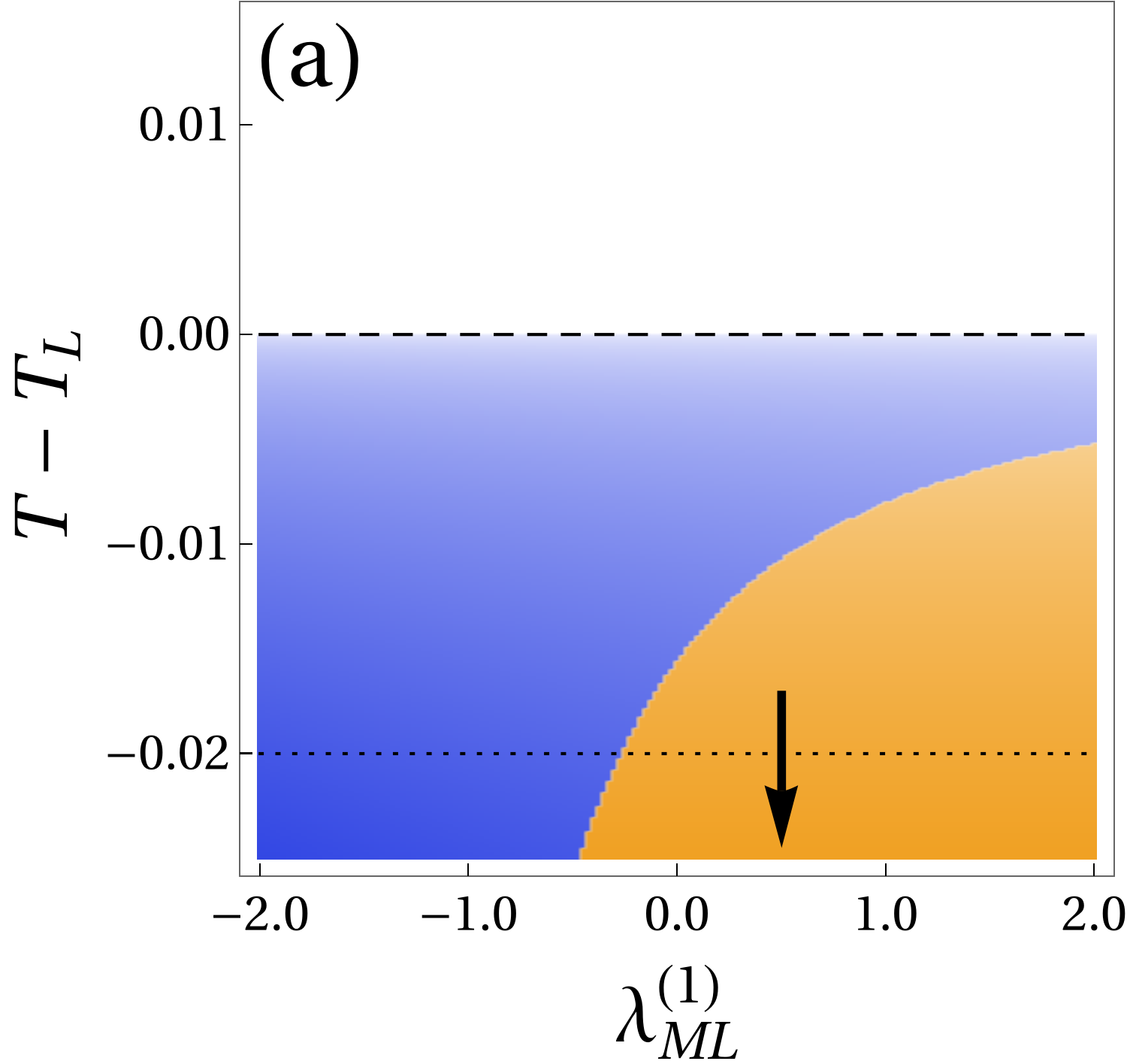}\includegraphics[width=0.48\columnwidth]{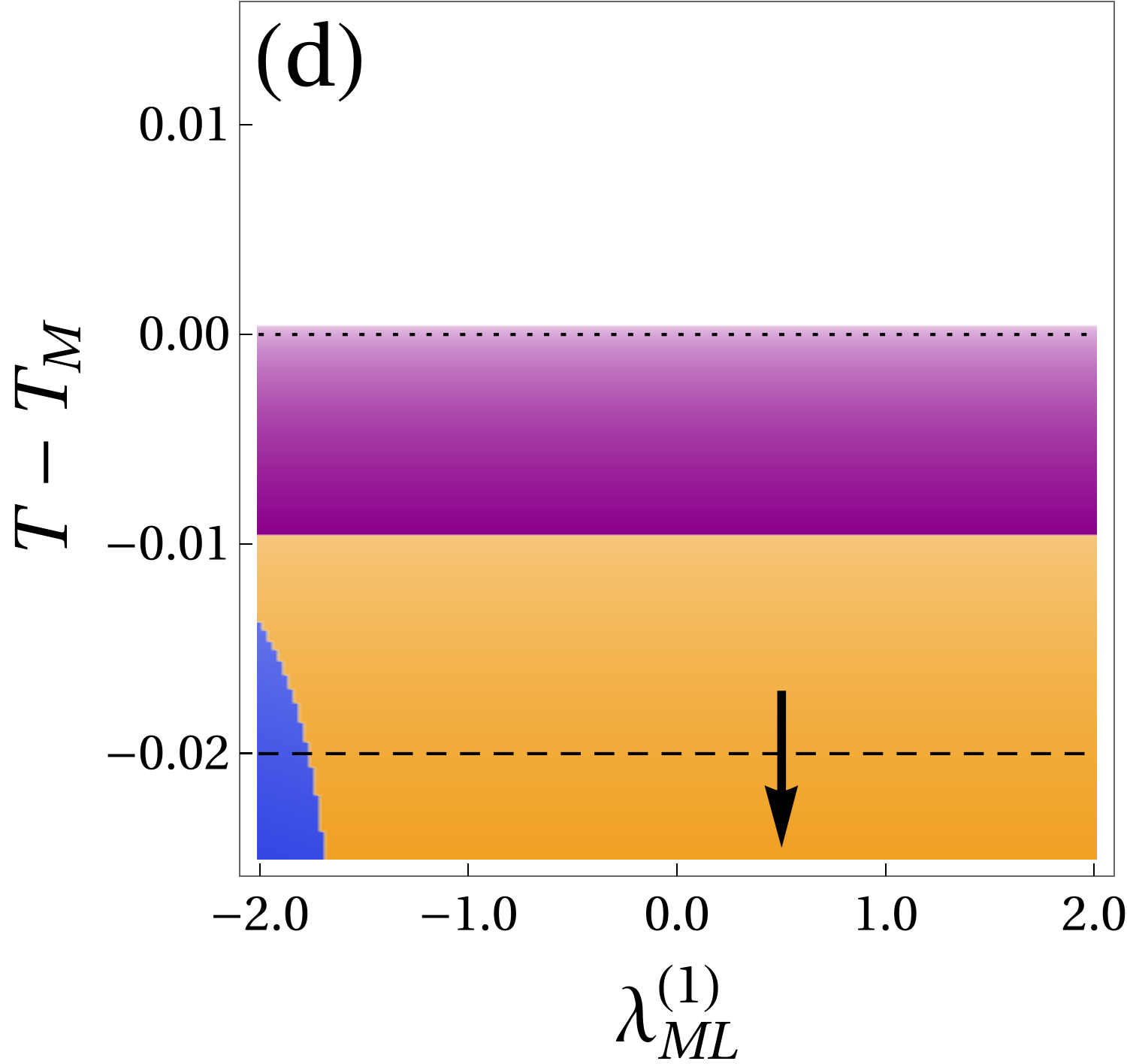}\\
\vspace{3mm}
\includegraphics[width=0.48\columnwidth]{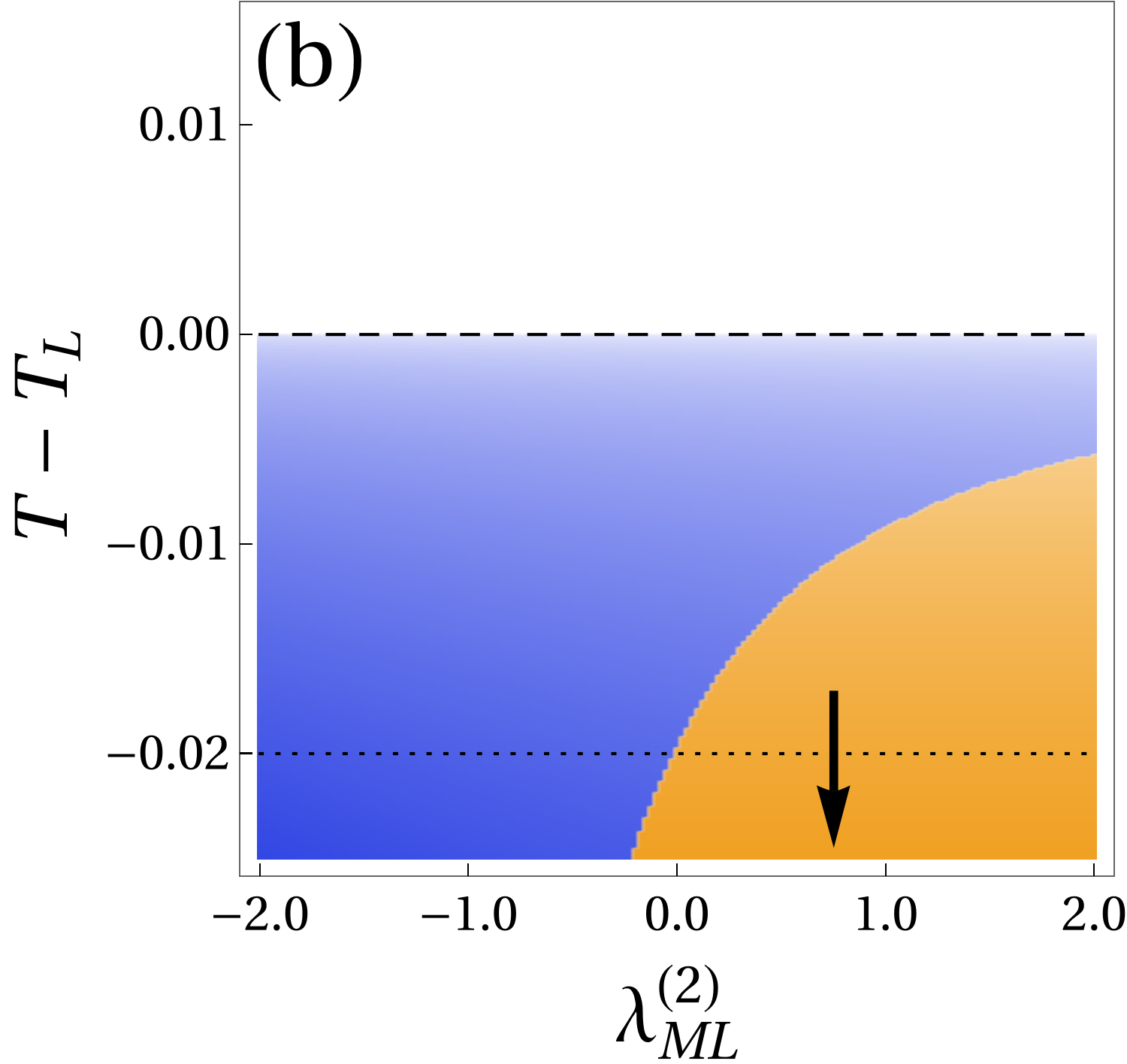}\includegraphics[width=0.48\columnwidth]{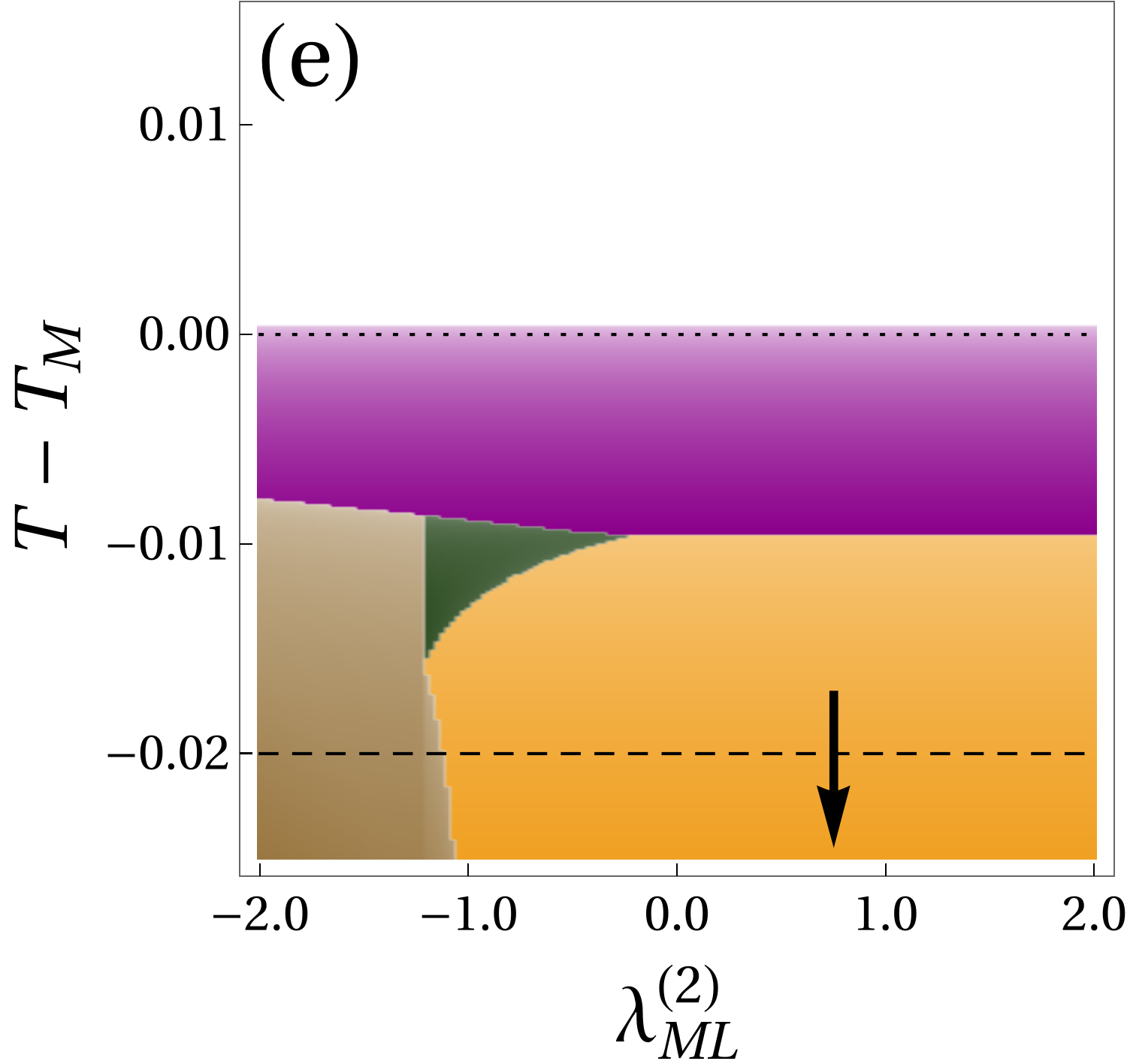}\\
\vspace{3mm}
\includegraphics[width=0.48\columnwidth]{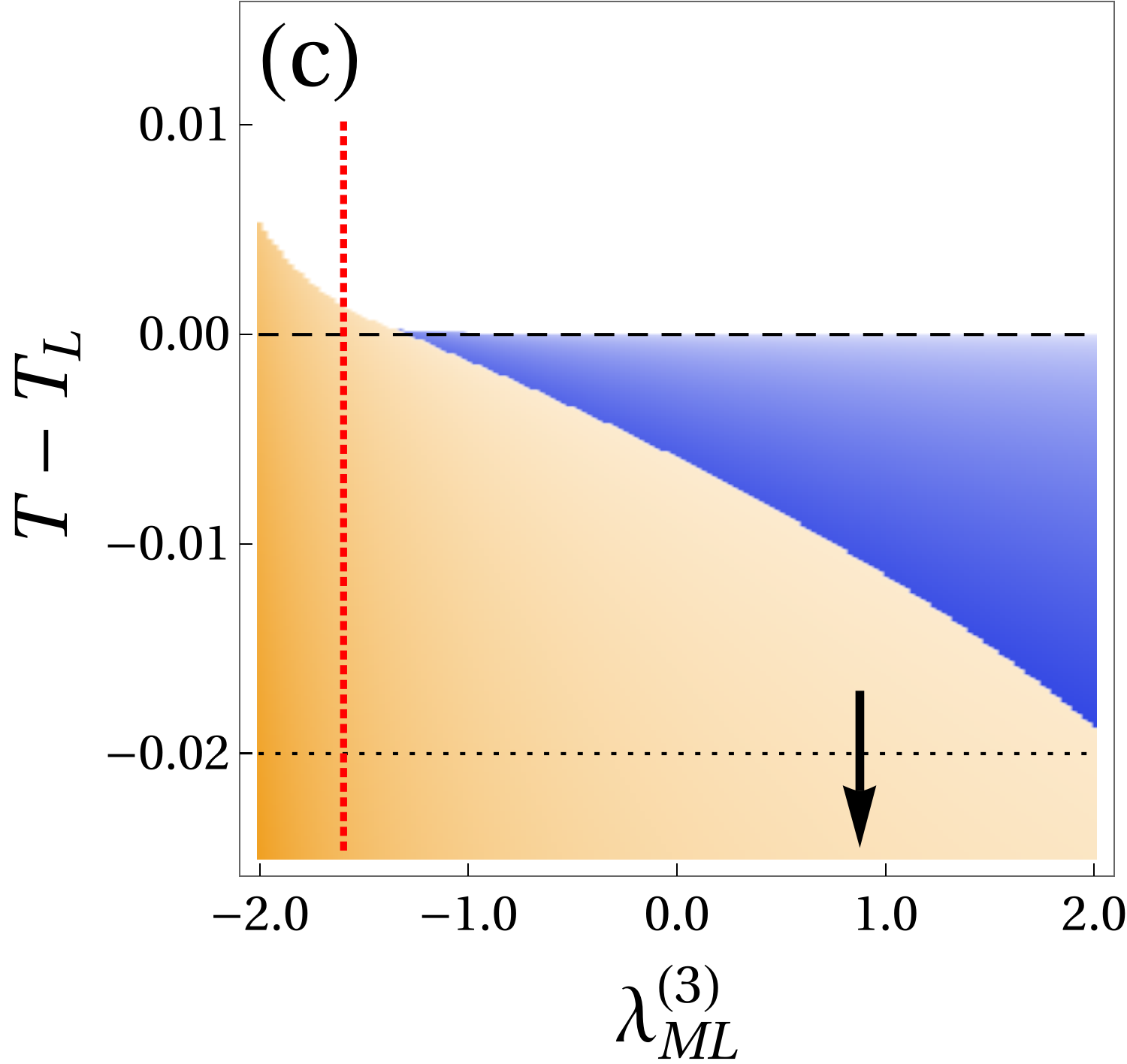}\includegraphics[width=0.48\columnwidth]{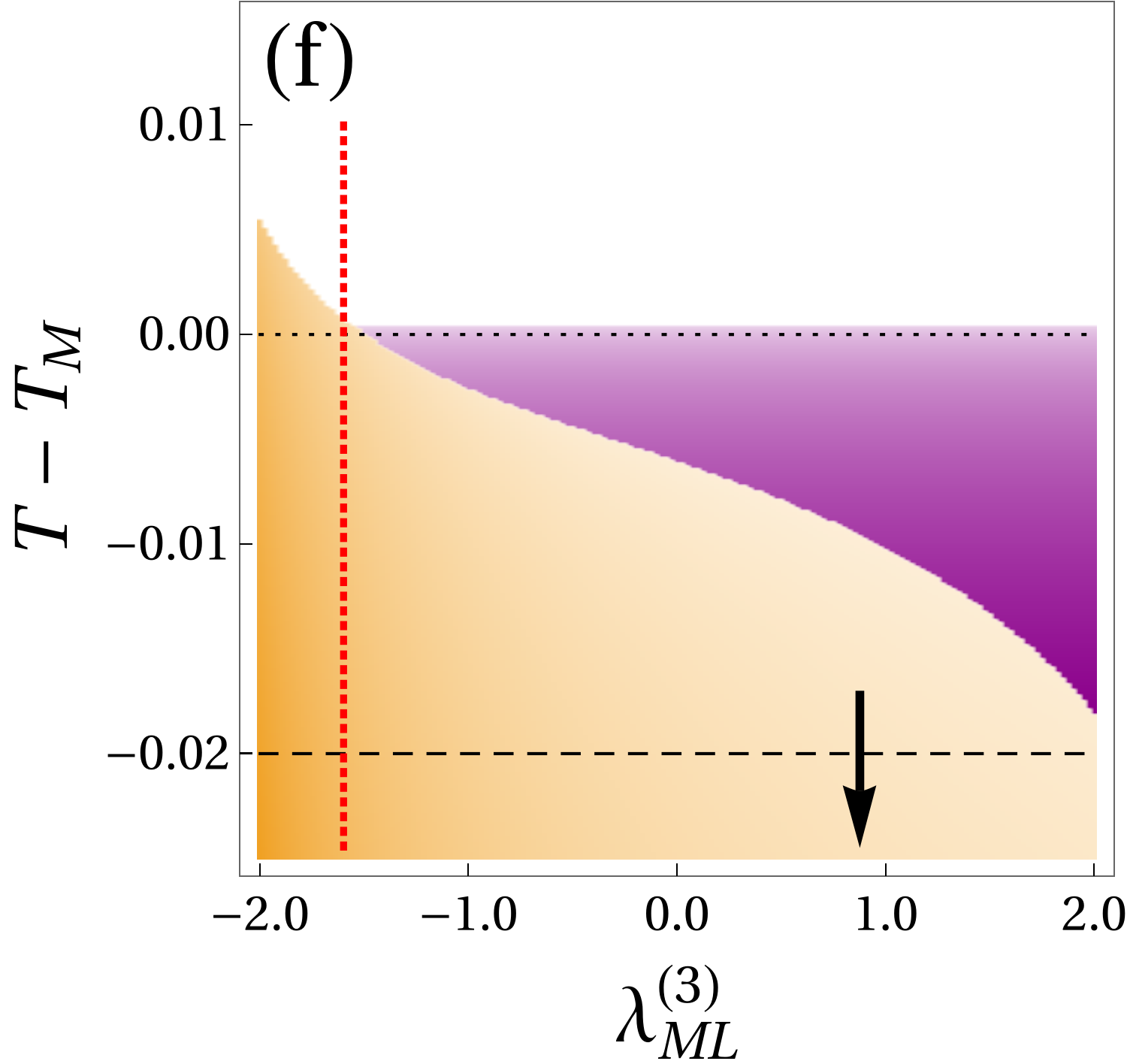}
\caption{\label{fig:phase_diagrams_ML_variation} Impact of changing the coefficients $\lambda_{ML}^{(1)}$, $\lambda_{ML}^{(2)}$, and $\lambda_{ML}^{(3)}$ of $\mathcal{F}_{ML}$, for fixed $\gamma_{ML}=-0.7$, on the phase diagrams of Figs.~\ref{fig:increasing_coupling_phase_diagrams}(b) and (e). Panels (a)--(c) correspond to $T_L > T_M$  and (d)--(f), to $T_M > T_L$. The arrows denote the parameters values corresponding to the dashed lines in Figs.~\ref{fig:increasing_coupling_phase_diagrams}(b) and (e) while the red dotted line denotes the value of $\lambda_{ML}^{(3)}$ chosen in Fig.~\ref{fig:phase_diagrams_MLL_pref}(e)--(h). The coefficients $\lambda_{ML}^{(1)}$ and $\lambda_{ML}^{(2)}$ do not change the leading instabilities, affecting only the secondary transitions at lower temperatures. On the other hand, when $\lambda_{ML}^{(3)}$ is negative, it can change the leading transition to the staggered tri-hexagonal phase (orange), regardless of whether $T_L > T_M$ [panel (c)] or $T_M > T_L$ [panel (f)].}
\end{figure}

We focus on the set of parameters given by the central row of Table \ref{tab:parameters}, marked by the vertical red dashed lines in Figs.~\ref{fig:increasing_coupling_phase_diagrams}(b) and (e). In particular, across Figs.~\ref{fig:phase_diagrams_M_variation}--\ref{fig:phase_diagrams_ML_variation}, we fix $\gamma_{ML}=-0.7$ and vary the other eight parameters that appear, respectively, in $\mathcal{F}_M$, $\mathcal{F}_L$, and $\mathcal{F}_{ML}$, for both $T_L > T_M$ (left panels) and $T_M > T_L$ (right panels). In these figures, the arrows denote the parameter values corresponding to the red dashed lines of Figs.~\ref{fig:increasing_coupling_phase_diagrams}(b) and (e), which are the same as those presented in the central row of Table \ref{tab:parameters}. 

We start by analyzing the impact of the changes in the coefficients of $\mathcal{F}_M$ in Fig.~\ref{fig:phase_diagrams_M_variation}. From simple power counting, the cubic coefficient $\gamma_M$ is expected to have a stronger impact on the phase diagram compared to the quartic coefficients $\lambda_M$ and $u_M$. This expectation is confirmed by the plots in Fig.~\ref{fig:phase_diagrams_M_variation}. Although changing $\gamma_M$ leads to the appearance of a variety of additional phases as secondary transitions, it does not alter the leading instabilities, except for the finely tuned case at $\gamma_M=0$. Therefore, $\gamma_{ML}$ remains the most important of the two cubic coefficients to determine the leading instability. Note that the superimposed tri-hexagonal Star-of-David phase (blue) becomes more robust when the signs of $\gamma_M$ and $\gamma_{ML}$ are the same, in agreement with what we discussed above.

The coefficients $\lambda_M$ and $u_M$, on the other hand, have a more limited effect, although $\lambda_M$ does lead to a few changes at lower temperatures when the $M_i$ order parameters are dominant, i.e. $T_M > T_L$ [Figs.~\ref{fig:phase_diagrams_M_variation}(d)--(f)]. Note that $u_M$ must be positive for the free energy to remain bounded, as argued in Sec.~\ref{sec:only_M}. In Figs.~\ref{fig:phase_diagrams_M_variation}(d) and (e), we observe the appearance of a state that mixes a triple-$\mbf{Q}_M$ phase with a double-$\mbf{Q}_L$ phase (dark green), described by three distinct order parameters, $(M_1 \overline{M}_1 M_2)+(L_1 L_1 0)$, as shown in Table \ref{tab:phase_summary_mixed}.

Moving on to Fig. \ref{fig:phase_diagrams_L_variation}, we consider the impact of the two coefficients of $\mathcal{F}_L$. While changing $\lambda_L$ does not lead to a qualitatively different phase diagram, it has a minor impact on the onset temperature of the secondary transition in the case $T_L > T_M$, as seen in panel (a). On the other hand, $u_L$ prominently impacts the phase diagrams, inducing a transition between the staggered tri-hexagonal phase (orange) and the superimposed tri-hexagonal Star-of-David phase (blue) as the leading instability of the system. Such a transition occurs for a very small value of $u_L$. Although the transition is only seen for $T_L > T_M$ [panel (b)], a similar trend is seen in the case $T_M > T_L$ [panel (d)].

Fig.~\ref{fig:phase_diagrams_ML_variation} shows the effect of the coupling constants $\lambda_{ML}^{(i)}$. Generally, their main impact is on the shape of the phase diagrams below the leading transition temperature. The only exception is the case of $\lambda_{ML}^{(3)}$: negative values of this coefficient stabilize the staggered tri-hexagonal phase (orange) as the leading instability of the system. We emphasize that the free energy remains bounded in all cases studied here.

Overall, the analyses presented in Figs.~\ref{fig:phase_diagrams_M_variation}--\ref{fig:phase_diagrams_ML_variation} reveal that, at least in what concerns the leading instabilities, the phase diagrams of Fig. \ref{fig:increasing_coupling_phase_diagrams} are rather robust against independent variations of the other eight Landau coefficients. This confirms that it is the cubic coefficient $\gamma_{ML}$ of $\mathcal{F}_{ML}$ which is responsible for promoting a coupled $M$-$L$ CDW state, with the quartic coefficients $\lambda_{ML}^{(i)}$ and the relative sign of $\gamma_M$ with respect to $\gamma_{ML}$ selecting between the triple-$\mbf{Q}_M$/triple-$\mbf{Q}_L$ phase and the single-$\mbf{Q}_M$/double-$\mbf{Q}_L$ phase. We once again emphasize that the nine-dimensional parameter space is vast, and thus we cannot rule out the appearance of other states.

\subsection{Sixfold-rotational symmetry breaking phases}\label{sec:rotational_sym}

Recent experiments on kagome metals are consistent with a CDW state that doubles the unit cell in all directions ($2\times2\times2$ unit cell) and lowers the sixfold-rotational symmetry of the lattice to twofold \cite{Ratcliff2021,Jiang2020,Zhao2021b,Chen2021,Li2021_rotation}. In this subsection, we further explore the possibility of obtaining such a CDW phase as the leading instability of the system. The breaking of the sixfold rotational symmetry is signaled by the onset of three nematic order parameters:
\begin{align}
    \Phi_M &= \begin{pmatrix}
        M_1^2 + M_3^2-2 M_2^2 \\
        \sqrt{3}\left(M_3^2 - M_1^2 \right)
    \end{pmatrix}, \label{eq:nematics} \\
    \Phi_L &= \begin{pmatrix}
        L_1^2 + L_3^2-2 L_2^2 \\
        \sqrt{3}\left(L_3^2 - L_1^2 \right)
    \end{pmatrix},\\
    \Phi_{ML} &= \begin{pmatrix}
        M_1 L_2 L_3 + M_3 L_1 L_2 - 2 M_2 L_1 L_3  \\
        \sqrt{3}\left( M_3 L_1 L_2  - M_1 L_2 L_3 \right)
    \end{pmatrix}\,.
\end{align}
Each of them, obtained using the INVARIANTS tool~\cite{Hatch2003}, transforms as the $\Gamma_5^+$ irreducible representation of the space group $P6/mmm$. $\Gamma_5^+$ is the irreducible representation corresponding to three-state Potts-nematic order \cite{Hecker2018,Fernandes2020} in a lattice with sixfold or threefold rotational symmetry, and it is subducted to the $E_{2g}$ irreducible representation of the $D_{6h}$ point group~\cite{footnote_1}.

\begin{figure*}
\includegraphics[width=0.245\textwidth]{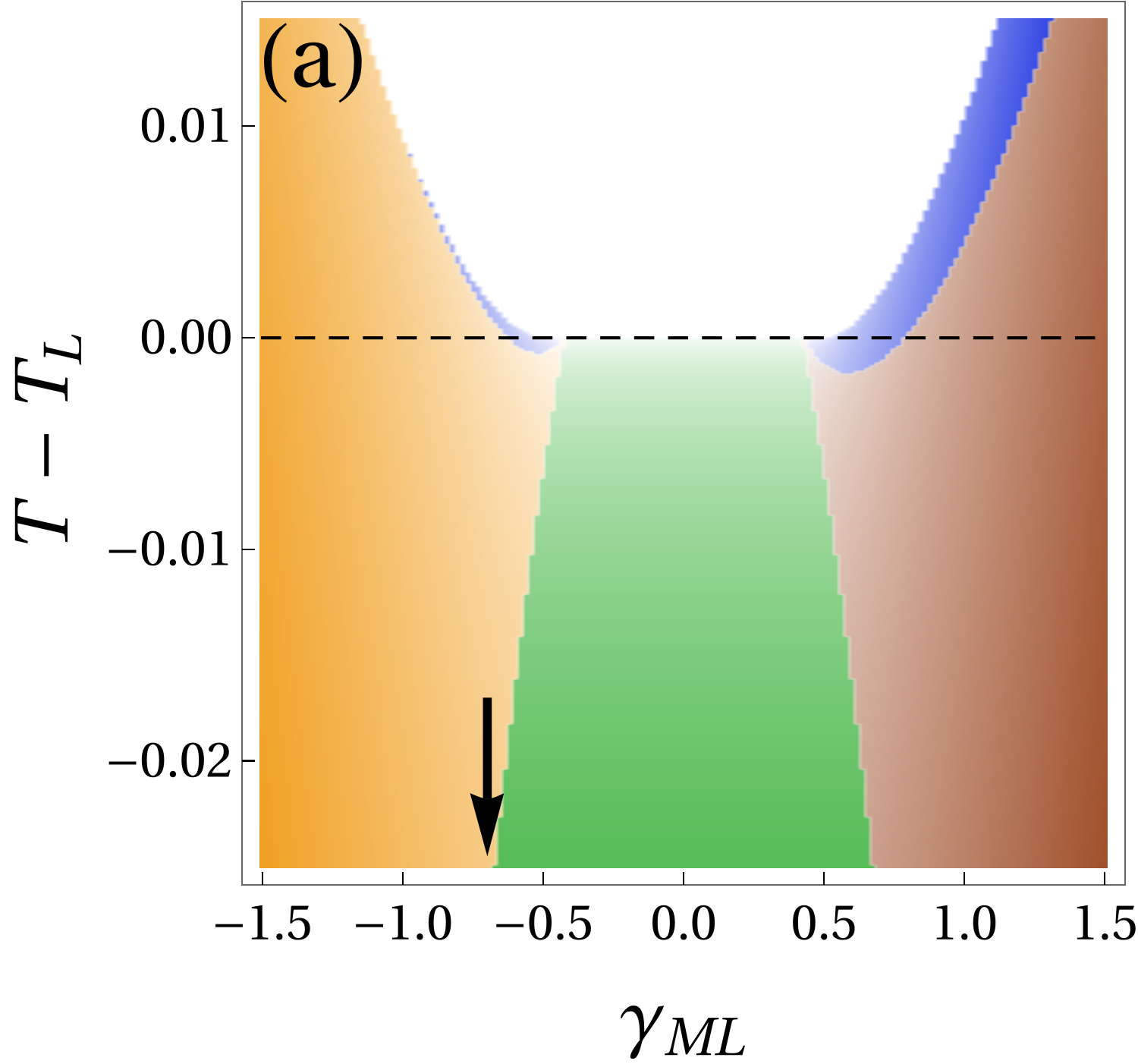}\includegraphics[width=0.245\textwidth]{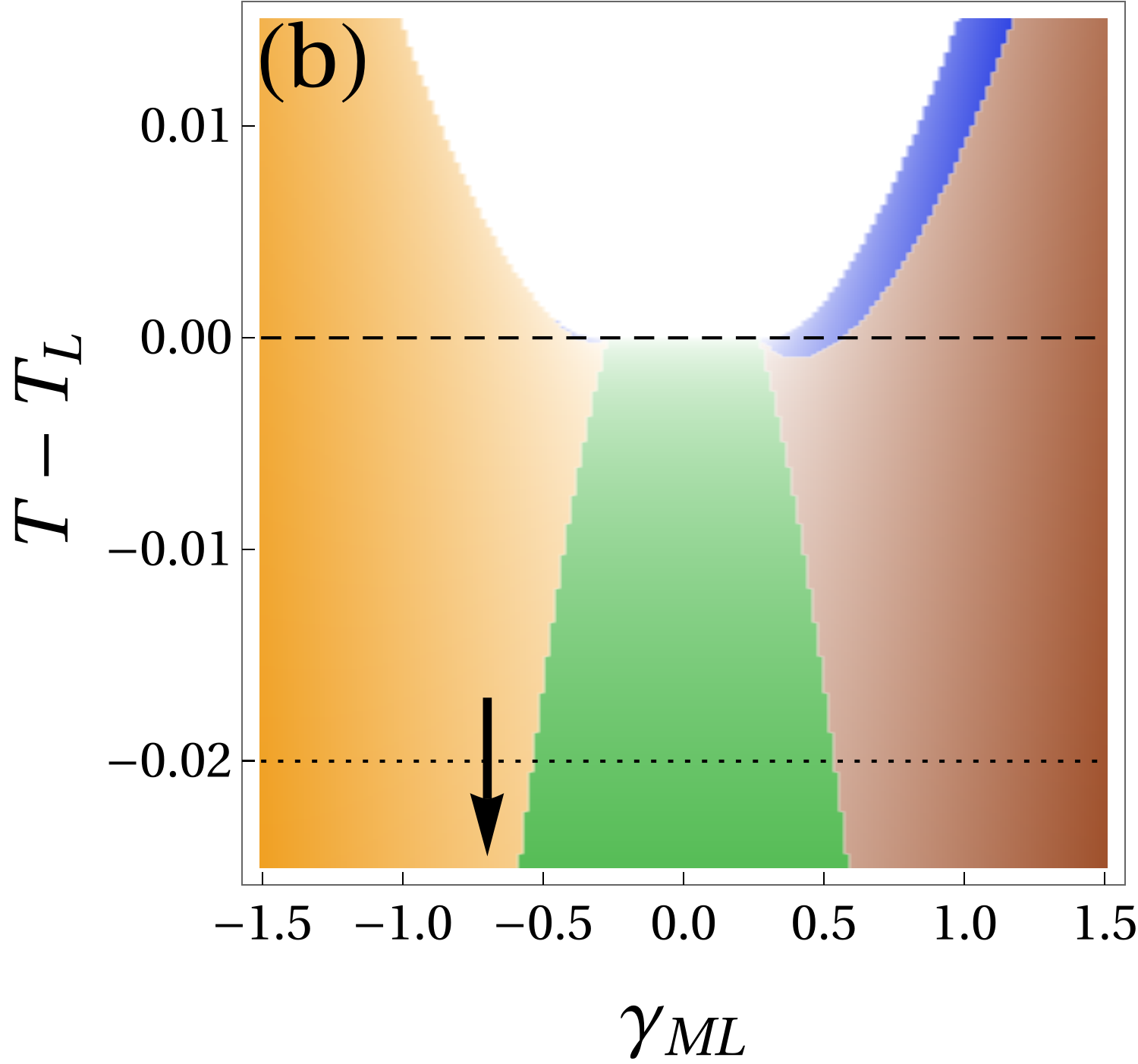}\includegraphics[width=0.245\textwidth]{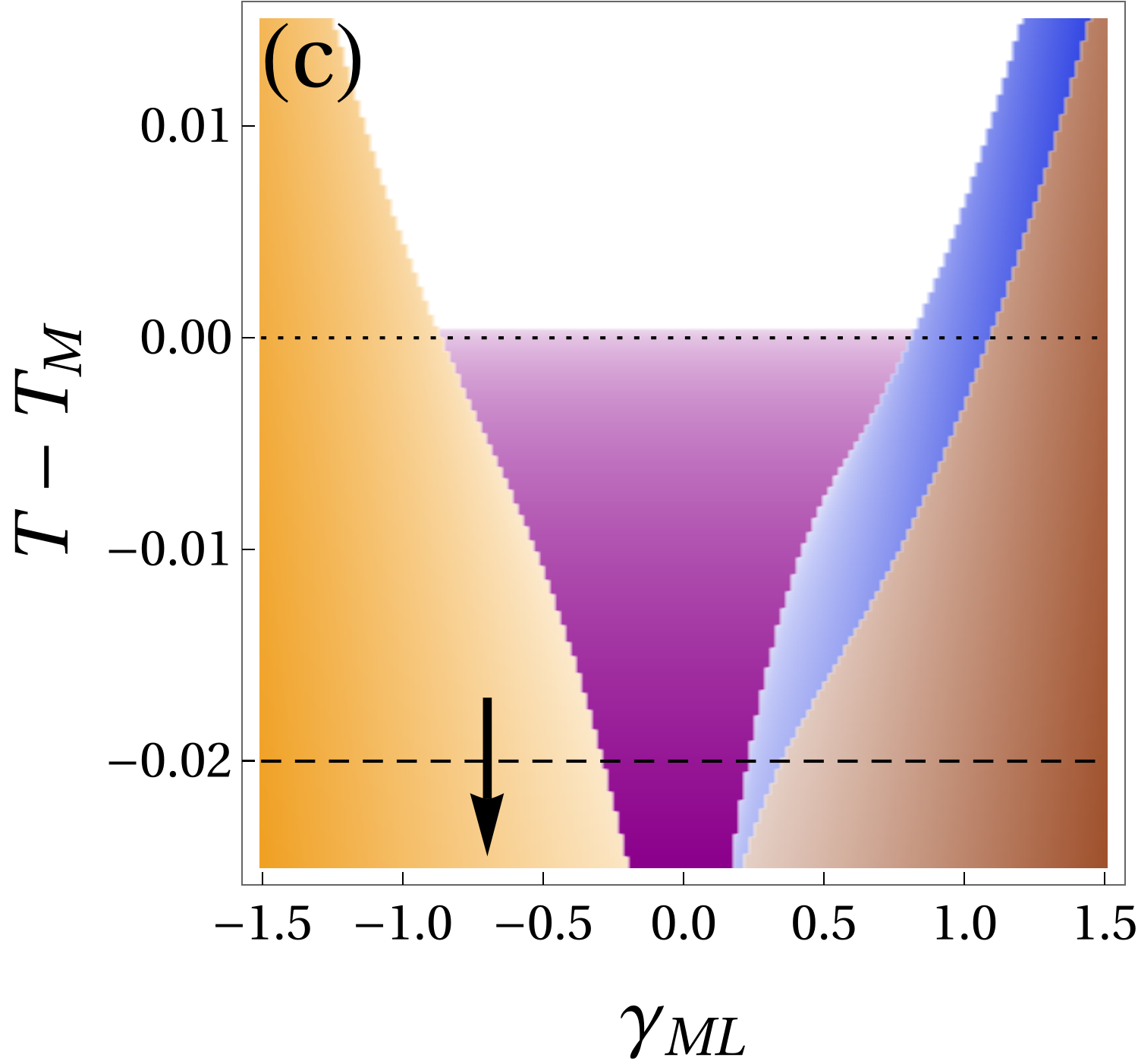}\includegraphics[width=0.245\textwidth]{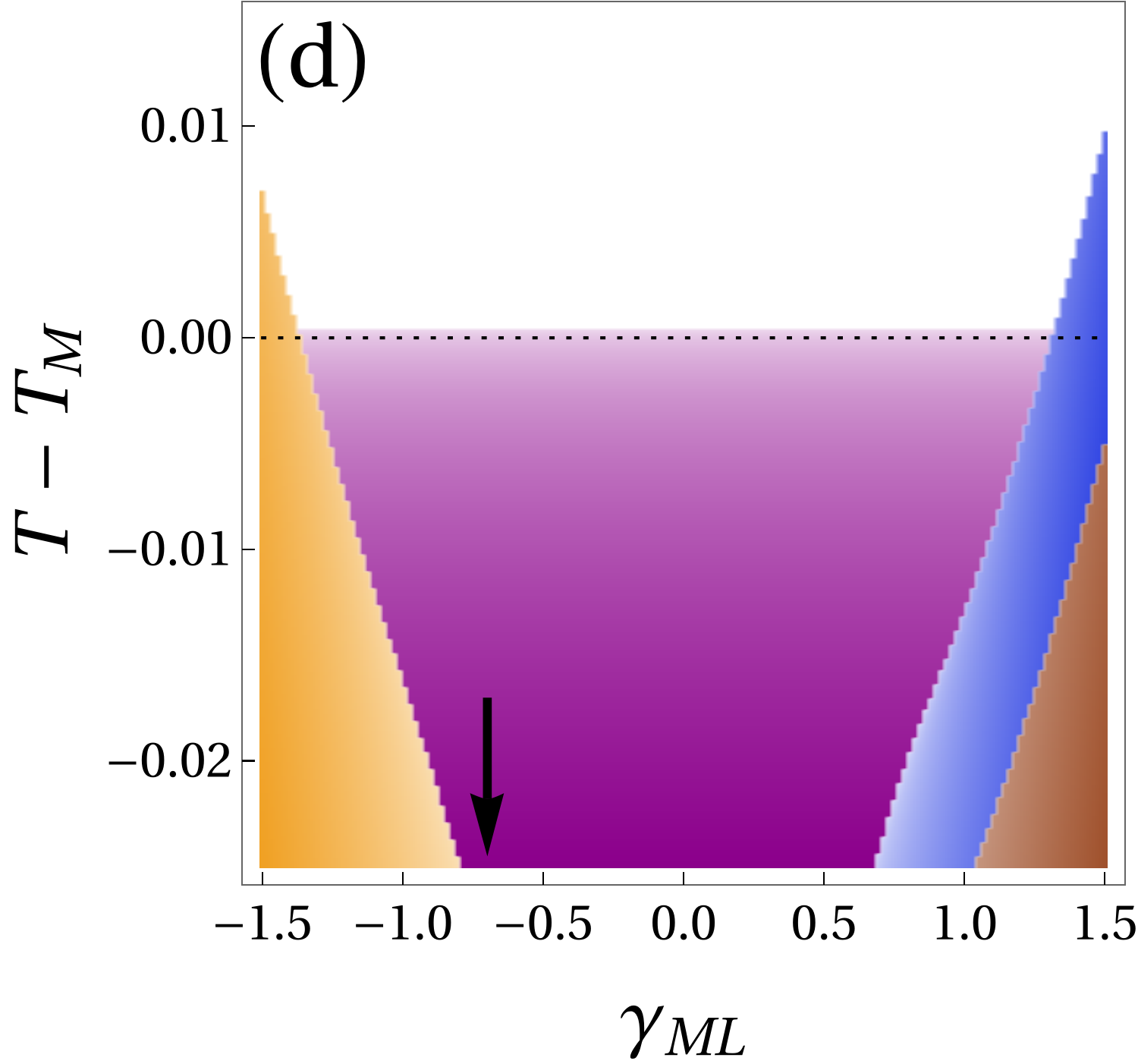}\\
\vspace{3mm}
\includegraphics[width=0.245\textwidth]{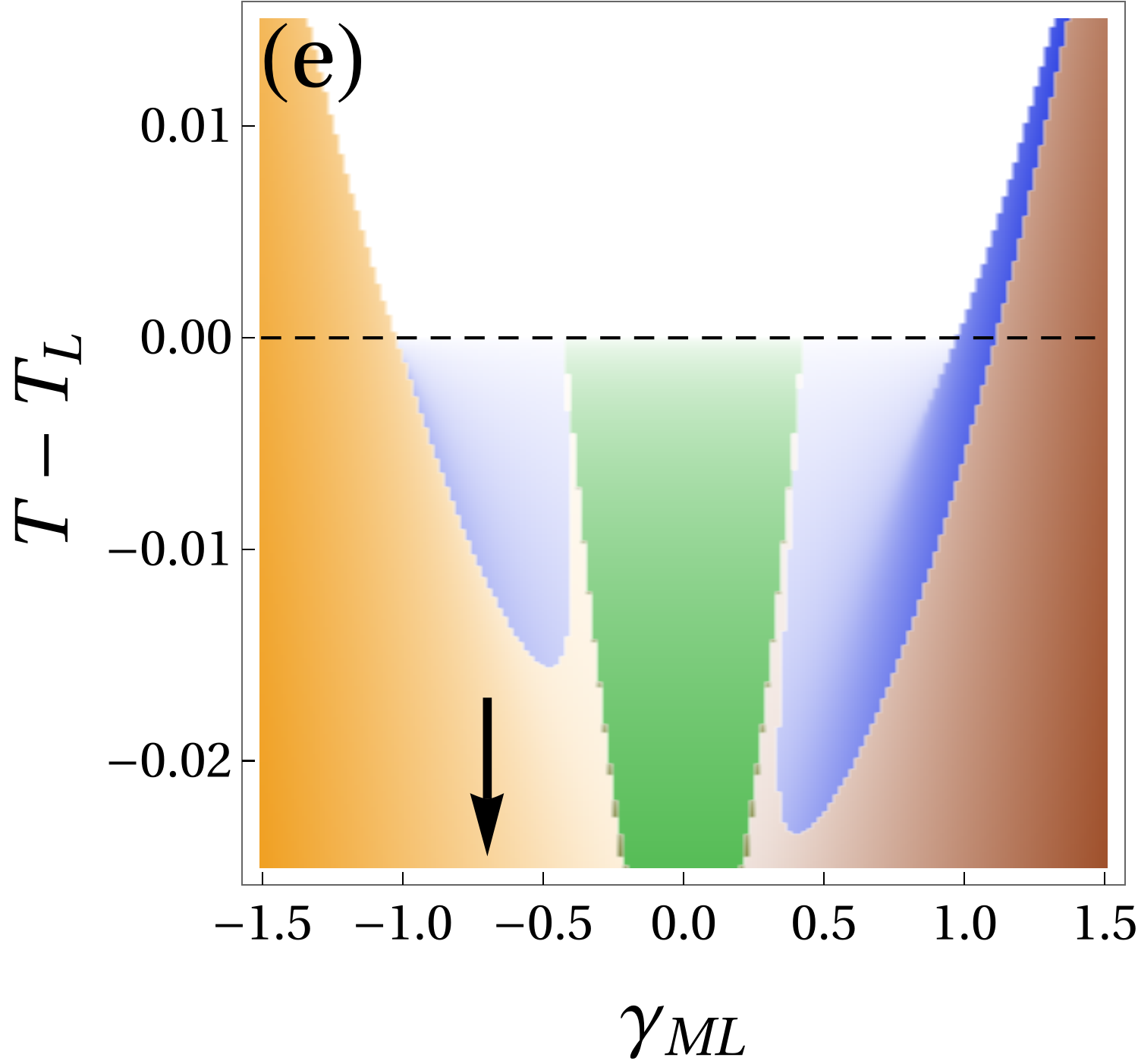}\includegraphics[width=0.245\textwidth]{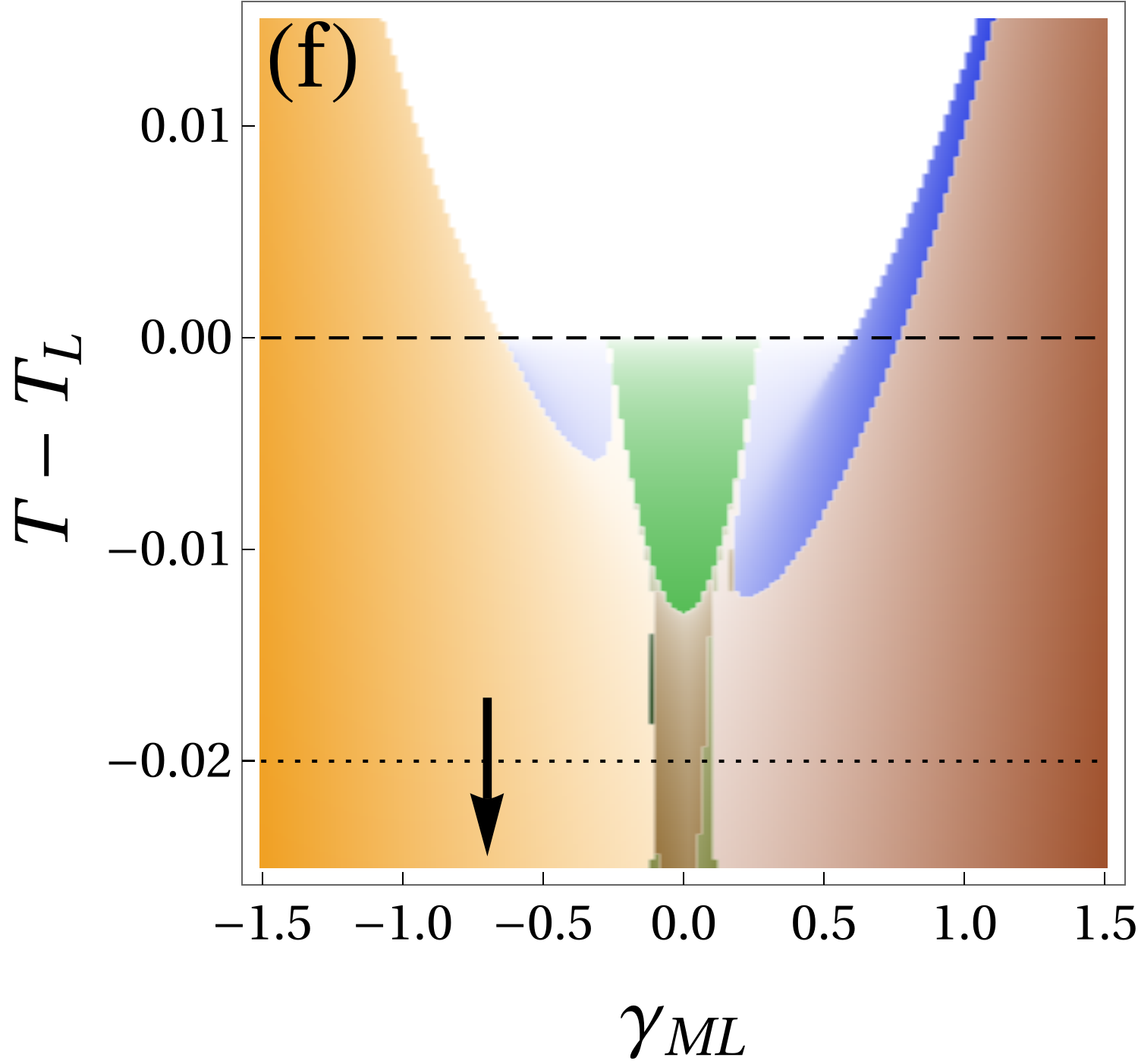}\includegraphics[width=0.245\textwidth]{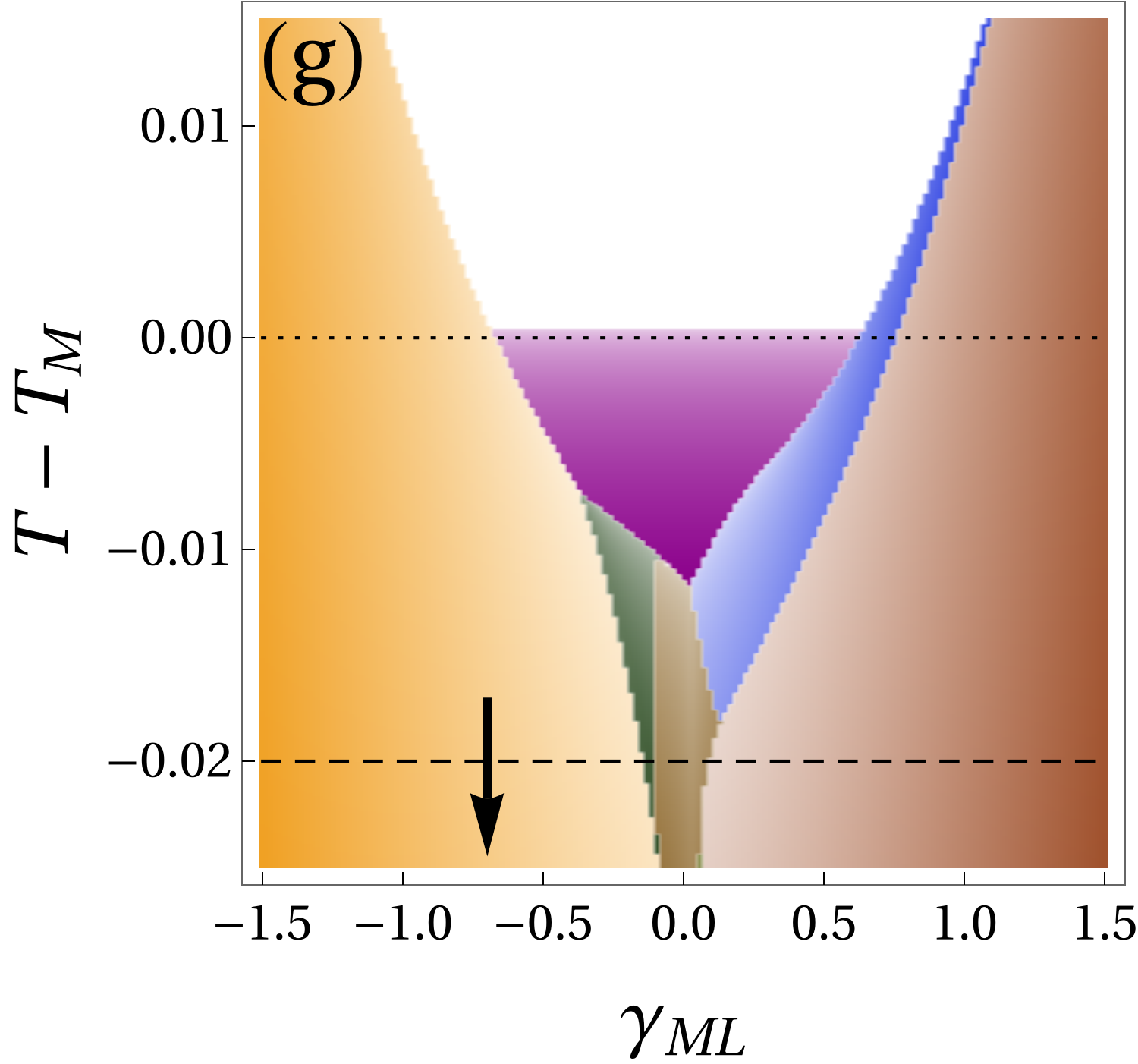}\includegraphics[width=0.245\textwidth]{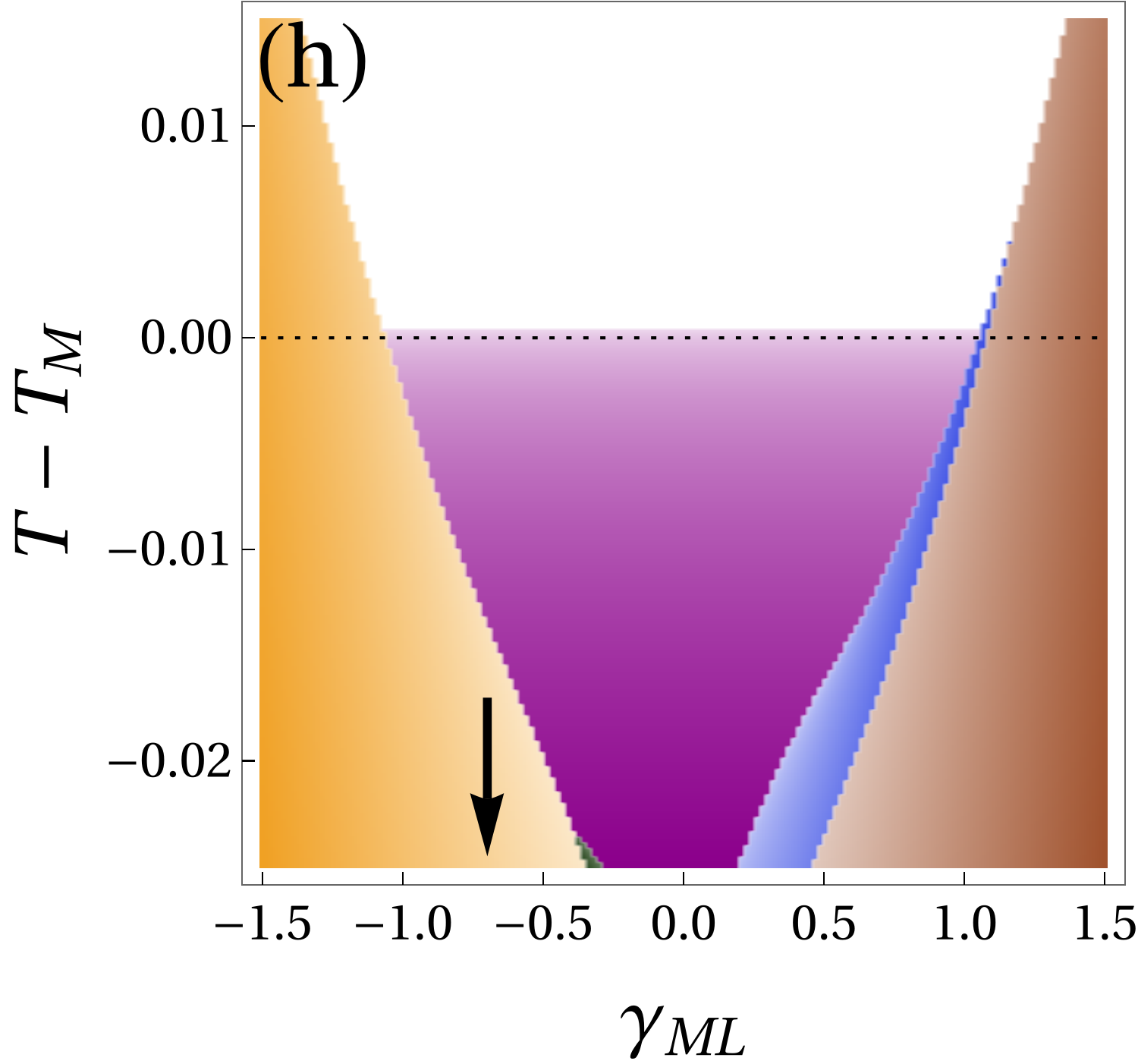}
\caption{\label{fig:phase_diagrams_MLL_pref} Phase diagrams for $u_L=0.1$ (a)--(d) [see red dotted line in Fig.~\ref{fig:phase_diagrams_L_variation}(b) and (d)] and $\lambda_{ML}^{(3)}=-1.6$ (e)--(h) [see red dotted line in Figs.~\ref{fig:phase_diagrams_ML_variation}(c) and (f)] for different values of $T_L - T_M$, from $T_L-T_M=-0.05$ (a), (e) to $T_L - T_M = 0.05$ (d), (h). $T_M$ and $T_L$ are denoted by the dotted and dashed black lines respectively. The remaining parameters are as presented in Table~\ref{tab:parameters}.}
\end{figure*}

The above composite order parameters are useful to determine whether a specific CDW phase breaks the sixfold rotational symmetry. Referring to Table \ref{tab:phase_summary_uncoupled}, as one might have expected, the stripe single-$\mbf{Q}_M$ and single-$\mbf{Q}_L$ phases lead to finite $\Phi_M$ and $\Phi_L$, respectively; however, the unit cell does not double in all three directions. On the other hand, the ``pure" triple-$\mbf{Q}_M$ and triple-$\mbf{Q}_L$ phases -- planar tri-hexagonal, planar Star-of-David, and superimposed tri-hexagonal Star-of-David -- do not result in a finite $\Phi_M$ or $\Phi_L$, and thus do not break the six-fold rotational symmetry of the system. 

Meanwhile, the coupling between $M_i$ and $L_i$ allows for coupled CDW phases that do break the six-fold rotational symmetry and double the unit cell in all directions, being consistent with experimental observations. Referring to Table \ref{tab:phase_summary_coupled}, $\Phi_M$, $\Phi_L$, and $\Phi_{ML}$ are all nonzero for the two types of single-$\mbf{Q}_M$/double-$\mbf{Q}_L$ phases, described by $(M 0 0)+(0 L L)$ (staggered tri-hexagonal) and $(\overline{M} 0 0)+(0 L L)$ (staggered star-of-David). On the other hand, the triple-$\mbf{Q}_M$/triple-$\mbf{Q}_L$ superimposed tri-hexagonal Star-of-David phase, given by  $(M M M)+(L L L)$, preserves the sixfold rotational symmetry.

There are other potential coupled CDW phases that also break $C_6$ symmetry, but we either do not see them or only see them as subleading instabilities that emerge inside another ordered state. Consider, for instance, the double-$\mbf{Q}_M$/single-$\mbf{Q}_L$ phase with $(M M 0)+(0 0 L)$ discussed in Ref.~\onlinecite{Ratcliff2021}; it has $\Phi_M, \Phi_L \neq 0$ but $\Phi_{ML}=0$. The reason we believe it does not show up in the phase diagrams is because it does not gain any energy from the trilinear term of $\mathcal{F}_{ML}$. Conversely, the triple-$\mbf{Q}_M$/triple-$\mbf{Q}_L$ phases described by $(M M \overline{M})+(L L L)$ and $(M M M)+(L L \overline{L})$ have $\Phi_M, \Phi_L = 0$ but $\Phi_{ML}\neq0$. While it gains energy from the cubic and quartic terms of $\mathcal{F}_{ML}$, it seems to not be able to compete with the other phases seen in our phase diagrams. This analysis also indicates that neither of these phases are likely to be leading instabilities of the system. The reason is because only a subset of the composite nematic order parameters $\Phi_{i}$ are nonzero. Symmetry imposes that, once one of the $\Phi_i$ is non-zero, all the other ones will become nonzero as well, since they all belong to the same irreducible representation $\Gamma_5^+$. But, clearly, the only way this can happen for, say, the $(M M \overline{M})+(L L L)$ phase, is by further making at least one of the three $M_i$ or $L_i$ components different from the others. This indicates that this phase is actually one of the mixed states shown in Table \ref{tab:phase_summary_mixed}, i.e. the $(M M \overline{M})+(L L L)$ phase necessarily mixes with other phase(s) to become $(M_1 M_1 \overline{M}_2)+(L_1 L_1 L_2)$.

Therefore, we conclude that the most promising candidates to explain the experimental observation of $C_6$ symmetry-breaking and $2\times2\times2$ unit cell enhancement are the staggered tri-hexagonal phase (orange) and staggered Star-of-David phase (brown) depicted in Fig. \ref{fig:mixed_order_examples}. The main question is whether they appear as a secondary instability inside the superimposed tri-hexagonal Star-of-David phase (blue) or as the leading instability of the system. Based on the results from Figs.~\ref{fig:increasing_coupling_phase_diagrams}--\ref{fig:phase_diagrams_ML_variation}, besides large and positive $\lambda_{ML}^{(1)}$ and $\lambda_{ML}^{(2)}$, two coefficients stand out as promoting the onset of the staggered tri-hexagonal phase as the leading instability: $u_L$ [see red dotted lines in Figs.~\ref{fig:phase_diagrams_L_variation}(b) and (d)] and $\lambda_{ML}^{(3)}$  [see red dotted lines in Figs. \ref{fig:phase_diagrams_ML_variation}(c) and (f)]. To further elucidate whether the parameter range where this state is the leading instability of the system can be enhanced, in Fig.~\ref{fig:phase_diagrams_MLL_pref} we show $\gamma_{ML}$-temperature phase diagrams starting with parameters corresponding to the red dotted lines of Figs.~\ref{fig:phase_diagrams_L_variation}(b) and (d) [arrows in panels (a)--(d)] and the red dotted lines of Figs. \ref{fig:phase_diagrams_ML_variation}(c) and (f) [arrows in panels (e)--(h)]. In particular, we interpolate between the $L$-dominated case (left most panels, $T_L \gg T_M$) and the $M$-dominated case (right most panels, $T_M \gg T_L$). While these parameter choices are obviously not exhaustive, they do represent the cases that we found to be the most favorable for the staggered tri-hexagonal phase.

From Fig.~\ref{fig:phase_diagrams_MLL_pref}, it is clear that the staggered tri-hexagonal phase (orange) only occurs as a leading instability for large (in magnitude) negative values of $\gamma_{ML}$. For smaller (in magnitude) negative values, it can appear only as a secondary transition inside another phase. The center of the phase diagrams remain dominated by the ``pure" alternating stripe (green) and the planar Star-of-David (dark purple) phases. Interestingly, there is an asymmetry in the phase diagram since even for large positive $\gamma_{ML}$, the staggered Star-of-David phase (brown) does not become the leading instability of the system. This is a consequence of the fact that we chose $\gamma_M > 0$. Had we considered $\gamma_M < 0$ instead, the superimposed tri-hexagonal Star-of-David phase (blue) envelope appearing above the staggered Star-of-David phase (brown) would cover the staggered tri-hexagonal phase (orange) instead, such that the staggered Star-of-David phase would become the leading instability for sufficiently large positive $\gamma_{ML}$. 

It is interesting to note how the difference in the bare transition temperatures, $T_L$ and $T_M$, affects the shape of the phase diagrams. As they correspond to the condensation of order parameters belonging to different irreducible representations, they are not guaranteed to be close. In both cases ($T_L > T_M$ and $T_M > T_L$), upon moving to the left of the phase diagram along the negative $\gamma_{ML}$ axis, the staggered tri-hexagonal phase (orange) first emerges as a secondary instability before becoming the leading instability. However, when $T_L > T_M$ [panels (a)--(b) and (e)--(f)], it is generally preceded by the superimposed tri-hexagonal Star-of-David phase (blue), whereas when $T_M > T_L$ [panels (c)--(d) and (g)--(h)], it is achieved via a direct transition from the ``pure" Star-of-David phase (dark purple). More generally, it is clear that for both $M$ and $L$ to condense simultaneously, the cubic coefficient $\gamma_{ML}$ must overcome the splitting between $T_L$ and $T_M$. Whether this is facilitated in $A$V$_3$Sb$_5$ by an intrinsically large trilinear coefficient or by an accidental near-degeneracy of $T_L$ and $T_M$ requires a microscopic model and is beyond the scope of this work. 

We finish this section by noting that it is possible for CDW fluctuations to lead to the condensation of the composite nematic order parameters in Eqs. (\ref{eq:nematics}) even when $M_i = L_i = 0$, i.e. above the CDW transition \cite{Fernandes2019}. Such a vestigial nematic phase arising from partially-melted CDW order has been recently proposed in the case of tetragonal Ni-based pnictide superconductors \cite{Eckberg2020}. To the best of our knowledge, no evidence of a separate nematic transition has been reported in $A$V$_3$Sb$_5$.

\section{Summary and Conclusions}\label{sec:conclusions}

In this paper, we derived the Landau free energy appropriate for studying coupled CDWs with wave-vectors at both the $M$ and $L$ points of the hexagonal BZ. In the absence of coupling between the $M_i$ and $L_i$ order parameters, the phase diagrams feature very little variation. For the $M$-point CDW, due to the presence of a trilinear term, only a triple-$\mbf{Q}_M$ phase, corresponding to either the planar tri-hexagonal or planar Star-of-David bond configuration, occurs as a leading instability. Because a cubic term is not allowed for the $L$-point CDW, the phase diagram features both a single-$\mbf{Q}_L$ (alternating stripe) and a triple-$\mbf{Q}_L$ (alternating tri-hexagonal Star-of-David) phase. This simple picture changes dramatically once the $M$- and $L$-point CDWs are coupled. Crucially, owing to the specific wave-vectors of the order parameters at $M$ and $L$, both cubic and quartic coupling terms are allowed. The trilinear coupling, $\gamma_{ML}$, plays a primary role in determining the leading instability as evidenced from Figs.~\ref{fig:increasing_coupling_phase_diagrams}--\ref{fig:phase_diagrams_MLL_pref}.

Importantly, even if the transition temperatures $T_L$ and $T_M$ of the pure CDW states are not extremely close, a large enough (in magnitude) trilinear coupling $\gamma_{ML}$ is able to stabilize a coupled $M$-$L$ CDW state. In this case, while for most of the parameter space studied here the leading coupled instability is the triple-$\mbf{Q}_M$/triple-$\mbf{Q}_L$ state dubbed superimposed tri-hexagonal Star-of-David phase, we also find a range of parameters for which the leading instability is the single-$\mbf{Q}_M$/double-$\mbf{Q}_L$ states dubbed staggered tri-hexagonal and staggered Star-of-David phases. These phases are interesting because they not only double the size of the unit cell in all directions, but they also break the sixfold rotational symmetry of the lattice via the condensation of secondary composite nematic order parameters. These two features are consistent with several experimental observations regarding the CDW state of $A$V$_3$Sb$_5$ kagome metals \cite{Jiang2020,Zhao2021b,Chen2021,Li2021_rotation}.

This phenomenological analysis raises interesting questions that deserve further experimental investigation. For instance, a direct transition to one of the single-$\mbf{Q}_M$/double-$\mbf{Q}_L$ states seems to require specially tuned Landau parameters. Most commonly, this state onsets inside either the triple-$\mbf{Q}_M$/triple-$\mbf{Q}_L$ state or one of the ``pure" triple-$\mbf{Q}_M$ states (i.e. the planar tri-hexagonal and planar Star-of-David phases). Therefore, it is crucial to experimentally establish whether there is a single or multiple CDW transitions in $A$V$_3$Sb$_5$. This will provide important constraints on the Landau parameters. Similarly, it will be important to determine whether the breaking of sixfold rotational symmetry takes place above, simultaneously with, or below the first CDW transition. Which of these scenarios is realized depends roughly on the size of the trilinear coupling $\gamma_{ML}$ relative to the energy difference between the pure $M$-point and pure $L$-point CDW states. Given the sensitivity of the phase diagram to these two parameters, it is conceivable that $A$V$_3$Sb$_5$ compounds with different alkali metals $A$ may show distinct CDW phase diagrams.

Our results also provide insights about the microscopic mechanism responsible for the onset of CDW order. Our DFT calculations reveal two interesting features associated with the CDW state: (i) three different phonon modes along the $U$ line (which includes the $M$ and $L$ points) are unstable; and (ii) the corresponding imaginary-valued frequencies display a strong dependence on the electronic temperature (as signaled by the Fermi surface smearing). Taken together, they provide strong support for the scenario in which the CDW is not the consequence of a pure lattice instability, but is driven by the electronic degrees of freedom. More specifically, because all points along the $U$ line share the same in-plane momentum, this suggests a primary role of the in-plane electronic dispersion and/or interactions in promoting the CDW transition. Importantly, an electronically-driven CDW instability is likely to have a more significant intertwining with the superconducting state. This is supported by recent transport measurements indicating a strong competition between the CDW and SC phases~\cite{Ni2021}.

\begin{acknowledgments}

We acknowledge fruitful discussions with N. Ni. M.H.C acknowledges support from the Carlsberg foundation. T.B. was supported by the NSF CAREER grant DMR-2046020. B. M. A. acknowledges support from the Independent Research Fund Denmark grant number 8021-00047B. R.M.F. was supported by the U.S. Department of Energy, Office of Science, Basic Energy Sciences, Materials Science and Engineering Division, under Award No. DE-SC0020045.
	
\end{acknowledgments}


\begin{thebibliography}{00}

\bibitem{Balents2010} L. Balents. \textit{Spin liquids in frustrated magnets}. Nature {\bf 464}, 199 (2010).

\bibitem{Kang2019} Mingu Kang, Linda Ye, Shiang Fang, Jhih-Shih You, Abe Levitan, Minyong Han, Jorge I. Facio, Chris Jozwiak, Aaron Bostwick, Eli Rotenberg, Mun K. Chan, Ross D. McDonald, David Graf, Konstantine Kaznatcheev, Elio Vescovo, David C. Bell, Efthimios Kaxiras, Jeroen van den Brink, Manuel Richter, Madhav Prasad Ghimire, Joseph G. Checkelsky, and Riccardo Comin. \textit{Dirac fermions and flat bands in the ideal kagome metal FeSn}. Nat. Mater. {\bf 19}, 163 (2019).

\bibitem{Kang2020} Mingu Kang, Shiang Fang, Linda Ye, Hoi Chun Po, Jonathan Denlinger, Chris Jozwiak, Aaron Bostwick, Eli Rotenberg, Efthimios Kaxiras, Joseph G. Checkelsky, and Riccardo Comin. \textit{Topological flat bands in frustrated kagome lattice CoSn}. Nat. Commun. {\bf 11}, 4004 (2020).

\bibitem{Meier2020} William R. Meier, Mao-Hua Du, Satoshi Okamoto, Narayan Mohanta, Andrew F. May, Michael A. McGuire, Craig A. Bridges, German D. Samolyuk, and Brian C. Sales. \textit{Flat bands in the CoSn-type compounds}. Phys. Rev. B {\bf 102}, 075148 (2020).

\bibitem{Guo2009} H.-M. Guo and M. Franz. \textit{Topological insulator on the kagome lattice}. Phys. Rev. B {\bf 80}, 113102 (2009). 

\bibitem{Mazin2014} I. I. Mazin, Harald O. Jeschke, Frank Lechermann, Hunpyo Lee, Mario Fink, Ronny Thomale, and Roser Valent{\'i}. \textit{Theoretical prediction of a strongly correlated Dirac metal}. Nat. Commun. {\bf 5}, 4261 (2014).

\bibitem{Nandkishore2012} R. Nandkishore, L. S. Levitov, and A. V. Chubukov. \textit{Chiral superconductivity from repulsive interactions in doped graphene}. Nat. Phys. {\bf 8}, 158 (2012).

\bibitem{Kiesel2012} Maximilian L. Kiesel and Ronny Thomale. \textit{Sublattice interference in the kagome Hubbard model}. Phys. Rev. B {\bf 86}, 121105(R) (2012).

\bibitem{Kiesel2013} Maximilian L. Kiesel, Christian Platt, and Ronny Thomale. \textit{Unconventional Fermi Surface Instabilities in the Kagome Hubbard Model}. Phys. Rev. Lett. {\bf 110}, 126405 (2013).

\bibitem{Nandkishore2014} Rahul Nandkishore, Ronny Thomale, and Andrey V. Chubukov. \textit{Superconductivity from weak repulsion in hexagonal lattice systems}. Phys. Rev. B {\bf 89}, 144501 (2014).

\bibitem{Ortiz2019} Brenden R. Ortiz, L{\'i}dia C. Gomes, Jennifer R. Morey, Michal Winiarski, Mitchell Bordelon, John S. Mangum, Iain W. H. Oswald, Jose A. Rodriguez-Rivera, James R. Neilson, Stephen D. Wilson, Elif Ertekin, Tyrel M. McQueen, and Eric S. Toberer. \textit{New kagome prototype materials: discovery of KV$_3$Sb$_5$, RbV$_3$Sb$_5$, and CsV$_3$Sb$_5$}. Phys. Rev. Mat. {\bf 3}, 094407 (2019).

\bibitem{Ortiz2020} Brenden R. Ortiz, Samuel M. L. Teicher, Yong Hu, Julia L. Zuo, Paul M. Sarte, Emily C. Schueller, A. M. Milinda Abeykoon, Matthew J. Krogstad, Stephan Rosenkranz, Raymond Osborn, Ram Seshadri, Leon Balents, Junfeng He, and Stephen D. Wilson. \textit{CsV$_3$Sb$_5$: A $\mathbb{Z}_2$ Topological Kagome Metal with a Superconducting Ground State}. Phys. Rev. Lett. {\bf 125}, 247002 (2020).

\bibitem{Ortiz2021} Brenden R. Ortiz, Paul M. Sarte, Eric M. Kenney, Michael J. Graf, Samuel M. L. Teicher, Ram Seshadri, and Stephen D. Wilson. \textit{Superconductivity in the Z2 kagome metal KV$_3$Sb$_5$}. Phys. Rev. Mat. {\bf 5}, 034801 (2021).

\bibitem{Yin2021} Qiangwei Yin, Zhijun Tu, Chunsheng Gong, Yang Fu, Shaohua Yan, and Hechang Lei. \textit{Superconductivity and normal-state properties of kagome metal RbV$_3$Sb$_5$ single crystals}. Chin. Phys. Lett. {\bf 38}, 037403 (2021).

\bibitem{Song2021} Yanpeng Song, Tianping Ying, Xu Chen, Xu Han, Yuan Huang, Xianxin Wu, Andreas P. Schnyder, Jian-Gang Guo, and Xiaolong Chen. \textit{Enhancement of superconductivity in hole-doped CsV$_3$Sb$_5$ thin flakes}. arXiv:2105.09898 (2021).

\bibitem{Du2021} Feng Du, Shuaishuai Luo, Brenden R. Ortiz, Ye Chen, Weiyin Duan, Dongting Zhang, Xin Lu, Stephen D. Wilson, Yu Song, and Huiqiu Yuan. \textit{Interplay between charge order and superconductivity in the kagome metal KV$_3$Sb$_5$}. arXiv:2102.10959 (2021).

\bibitem{Zhao2021a} C. C. Zhao, L. S. Wang, W. Xia, Q. W. Yin, J. M. Ni, Y. Y. Huang, C. P. Tu, Z. C. Tao, Z. J. Tu, C. S. Gong, H. C. Lei, Y. F. Guo, X. F. Yang, and S. Y. Li. \textit{Nodal superconductivity and superconducting domes in the topological Kagome metal CsV$_3$Sb$_5$}. arXiv:2102.08356 (2021).

\bibitem{Zhang2021_pressure} Zhuyi Zhang, Zheng Chen, Ying Zhou, Yifang Yuan, Shuyang Wang, Jing Wang, Haiyang Yang, Chao An, Lili Zhang, Xiangde Zhu, Yonghui Zhou, Xuliang Chen, Jianhui Zhou, and Zhaorong Yang. \textit{Pressure-induced reemergence of superconductivity in the topological kagome metal CsV$_3$Sb$_5$}. Phys. Rev. B {\bf 103}, 224513 (2021).

\bibitem{Chen_pressure_2021} K.-Y. Chen, N.-N. Wang, Q.-W. Yin, Y.-H. Gu, K. Jiang, Z.-J. Tu, C.-S. Gong, Y. Uwatoko, J.-P. Sun, H.-C. Lei, J.-P. Hu, and J.-G. Cheng \textit{Double superconducting dome and triple enhancement of Tc in the kagome superconductor CsV$_3$Sb$_5$ under high pressure}. Phys. Rev. Lett. {\bf 126}, 247001 (2021).

\bibitem{Zhu_pressure_2021} C. C. Zhu, X. F. Yang, W. Xia, Q. W. Yin, L. S. Wang, C. C. Zhao, D. Z. Dai, C. P. Tu, B. Q. Song, Z. C. Tao, Z. J. Tu, C. S. Gong, H. C. Lei, Y. F. Guo, S. Y. Li. \textit{Double-dome superconductivity under pressure in the V-based Kagome metals AV3Sb5 (A = Rb and K)}. arXiv:2104.14487  (2021).

\bibitem{Uykur_2021} E. Uykur, B. R. Ortiz, S. D. Wilson, M. Dressel, A. A. Tsirlin. \textit{Optical detection of charge-density-wave instability in the non-magnetic kagome metal KV3Sb5}. arXiv:2103.07912 (2021).


\bibitem{Zhou_HHWen_2021} Xiaoxiang Zhou, Yongkai Li, Xinwei Fan, Jiahao Hao, Yaomin Dai, Zhiwei Wang, Yugui Yao, Hai-Hu Wen. \textit{Origin of the Charge Density Wave in the Kagome Metal CsV3Sb5 as Revealed by Optical  Spectroscopy}. arXiv:2104.01015 (2021).

\bibitem{Jiang2020} Yu-Xiao Jiang \textit{et al}. \textit{Unconventional chiral charge order in kagome superconductor KV$_3$Sb$_5$}. Nat. Mater. (2021).

\bibitem{Zhao2021b} He Zhao, Hong Li, Brenden R. Ortiz, Samuel M. L. Teicher, Taka Park, Mengxing Ye, Ziqiang Wang, Leon Balents, Stephen D. Wilson, Ilija Zeljkovic. \textit{Cascade of correlated electron states in a kagome superconductor CsV$_3$Sb$_5$}. arXiv:2103.03118 (2021).

\bibitem{Liang2021} Zuowei Liang, Xingyuan Hou, Fan Zhang, Wanru Ma, Ping Wu, Zongyuan Zhang, Fanghang Yu, J. -J. Ying, Kun Jiang, Lei Shan, Zhenyu Wang, and X. -H. Chen. \textit{Three-dimensional charge density wave and robust zero-bias conductance peak inside the superconducting vortex core of a kagome superconductor CsV$_3$Sb$_5$}. arXiv:2103.04760 (2021).

\bibitem{Chen2021} Hui Chen, Haitao Yang, Bin Hu, Zhen Zhao, Jie Yuan, Yuqing Xing, Guojian Qian, Zihao Huang, Geng Li, Yuhan Ye, Qiangwei Yin, Chunsheng Gong, Zhijun Tu, Hechang Lei, Shen Ma, Hua Zhang, Shunli Ni, Hengxin Tan, Chengmin Shen, Xiaoli Dong, Binghai Yan, Ziqiang Wang, and Hong-Jun Gao. \textit{Roton pair density wave and unconventional strong-coupling superconductivity in a topological kagome metal}. arXiv:2103.09188 (2021).


\bibitem{Li2021_observation} H. X. Li, T. T. Zhang, Y.-Y. Pai, C. Marvinney, A. Said, T. Yilmaz, Q. Yin, C. Gong, Z. Tu, E. Vescovo, R. G. Moore, S. Murakami, H. C. Lei, H. N. Lee, B. Lawrie, and H. Miao. \textit{Observation of Unconventional Charge Density Wave without Acoustic Phonon Anomaly in Kagome Superconductors AV$_3$Sb$_5$ (A=Rb,Cs)}. arXiv:2103.09769 (2021).

\bibitem{Ortiz2021b} Brenden R. Ortiz, Samuel M. L. Teicher, Linus Kautzsch, Paul M. Sarte, Jacob P. C. Ruff, Ram Seshadri, and Stephen D. Wilson. \textit{Fermi surface mapping and the nature of charge density wave order in the kagome superconductor CsV$_3$Sb$_5$}. arXiv:2104.07230 (2021).

\bibitem{Luo2021} Yang Luo, Shuting Peng, Samuel M. L. Teicher, Linwei Huai, Yong Hu, Brenden R. Ortiz, Zhiyuan Wei, Jianchang Shen, Zhipeng Ou, Bingqian Wang, Yu Miao, Mingyao Guo, M. Shi, Stephen D. Wilson, J.-F. He. \textit{Distinct band reconstructions in kagome superconductor CsV3Sb}. arXiv:2106.01248 (2021).

\bibitem{Ni2021} T. Qian, M. H. Christensen, C. Hu, A. Saha, B. M. Andersen, R. M. Fernandes, T. Birol, and N. Ni. \textit{Revealing the competition between charge-density wave and superconductivity in CsV$_3$Sb$_5$ through uniaxial strain}. arXiv:2107.04545 (2021).

\bibitem{Li2021_rotation} Hong Li, He Zhao, Brenden R. Ortiz, Takamori Park, Mengxing Ye, Leon Balents, Ziqiang Wang, Stephen D. Wilson, Ilija Zeljkovic. \textit{Rotation symmetry breaking in the normal state of a kagome superconductor KV3Sb5}. arXiv:2104.08209 (2021).

\bibitem{Tan2021} Hengxin Tan, Yizhou Liu, Ziqiang Wang, Binghai Yan. \textit{Charge density waves and electronic properties of superconducting kagome metals}. arXiv:2103.06325 (2021).

\bibitem{Ratcliff2021} Noah Ratcliff, Lily Hallett, Brenden R. Ortiz, Stephen D. Wilson, and John W. Harter. \textit{Coherent phonon spectroscopy and interlayer modulation of charge density wave order in the kagome metal CsV$_3$Sb$_5$}. arXiv:2104.10138 (2021).

\bibitem{Shumiya2021} Nana Shumiya, Md Shafayat Hossain, Jia-Xin Yin, Yu-Xiao Jiang, Brenden R. Ortiz, Hongxiong Liu, Youguo Shi, Qiangwei Yin, Hechang Lei, Songtian S. Zhang, Guoqing Chang, Qi Zhang, Tyler A. Cochran, Daniel Multer, Maksim Litskevich, Zi-Jia Cheng, Xian P. Yang, Zurab Guguchia, Stephen D. Wilson, and M. Zahid Hasan. \textit{Tunable chiral charge order in kagome superconductor RbV$_3$Sb$_5$}. arXiv:2105.00550 (2021).

\bibitem{Setty2021} Chandan Setty, Haoyu Hu, Lei Chen, and Qimiao Si. \textit{Electron correlations and T-breaking density wave order in a $\mathbb{Z}_2$ kagome metal}. arXiv:2105.15204 (2021).

\bibitem{Mielke2021} C. Mielke III, D. Das, J.-X. Yin, H. Liu, R. Gupta, C.N. Wang, Y.-X. Jiang, M. Medarde, X. Wu, H.C. Lei, J.J. Chang, P. Dai, Q. Si, H. Miao, R. Thomale, T. Neupert, Y. Shi, R. Khasanov, M.Z. Hasan, H. Luetkens, and Z. Guguchia. \textit{Time-reversal symmetry-breaking charge order in a correlated kagome superconductor}. arXiv:2106.13443 (2021).

\bibitem{Park2021} Takamori Park, Mengxing Ye, Leon Balents. \textit{Electronic instabilities of kagome metals: saddle points and Landau theory}. arXiv:2104.08425 (2021).


\bibitem{Kresse1993} G. Kresse and J. Hafner. \textit{Ab initio molecular dynamics for liquid metals}. Phys. Rev. B {\bf 47}, 558(R) (1993).

\bibitem{Kresse1996CMS} G. Kresse, J. Furthm{\"u}ller. \textit{Efficiency of ab-initio total energy calculations for metals and semiconductors using a plane-wave basis set}. Comp. Mat. Sci. {\bf 6}, 15 (1996).

\bibitem{Kresse1996PRB} G. Kresse and J. Furthm{\"u}ller. \textit{Efficient iterative schemes for ab initio total-energy calculations using a plane-wave basis set}. Phys. Rev. B {\bf 54}, 11169 (1996).

\bibitem{Perdew2008} John P. Perdew, Adrienn Ruzsinszky, G{\'a}bor I. Csonka, Oleg A. Vydrov, Gustavo E. Scuseria, Lucian A. Constantin, Xiaolan Zhou, and Kieron Burke. \textit{Restoring the Density-Gradient Expansion for Exchange in Solids and Surfaces}. Phys. Rev. Lett. {\bf 100}, 136406 (2008).

\bibitem{Cho2021} Soohyun Cho, Haiyang Ma, Wei Xia, Yichen Yang, Zhengtai Liu, Zhe Huang, Zhicheng Jiang, Xiangle Lu, Jishan Liu, Zhonghao Liu, Jinfeng Jia, Yanfeng Guo, Jianpeng Liu, and Dawei Shen. \textit{Emergence of new van Hove singularities in the charge density wave state of a topological kagome metal RbV$_3$Sb$_5$}. arXiv:2105.05117 (2021).

\bibitem{Kang2021} Mingu Kang, Shiang Fang, Jeong-Kyu Kim, Brenden R. Ortiz, Jonggyu Yoo, Byeong-Gyu Park, Stephen D. Wilson, Jae-Hoon Park, and Riccardo Comin. \textit{Twofold van Hove singularity and origin of charge order in topological kagome superconductor CsV$_3$Sb$_5$}. arXiv:2105.01689 (2021).

\bibitem{Hu2021} Yong Hu, Xianxin Wu, Brenden R. Ortiz, Sailong Ju, Xinlong Han, J. Z. Ma, N. C. Plumb, Milan Radovic, Ronny Thomale, S. D. Wilson, Andreas P. Schnyder, and M. Shi. \textit{Rich Nature of Van Hove Singularities in Kagome Superconductor CsV$_3$Sb$_5$}. arXiv:2106.05922 (2021).

\bibitem{Hatch2003} D. M. Hatch and H. T. Stokes. \textit{INVARIANTS: program for obtaining a list of invariant polynomials of the order-parameter components associated with irreducible representations of a space group}. J. Appl. Cryst. {\bf 36}, 951 (2003).

\bibitem{Toledano1987} Pierre Toledano, and Jean-Claude Toledano. \textit{Landau Theory Of Phase Transitions, The: Application To Structural, Incommensurate, Magnetic And Liquid Crystal Systems}. World Scientific Publishing Company (1987).

\bibitem{Bousquet2008} Eric Bousquet, Matthew Dawber, Nicolas Stucki, C{\'e}line Lichtensteiger, Patrick Hermet, Stefano Gariglio, Jean-Marc Triscone, and Philippe Ghosez. \textit{Improper ferroelectricity in perovskite oxide artificial superlattices}. Nature {\bf 452}, 732 (2008).

\bibitem{Benedek2011} Nicole A. Benedek and Craig J. Fennie. \textit{Hybrid Improper Ferroelectricity: A Mechanism for Controllable Polarization-Magnetization Coupling}. Phys. Rev. Lett. {\bf 106}, 107204 (2011).

\bibitem{Etxebarria2010} I. Etxebarria, J. M. Perez-Mato, and P. Boullay. \textit{The Role of Trilinear Couplings in the Phase Transitions of Aurivillius Compounds}. Ferroelectrics {\bf 401}, 17 (2010).


\bibitem{Wang_Chubukov} Y. Wang and A. Chubukov. \textit{Charge-density-wave order with momentum $(2Q,0)$ and $(0,2Q)$ within the spin-fermion model: Continuous and discrete symmetry breaking, preemptive composite order, and relation to pseudogap in hole-doped cuprates}. Phys. Rev. B {\bf 90}, 035149 (2014).

\bibitem{Hecker2018} M. Hecker and J. Schmalian. \textit{Vestigial nematic order and superconductivity in the doped topological insulator Cu$_x$Bi$_2$Se$_3$}. npj Quant. Mat. {\bf 3}, 26 (2018).

\bibitem{Fernandes2020} R. M. Fernandes and J. W. F. Venderbos. \textit{Nematicity with a twist: Rotational symmetry breaking in a moir{\'e} superlattice}. Sci. Adv. {\bf 6}, eaba8834 (2020).


\bibitem{footnote_1} In other words, the matrices for the two irreducible representations of the space and the point groups are the same, except for the translational symmetry elements, which do not exist in the point group.


\bibitem{Fernandes2019} R. M. Fernandes, P. P. Orth, and J. Schmalian. \textit{Intertwined Vestigial Order in Quantum Materials: Nematicity and Beyond}. Ann. Rev. Cond. Mat. Phys. {\bf 10}, 133 (2019).

\bibitem{Eckberg2020} C. Eckberg \textit{et al.} \textit{Sixfold enhancement of superconductivity in a tunable electronic nematic system}. Nat. Phys. {\bf 16}, 346 (2020).




\end{thebibliography}
\end{document}